\documentclass{aa}  
\usepackage{graphicx}
\usepackage{caption}
\usepackage{subcaption}
\usepackage{placeins}
\usepackage{booktabs}
\usepackage{stfloats}
\usepackage{xcolor}
\usepackage[normalem]{ulem}
\usepackage{orcidlink}
\usepackage{txfonts}
\usepackage{xspace}
\usepackage{gensymb}
\usepackage{hyperref}
\usepackage{lipsum}
\hypersetup{
    colorlinks=true,
    linkcolor=blue,
    filecolor=magenta,      
    urlcolor=blue,
    citecolor=blue,
}

\def\hess{H.E.S.S.\xspace}

\def\ssftt{SS~433\xspace}
\def\xray{X-ray\xspace}
\def\vsgr{V4641~Sgr\xspace}
\def\grs{GRS~1915+105\xspace}
\def\maxi{MAXI~J1820+070\xspace}
\def\cyg{Cyg~X-1\xspace}
\def\cygt{Cyg~X-3\xspace}

\begin{document} 

   \title{The link between extreme particle acceleration and X-ray binary jets}

   \author{Laura Olivera-Nieto \inst{\ref{inst:API}, \ref{inst:GRAPPA}}\orcidlink{0000-0002-9105-0518}  
   \and Fraser J. Cowie \inst{\ref{inst:OXF}}\orcidlink{0009-0009-0079-2419}
   \and Sera Markoff\inst{\ref{inst:API}, \ref{inst:GRAPPA}, \ref{inst:CAM}}\orcidlink{0000-0001-9564-0876}
   \and Rob Fender\inst{\ref{inst:OXF}, \ref{inst:SA}}\orcidlink{0000-0002-5654-2744} 
   \and Justine Crook-Mansour\inst{\ref{inst:OXF}}\orcidlink{0000-0001-7466-1192}
   }
    \institute{Anton Pannekoek Institute for Astronomy, University of Amsterdam, Science Park 904, 1098 XH Amsterdam, The Netherlands\label{inst:API}
    \and Gravitation and Astroparticle Physics Amsterdam Institute, University of Amsterdam, Science Park 904, 1098 XH Amsterdam, The Netherlands\label{inst:GRAPPA}
    \and Department of Physics, University of Oxford, Denys Wilkinson Building, Keble Road, Oxford OX1 3RH, United Kingdom\label{inst:OXF}
    \and Institute of Astronomy, University of Cambridge, Madingley Road, Cambridge CB3 0HA, United Kingdom  \label{inst:CAM}
    \and Department of Astronomy, University of Cape Town, Private Bag X3, 7701 Rondebosch, South Africa \label{inst:SA}
    }

   \date{Received July 7th, 2026; accepted October 9th, 2026}

\offprints{
\protect\\\email{\href{mailto:l.oliveranieto@uva.nl}{l.oliveranieto@uva.nl}}
}
\abstract
{The detection of multi-TeV radiation from a handful of black-hole \xray binaries has positioned the source class as a promising candidate to explain the observed cosmic-ray flux in the PeV energy range.}
{ We aim to determine what distinguishes the systems detected in the gamma-ray range from the rest of the population.  }
{We build on existing catalogues of black-hole \xray binaries by including information on the radio and gamma-ray emission from these systems. We then compare the $>100$~TeV gamma-ray luminosity to different properties of the systems and evaluate the strength of the (rank) correlation ($\tau_K$) between them. Additionally, we compare the distribution of sources in our sample with the positions of unassociated gamma-ray sources in the 1LHAASO and 1HGPS catalogues to determine whether a correlation exists between them.}
{We find no correlation with the mass of the compact object, the companion mass or the inclination angle. 
Instead, we find evidence for a strong correlation ($\tau_K>0.5$) between the gamma-ray luminosity and the radio and hard-state \xray luminosities in our sample. The limited number of gamma-ray detections means that this correlation cannot yet be claimed as statistically significant.
We find indications that a high ($\gtrsim 20$\%) duty cycle (defined as fraction of time spent in outburst) might be a requirement for $>100$~TeV gamma-ray loudness. We find significant evidence ($>3\sigma$) that black-hole \xray binaries are located near unassociated gamma-ray sources in the region of the sky not covered by LHAASO. This evidence is only significant when comparing to \xray binaries detected in the radio band, not for the sample without a reported radio detection. }
{We conclude that radio-bright black-hole \xray binaries with a high duty cycle are more likely to be detected as gamma-ray sources than sources which are not detected in the radio band or those with low duty cycles. This result suggests that the particles responsible for the gamma-ray emission are associated with the jets, either accelerated within them or via their interaction with their surroundings. }


\maketitle
\nolinenumbers


\section{Introduction}
\label{Intro}
Stellar binary systems hosting a single compact object (\xray binaries) allow us to study the physics of accretion on directly observable timescales. Accretion onto the compact object in these systems produces bright \xray radiation as matter is accreted inwards, which in turn results in the launching of powerful, collimated outflows ("jets"), which are usually identified by their radio synchrotron emission. \xray emission from \xray binaries is highly variable and usually split into two main \xray spectral states: a soft state, dominated by thermal radiation from an optically thick accretion disk, and a hard state, characterized by a non-thermal spectrum usually explained by either jet emission~\citep[e.g.][]{Markoff2001} or inverse Compton scattering in a hot, optically thin corona~\citep[e.g.][]{Frontera2003}. Evidence of steady jet activity is typically found when sources are in the hard spectral state and during the transition to the soft state~\citep{Fender2001a,Fender2004a, Homan2005,Remillard2006, Belloni2016}.

The recent detection of gamma-ray photons with energies greatly surpassing tens of TeV from a handful of \xray binaries has firmly established their nature as extreme particle accelerators~\citep{LhaasoCollaboration2025}, making \xray binaries an ideal laboratory to study particle acceleration in astrophysical jets.

Despite the rapid progress, many questions remain regarding how and where particles are accelerated and what sets the detected objects apart from the rest of the class. A total of six \xray binaries have so far been associated with gamma-ray emission, out of a total of $\sim 70-100$ known objects~\citep[see Section~\ref{sec:the_cat}, ][]{Tetarenko2016,CorralSantana2016, Neumann2023, Avakyan2023}. There are a few common properties of the gamma-ray emission across the different sources: for all but one object (see Section~\ref{subsec:x3}), the multi-TeV emission is persistent and either extends out to large distances from the binary (up to $\sim 100$~pc) or is offset in the direction of the jets by $\sim 10-40$~pc. There are no obvious commonalities between the detected sources. Three out of the six have a massive ($>10M_{\odot}$) stellar companion (high-mass \xray binaries, HMXB), producing strong winds which go on to feed the compact object. Another two have companions of mass $<1M_{\odot}$ (low-mass \xray binaries, LMXB), and accretion is expected to occur in a completely different regime through Roche‑lobe overflow of the stellar companion. Finally, the most powerful accelerator among them has a companion of ~3$M_{\odot}$, sitting awkwardly in the middle of these two regimes. Their observed power output is also diverse. Three of the detected systems are known to currently be highly powerful systems, either often or persistently accreting matter at the Eddington limit (L$_{\mathrm{Edd}} = 1.26\times 10^{38}\left(\mathrm{M}_{\mathrm{BH}}/M_\odot\right) $~erg~s$^{-1}\approx 10^{39}$~erg~s$^{-1})$. Conversely, the other three have only been observed to reach this limit occasionally and for relatively short transient outbursts. It is worth noting, however, that existing observations of black-hole \xray binaries span at most several decades, much shorter than the lifetime of the objects (of up to billion years for LMXB). As a result, little is known about their long-term history and behaviour on longer time-scales.

While jetted \xray binaries have long been theorized to be sources of cosmic rays of a wide range of energies~\citep[e.g.][]{Heinz2002a,Aharonian2004,Fender2005a,Romero2008, Escobar2021, Escobar2022, Cooper2020, Kantzas2023}, the detection of multi-TeV emission from some \xray binaries has catapulted the source class into the category of prime candidate to explain the origin of the most energetic Galactic cosmic rays, in and above the PeV range~\citep[e.g.][]{Kaci2025, Zhang2026a, Abaroa2026}. The observational evidence provided by LHAASO makes it indisputable that \xray binaries can contribute to the cosmic ray spectrum in that energy range. The question then becomes: are they the dominant contributor as a class? In order to address this question, one has to understand the population of TeV-emitting \xray binaries, as well as what determines the maximum energy reached by each source, how much power they provide to particles, and what kind of particles (electrons, protons and/or nuclei) dominates their output.

In this paper, we attempt to answer the question of whether there is any commonality between the detected systems when investigating their properties in detail. We rely on existing source catalogues and decades of observational coverage in the \xray and radio bands to compare the population of black-hole \xray binaries which have been detected in the TeV band to those which have not. The paper is organised as follows: in Section~\ref{sec:the_guys} we describe the construction of our source sample through existing catalogues. The sample is then expanded to include constraints on the gamma-ray and radio emission from these sources, detailed in Sections~\ref{sec:gamma-ray} and~\ref{sec:jets}, respectively. In Section~\ref{sec:the_corr} we derive the (rank) correlations between different properties of the sources in our sample and their gamma-ray luminosity. We then compare the positions of the sources in our samples with those of unassociated gamma-ray sources in Section~\ref{sec:the_cat} and compile our conclusions in Section~\ref{sec:the_conclusions}.

\section{Source selection}
\label{sec:the_guys}
\subsection{\xray binary catalogues}
We combine three recent catalogues of black-hole \xray binaries to construct our source sample. The different catalogues have different focuses, aims, and cut-off dates, as detailed below.
\begin{itemize}
    \item WATCHDOG~\citep{Tetarenko2016}: collects information on the activity of black hole (and black hole candidate) \xray binaries between 1996 and 2015, as revealed by \xray observations. It includes 77 sources and splits them between dynamically confirmed black holes (class A, N=21), sources with black-hole-like spectra (class B, N=30) and other \xray binaries with weak evidence for a black hole (class C, N=26). The catalogue includes basic information on each binary such as distance, inclination or orbital period, but also a detailed characterization of every \xray outburst detected from each of the sources included. Because of the cut-off in 2015, a handful of sources and outbursts discovered since then are missing from the catalogue (such as \maxi or the 2015 outburst of V404~Cyg). 
    \vskip 1mm
    \item BlackCAT~\citep{CorralSantana2016}: is a continuously updated database of \xray transients from black hole or black-hole candidate \xray binaries. It includes 73 sources and includes their name identifiers, sky coordinates, optical magnitude in outburst and quiescence, their distance, orbital period, and finally the peak observed flux in the 2 to 10~keV range.
    \item XRBCat~\citep{Neumann2023,Avakyan2023}: is slightly different from the two catalogues above in that it focuses instead on \xray binaries in general, and not just those hosting a (likely) black hole. The catalogue provides an \texttt{Xray\_Type} keyword which indicates what type of \xray binary an entry is. We select all sources  with "BH" (black holes and candidates) in that column. Doing so yields 6 HMXB and 80 LMXB. Note that the catalogue also provides a "MQ" keyword for jetted sources - we chose to not use it because doing so only excludes Cir~X~1 and Sco~X-1, which are "Z-type" neutron star \xray binaries which will be, as a population, subject to future work.
    
\end{itemize}

We merged the catalogues by source coordinates, which initially yielded 141 objects. We then merged all objects with slightly different coordinates but separated by less than 0.01$\degree$. We chose this size because it ensures that all the merged objects share a common name identifier. This step reduces the source list to 110 objects. We then inspect their name identifiers and alternative identifiers and find four sources with matching names but slightly distant coordinates: GRS~1730-312,  Swift~J1357.2-0933, XTE~J1856+053 and MAXI~J1803-298. We inspected the literature and adopted the most recent localization. Finally, we remove a repeated entry for 3A~0620-003, as it was also included under a different identifier (1A~0620-00) with slightly different coordinates. 
The final selection has 105 unique objects. 52 of these are listed in all three catalogues. There are  10 objects listed only in WATCHDOG, 7 objects listed only in BlackCat, and 10 objects listed only in XRBCat. We inspect these 105 sources and remove five from our sample because further investigation indicates that they are no longer considered to be a black hole \xray binary. The removed sources are HD~96670~\citep[which does not host a compact object, as revealed by][]{Naze2025},  IGR J17361-4441 ~\citep[which has been associated with a tidal disruption event, see][]{DelSanto2014}, IGR J17379-3747~\citep[most likely an accreting millisecond pulsar, see][]{Bult2019} and GRS 1736-297~\citep[a possible accreting millisecond pulsar, see][]{Tetarenko2016a}. We also remove Cen~X-2, which was detected only once in the 1960s but never again, because it might be the same object as BW~Cir~\citep{Kitamoto1990}, but note that the association between the two objects has been questioned~\citep{Casares2009}. This leaves 100 sources in the list. Following the A-B-C classification from WATCHDOG, we add a class entry for the 24 binaries which were not included in that catalogue. For this, we use a combination of the "Dynamically confirmed BH" label from BlackCAT and \texttt{Xray\_Type} entry from XRBCat. This process yields 4 additional class A objects (Swift J1727.8-1613, MAXI J1820+070, MAXI J1348-630 and Swift J1728.9-3613),  16 additional class B objects and three additional class "C" objects (KS 1739-304, CXOU J174805.0-244643 and Swift J1729.5-3223).

For each of the objects, the system properties from each catalogue such as distance, orbital period, black hole mass, inclination and companion mass/type are merged. When different catalogues report different values for any of these parameters, we inspected the literature and chose the most recently reported value. For the distance estimates, we adopted those reported by BlackCat and WATCHDOG rather than by XRBCat as the latter generally relies on Gaia estimates which might be inaccurate for binary systems~\citep[e.g.][]{Gandhi2019}. Additionally, when selecting a distance estimate, we also favoured estimates consistent with those assumed when deriving the other system properties. For the vast majority of objects, the estimates provided in the different catalogues were in agreement within uncertainties. The handful of cases for which that was not the case are discussed in further detail in Appendix~\ref{app:cat}. Following the approach of WATCHDOG, when no reliable distance estimate exists, we sample from a uniform distribution ranging between 2 and 8~kpc. The sampling uses a different random seed per source. When the chosen distance estimate differs from the one from WATCHDOG, we re-scaled the luminosities provided in that catalogue to the adopted distance. Dedicated references for each source are collected in Table~\ref{tab:sources}.

\subsection{The Gaia dormant black holes}
If the LHAASO observatory had started operating some time in the 2010s (less than ten years before it did), it would have reported the gamma-ray source now associated to \maxi as an unassociated multi-TeV source in an otherwise empty field. That is, until \maxi went into outburst in 2018 and revealed itself as the most likely counterpart. Similarly, gamma-ray emission from now-dormant black hole binaries might be detected but not identified as such because its binary counterpart has not gone into outburst in the past few decades. A promising avenue to explore this possibility is the small but steadily increasing population of "dormant" black holes discovered through precise astrometrical measurements of their companion. There are three such objects known to date: Gaia BH1~\citep{ElBadry2022}, BH2~\citep{Tanikawa2023} and BH3~\citep{GaiaColl2024}. While we do not include them in our sample directly, we will use their reported positions to perform an additional comparison to existing gamma-ray catalogues.

\section{Gamma-ray measurements and constraints}
\label{sec:gamma-ray}
 Detecting an object in the gamma-ray band requires particle acceleration to take place either within or near that region because gamma-ray emission is only produced by very energetic particles. If the particles are of hadronic nature (protons, heavier nuclei), gamma-ray emission is produced by interactions of the hadrons with nearby target gas (pp-interactions) or ambient radiation fields (p$\gamma$-interactions). Because of the relatively small cross-section of the p$\gamma$ process, this channel is usually subdominant in the absence of a nearby source of additional radiation other than the average Galactic and extra-Galactic diffuse radiation components. Consequently the p$\gamma$ process can usually be neglected when discussing the extended, several-pc scale emission discussed here. If instead the particles producing the radiation are leptonic (i.e. electrons and/or positrons), the relevant mechanism is inverse Compton (IC) up-scattering of low-energy photons from an ambient radiation field into gamma-ray energies. 

 Considering the conditions present in the average interstellar medium (ISM), with a relatively low average gas density ($n$$\sim$1~cm$^{-3}$), and diffuse radiation fields such as the cosmic microwave background (CMB) or diffuse interstellar radiation fields~\citep[ISRFs,][]{Popescu2017}, IC is generally more efficient than either hadronic mechanism since the particle luminosity required to provide the same gamma-ray luminosity is much smaller in the leptonic case. However, the Klein-Nishina suppression (which reduces the IC scattering cross-section at high electron energies for high target photon energies) and competing losses with synchrotron radiation (which produces photons ranging from radio to \xray energies) reduce this advantage of the leptonic mechanisms for energies greater than $\sim100$~TeV~\citep{Blumenthal1970}. As a consequence, both the morphology of the observed emission and its inferred nature (leptonic or hadronic) might simply reflect the environment (i.e. whether high-density gas is available for hadronic interactions or not) instead of the intrinsic particle output of the source.
 
\subsection{Detected systems}
A total of six compact \xray binary systems have firmly been associated with TeV gamma-ray sources. We collected the properties of their emission, which we detail below. Table~\ref{tab:detected} summarizes the morphological and spectral properties of the gamma-ray emission associated to each of the systems.

\begin{table*}[]
 \caption[]{\label{tab:detected} Properties of the gamma-ray emission associated with detected \xray binaries.   }
\resizebox{\textwidth}{!}{\begin{tabular}{lccccc}
 \hline \hline
 \noalign{\vskip 1mm} 

  Source  & Offset from binary & Major axis extent & F$_{>\mathrm{1~TeV}}$ & F$_{>\mathrm{100~TeV}}$ \\
      & (deg) & (deg) & ($10^{-12}$~erg~cm$^{-2}$s$^{-1}$) & ($10^{-13}$~erg~cm$^{-2}$s$^{-1}$) & \\
      & (pc) & (pc) &  \\

\noalign{\vskip 1mm} 
\hline

\noalign{\vskip 1mm} 

\ssftt & 0.48$\pm$0.03 &  0.21$\pm$0.04  & 2.50$^{+0.33}_{-0.40}$ & 1.32$^{+0.19}_{-0.26}$ \\
& 45.91$\pm$2.49 & 19.68$\pm$3.36 & & \\
\noalign{\vskip 1mm} 

\vsgr &  0.37$\pm$0.04 &  0.18$\pm$0.05  & 11.08$^{+0.85}_{-0.90}$ & 33.92$^{+4.95}_{-6.63}$ \\
 & 40.17$\pm$3.83 & 19.83$\pm$5.10 & & \\
\noalign{\vskip 1mm} 

\grs &  0.09$\pm$0.05 &  0.28$\pm$0.05  & 1.21$^{+0.20}_{-0.24}$ & 1.48$^{+0.25}_{-0.83}$ \\
 & 14.29$\pm$8.19 & 45.94$\pm$8.20 & & \\
\noalign{\vskip 1mm} 

MAXI~J1820+070 &  0.24$\pm$0.06 &  $<$0.28  & $<$0.39 & 0.16$^{+0.05}_{-0.08}$ \\
 & 12.45$\pm$3.10&  $<$14.46 & & \\

Cyg~X-1&  0.19$\pm$0.08 &  $<$0.22  & $<$0.30 & 0.04$^{+0.01}_{-0.02}$ \\
 & 6.33$\pm$2.56&  $<$7.14 & & \\
\noalign{\vskip 1mm} 

Cyg~X-3 (flare)&  - &  -  & - & 1.75$^{+0.51}_{-0.68}$ \\
\noalign{\vskip 1mm} 
Cyg~X-3 (extended)&  ? &  ? & - & $<$23.51 \\
\noalign{\vskip 1mm} 
\hline

\hline
\end{tabular}}
\tablefoot{The offset from the binary is computed from the centroid of the reported spatial best-fit model in each case. The extent is the 39\% containment radius for a 2D Gaussian. The final two columns are the measured flux above 1 and 100~TeV, respectively. When emission from a system has been resolved into two components (\ssftt, \vsgr), we use the spatial properties of the largest one. In all cases, the flux reported in the table corresponds to the flux measured from the entire system and not just from one component. Since \maxi and \cyg are not detected below 25~TeV, the upper limit reported in the table for F$_{>\mathrm{1~TeV}}$ is derived using published upper limits, which assume a point-like or only slightly extended source morphology.}
\end{table*}

\subsubsection{SS 433}
\ssftt is a HMXB hosting a compact object of debated nature~\citep{Seifina2010, Bowler2018} which is persistently accreting in a super-Eddington regime, launching a precessing jet observed across the multi-wavelength spectrum~\citep{Margon1984}. \ssftt was the first system to be firmly detected in the TeV band, reported by HAWC Observatory~\citep{Abeysekara2018}. The TeV emission traces previously known parsec-scale \xray emission regions~\citep[e.g.][]{SafiHarb2022} aligned with the mean axis of the inner, precessing jets of \ssftt. Follow-up with the \hess telescopes~\citep{HESSCollaboration2024} led to the identification of the dominant particle acceleration sites, located at approximately 20~pc either side of the compact object and consistent with the expectations for shocks within the jets. While the properties of the emission below $\approx$100~TeV indicate a leptonic origin, the morphology revealed by LHAASO at higher energies~\citep{LhaasoCollaboration2025} suggests the existence of an additional emission component of hadronic nature, which reveals itself through interactions with a nearby neutral hydrogen cloud. 
LHAASO measured the gamma-ray spectrum of the entire \ssftt system above 1~TeV~\citep{LhaasoCollaboration2025}. While slight disagreements exist between the LHAASO measurement at low energies and the previously reported values~\citep{HESSCollaboration2024, Alfaro2024b}, we adopt this measurement for simplicity. To characterize the flux above 1 and 100~TeV, we fit a simple model (a power-law with an exponential cut-off). The measured spectrum and model are shown in Figure~\ref{fig:spectra}, and the integrated flux values reported in Table~\ref{tab:detected}.

\subsubsection{\vsgr}
\vsgr is a LMXB hosting a compact object of mass $\mathrm{M}_{\mathrm{BH}}\approx$7~M$_{\odot}$~\citep{Goranskij2003, Orosz2001, MacDonald2014}. The companion star mass is estimated to be $\mathrm{M}_{*}\approx$3~M$_{\odot}$, making it one of the most massive companions observed in a LMXB system. \vsgr was identified in 1999, when it underwent a strong outburst~\citep{Markwardt1999,Zand2000, Wijnands2000} during which its \xray luminosity increased by several orders of magnitude. A compact radio source was observed only briefly during this outburst~\citep{Hjellming2000}, which reveals the presence of a relativistic jet~\citep{Marti2026}. Unlike \ssftt, which is known to have been in a super-Eddington accretion regime since its discovery, the only instance where the output of \vsgr was thought to have reached Eddington levels was this 1999 outburst. Since then, the source has been comparatively quiet, with only smaller outbursts with average luminosities of $\approx 10^{36}$~erg~s$^{-1}$~\citep[e.g.][]{Shaw2022}. The gamma-ray emission around \vsgr is highly extended~\citep{Alfaro2024}, and is detected up to photon energies of 800~TeV~\citep{LhaasoCollaboration2025}. The emission was resolved into two extended components~\citep{Acharyya2026} with indistinguishable spectral properties. Due to the lack of dense nearby gas, the emission mechanism is most likely leptonic~\citep{Wan2026,Acharyya2026}. This interpretation is supported by the recent report of \xray counterparts to the gamma-ray emission sites~\citep{Tsuji2026,Chi2026}. On the radio band, \citet{Grollimund2026} report the discovery of a large-scale bow-tie shaped structure around \vsgr, but it is much smaller ($\sim$ 35 pc diameter) and has a different morphology than the gamma-ray emission region ($\sim$ 100~pc major axis diameter). To characterize the gamma-ray flux above 1 and 100~TeV, we fit a simple model (a power-law with an exponential cut-off) to all the available spectral measurements. The measured spectrum and model are shown in Figure~\ref{fig:spectra}, and the integrated flux values reported in Table~\ref{tab:detected}.

\subsubsection{\grs}
\grs is a LMXB hosting a black hole and an 0.5~M$_{\odot}$ star~\citep{Reid2014, Steeghs2013}. Since its detection more than 30 years ago, \grs  had been a consistently bright \xray source, implying intrinsic \xray luminosities in the range of 0.1 to 1 L$_{\mathrm{Edd}}$, until a sudden decrease in June 2018. This behaviour has been attributed to either an obscuration~\citep{Motta2021, Balakrishnan2021} or an intrinsic change into the hard state~\citep{Koljonen2021}. In the radio band, \grs is observed to be a highly variable source displaying apparently superluminal two-sided ejections~\citep{Mirabel1994,Rodriguez1999, Fender1999}, establishing the presence of a relativistic jet in the system. 
In the gamma-ray band LHAASO reports an extended source consistent with the position of \grs, with a spectrum extending to several hundred TeV. The extension of the LHAASO source includes the position of a bow-shock-like structure discovered in the direction of the \grs jet~\citep{Motta2025}. Notably, \grs is the only detected source for which a persistent point-like GeV counterpart has been reported~\citep{MartiDevesa2025}. The multi-TeV emission reported by LHAASO includes the best-fit position of the GeV source, although this might simply be due to the large extension of the LHAASO source (0.28$\degree$ radius). We tentatively assume that the two sources are associated and use the GeV flux to extrapolate the spectral shape in the 1-10~TeV gap. As before, to characterize the flux above 1 and 100~TeV, we fit an analytical model (a power-law with an exponential cut-off) to all the available spectral measurements. The measured spectrum and model are shown in Figure~\ref{fig:spectra}, and the integrated flux values reported in table~\ref{tab:detected}.

\subsubsection{MAXI~J1820+070}
The LMXB \maxi hosts a black hole and a $<$1~$M_{\odot}$ star~\citep{Torres2020}. It was discovered in 2018, when it underwent an exceptionally bright outburst, with \xray flux peaking at 4 times that of the Crab nebula. The relatively low distance to the source of around 3~kpc~\citep{Atri2020} and high intrinsic luminosity of the outburst enabled excellent multi-wavelength coverage of the event. Jet activity was detected from \maxi in the form of radio and infrared emission~\citep{Bright2020, Wood2021} and even \xray emission~\citep{Espinasse2020}, requiring the presence of $>10$~TeV electrons. Deep observations during the 2018 outburst by several gamma-ray telescopes did not yield a detection~\citep{Abe2022}. Instead, LHAASO reported the presence of gamma-ray emission associated with the system, which is persistent and slightly offset from the \xray binary in the direction of the receding jet~\citep{LhaasoCollaboration2025}. The outburst observations can be used to place upper limits on the spectrum in the 1$-$10~TeV range. Note that by using these upper limits, we assume that the emission is either point-like or only slightly extended. If the source extension is instead as high as allowed by the upper limit placed by LHAASO ($<$0.28$\degree$ radius), the MAGIC, \hess and VERITAS data would need to be re-analysed taking into account the large extension. Because of the non-detection at low energies, we use the best-fit model reported by LHAASO to derive the flux above 100~TeV, and report only an upper limit above 1~TeV using an analytical curved model. Note that this is just for reference and we will not use this upper limit to draw any conclusions. The measured spectrum and model are shown in Figure~\ref{fig:spectra}, and the integrated flux values reported in Table~\ref{tab:detected}.

\subsubsection{Cygnus~X-1}
Cygnus~X-1 (\cyg for short) is a HMXB hosting a black hole of mass $\sim21M_{\odot}$ which orbits a blue supergiant companion~\citep[e.g.][]{Ramachandran2025}. \cyg is detected as a persistent \xray source hosting a steady radio jet~\citep{Stirling2001} which bends due to its interaction with the stellar wind~\citep{Prabu2026}. Despite extensive studies with the MAGIC telescopes~\citep{Ahnen2017a}, no gamma-ray emission with energies below a few TeV has been detected from this system. Instead, \cite{LhaasoCollaboration2025} reports the detection of a faint point-like source located a few pc away from \cyg. The LHAASO source is spatially consistent with the location of a bow shock attributed to the interaction of the jet in \cyg with the ISM~\citep{Gallo2005, Russell2007}. As in the case of \maxi, we use the MAGIC upper limits to derive a reference upper limit above 1~TeV using a simple curved model, but once again note that this limit would not hold if the emission region is actually much larger than the MAGIC angular resolution.

\subsubsection{Cygnus~X-3}
\label{subsec:x3}
Cygnus~X-3 (\cygt for short) is a HMXB hosting a compact object of debated nature~\citep[e.g.][]{Zdziarski2013} orbiting a Wolf-Rayet star~\citep{VanKerkwijk1992}. \cygt appears as a persistent, highly variable \xray source, indicating a sustained high accretion rate. It is also the most luminous \xray binary in the radio band, where it has been observed to undergo strong flares, as well as produce compact resolved jets\citep[e.g.][]{Marti2001}. Unlike the other systems in this section, no extended or spatially offset gamma-ray emission has (yet) been attributed to \cygt. Instead,~\cite{LhaasoCollaboration2025a} reported the detection of a variable, point-like PeV source consistent with \cygt. The source is only detected at energies greater than 100~TeV, and is detected only during phases of GeV activity~\citep{Tavani2009a,FermiLATCollaboration2009, Dubus2010}. Within those times, the source displays as a hint of variability corresponding to the orbital period of the system. Consequently, \cygt is the only \xray binary detected so far for which the VHE gamma-ray emission can be unequivocally attributed to processes taking place near the binary itself. Early modelling of the PeV emission suggests that particles are accelerated in the jets of \cygt, with the variability arising from the periodic interaction of the jets with the wind of the Wolf-Rayet companion~\citep[e.g.][]{Zhang2026b,Zdziarski2026a}. 
It is unclear whether \cygt also produces spatially extended gamma-ray emission like the rest of the detected sources because it lies (in projection) in the middle of the star-forming association Cygnus OB2, which is itself a bright and extremely extended gamma-ray source~\citep{LhaasoCollaboration2024}, usually referred to as the Cygnus "bubble". We thus adopt the flux above 100~TeV from the entire bubble as an upper limit for the possible contribution of \cygt to this extended gamma-ray source. We note that this upper limit is rather conservative, as modelling of the region suggests that a possible contribution from \cygt would only become dominant at energies closer to 1~PeV~\citep{Harer2025, Kachelriess2025}.

\subsection{4U~1957+115}
\label{subsec:4u}
4U~1957+115 is a highly intriguing system hosting a compact object of unknown nature~\citep{Bayless2011,Gomez2015}. Since its discovery~\citep{Giacconi1974}, it has been persistently in outburst but also almost always in the soft state~\citep[e.g.][]{Yaqoob1993,Maitra2014,Barillier2023}. It has also not been detected in the radio band despite extensive observations~\citep{Russell2011}, resulting in the best existing constraints on core jet production in the soft state. Interestingly 4U~1957+115 is  located 0.23$\degree$ away from an unassociated gamma-ray source~\citep[1LHAASO~J1959+1129u, ][]{Cao2024a} with flux above 100~TeV measured to be F$_{>\mathrm{100~TeV}}= (0.92^{+0.27}_{-0.42}) \cdot 10^{-13}$~erg~cm$^{-2}$s$^{-1}$ (note that no spectral cut-off is assumed, so this value should be interpreted as an upper limit). The distance to 4U~1957+115 is not precisely known, with different estimates placing it at either around 5$-$8~kpc or much further at ~20~kpc~\citep[e.g.][]{Barillier2023}. Adopting a distance of 7.8~kpc, as preferred by~\citet{Barillier2023}, translates to a $\sim 30$~pc offset between 4U~1957+115 and the unassociated gamma-ray source. For the larger distance of ~20~kpc, this offset becomes 80~pc, around twice the largest offset observed in any of the other sources (see Table~\ref{tab:detected}). Such an offset between the \xray binary and the gamma-ray excess could be explained by either particle acceleration or fast transport within a relativistic jet, or by the interaction of escaped particles accelerated somewhere in the system with a serendipitous nearby gas cloud. Because a jet has never been observed from this system and no such cloud has been identified (the system lies $>800$~pc below the Galactic plane), it is unclear whether it can be linked to 4U~1957+115. For this reason, we consider 4U~1957+115 an undetected system going forward but note that further investigations of the surroundings of this system should be carried out to confirm or firmly rule out the association.

\subsection{Non-detected systems}
Most \xray binaries have not been detected in the gamma-ray band. We used published upper limits and instrumental sensitivity to place constraints on their gamma-ray flux. When a source falls within the field-of-view (FoV) of LHAASO~\citep[defined using a maximum zenith of 50$\degree$, as in][]{LhaasoCollaboration2025}, we can either use the dedicated upper limits provided by LHAASO~\citep{LhaasoCollaboration2025}, or estimate the sensitivity of LHAASO at the location of the \xray binary using the sensitivity curves provided by~\citet{Cao2024a}. For sources not in the LHAASO FoV, we have to rely on the upper limit map of the Galactic plane provided with the data release of the first \hess Galactic plane survey~\citep[1HGPS,][]{HESSCollaboration2018}. We note that a relatively soft spectral index ($\Gamma$=-2.3) was assumed when producing this map, which is softer than most of the detected sources in the \hess energy range (see Figure~\ref{fig:spectra}). If a source is neither visible by LHAASO nor covered by the 1HGPS, we do not consider any gamma-ray constraints.

\section{Jet energetics estimates}
\label{sec:jets}

\begin{figure*}[b]
	\centering
		\includegraphics[width=1\linewidth]{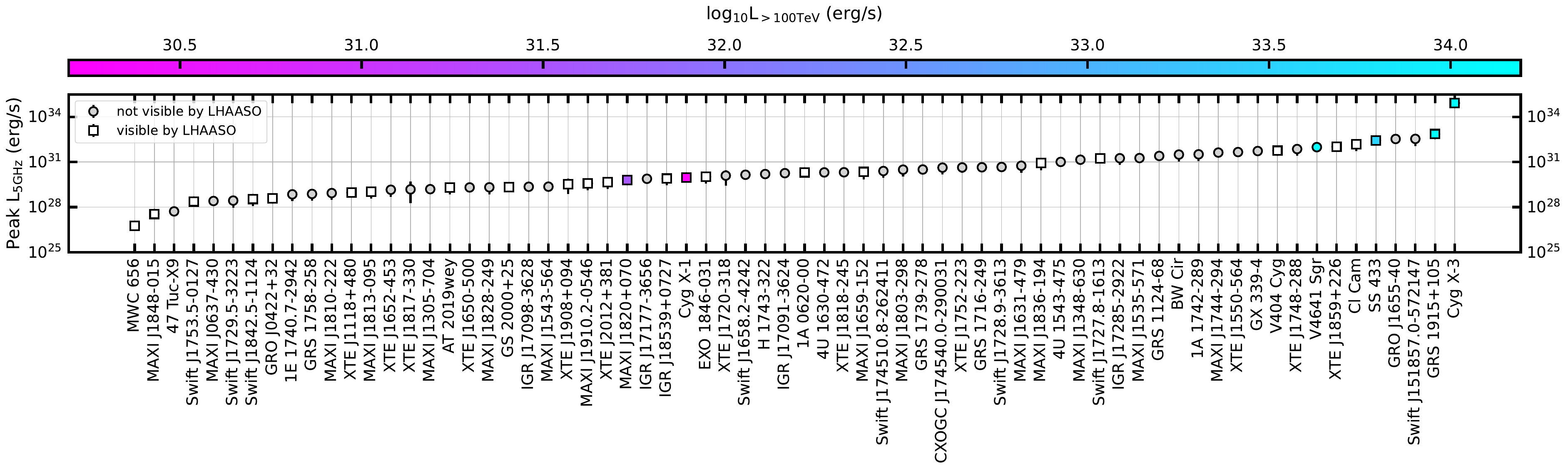} 
	\caption{\label{fig:radlum} Maximum observed radio luminosity at 5~GHz (see Section~\ref{sec:jets}) for every \xray binary for which a radio detection has been reported ordered from least to most luminous from left to right. Square and round symbols indicate systems included and not included in the LHAASO FoV, respectively. \vsgr is indicated as outside the LHAASO FoV because a dedicated analysis above 50$\degree$ zenith was exclusively deployed for this source in the LHAASO \xray binary survey~\citep{LhaasoCollaboration2025}. The colorbar indicates the measured gamma-ray luminosity above 100~TeV as collected in Table~\ref{tab:detected} (the "flare" value for \cygt is used). White circles represent sources which are visible by LHAASO but for which a detection has not been reported.}
\end{figure*}
Our source sample includes a detailed characterization of the energetics of the \xray outbursts for the sources included in the WATCHDOG catalogue, and peak \xray fluxes otherwise. However, the origin of \xray emission during an outburst can be hard to disentangle, as it could be attributed to the accretion disk, the corona, the jet, the disk wind, or a combination of the above~\citep[e.g.][]{Remillard2006}. Conversely, the detection of radio emission (usually with a flat spectral index) provides an unequivocal indication of the presence of a jet. In fact, the observed radio luminosity from an \xray binary ($L_{\mathrm{rad}}$) is the best observational proxy for the jet power for hard-state compact jets ($P_{\mathrm{jet}}$) available~\citep[see e.g.][]{Falcke1995,Fender2003, Heinz2005, Kording2006}, with a relation which simplifies to
\begin{equation}
    P_{\mathrm{jet}} \gtrsim  10^{36-37}\mathrm{ erg}\frac{L_{\mathrm{rad}}}{10^{30}\mathrm{erg/s}}.
\end{equation}
When jets are observed outside of the hard state during \xray state transitions, the relation above cannot be applied directly, and instead synchrotron self absorption needs to be taken into account. However, even in that case, the jet power is still expected to scale monotonically with the radio luminosity~\citep{Fender2019, Cowie2026a}. In any case, the instantaneous radio luminosity provides only a lower limit on the jet power. Observations of large-scale structures formed by the interaction of \xray binary jets with their environment indicate that a large fraction of the jet kinetic power can be carried in a "dark" component of the jet which is otherwise not detectable~\citep{Gallo2005, Motta2025, Cowie2026}.

Unfortunately, none of the available catalogues include any characterization of the radio properties of the sources. To remediate this, we surveyed the existing literature and the Astronomer's Telegram\footnote{\url{https://astronomerstelegram.org/}} database to collect the maximum observed radio flux for each of the sources in our sample. We also used the database provided by~\cite{Crook-Mansour2026} as a result of the ThunderKAT programme\footnote{\url{https://thunderkat.physics.ox.ac.uk/}} which conducted weekly radio monitoring of bright, active \xray binaries using the MeerKAT array.

We use 5~GHz as our reference frequency, and chose the maximum flux at the closest reported frequency. The minimum frequency included in our sample is 0.8~GHz and the maximum 10~GHz.
We assume a flat spectral index when converting flux values from other frequencies to 5~GHz. Figure~\ref{fig:radlum} shows the maximum radio luminosity\footnote{Derived as $L_{\mathrm{rad}} = \frac{\nu F_{\nu}}{4\pi D^2}$, where $\nu$ is taken to be 5~GHz and $D$ is the distance to the source.} derived from the collected fluxes and the distances in the catalogue. The error on the luminosity has been derived by sampling from the distributions (Gaussian or uniform) defined by the uncertainties in the flux and distance. 
Out of the 100 sources in our sample, we find that 68 have been detected in the radio band and 32 do not have a reported radio detection. We refer to these two sub-samples as "radio-detected" and "radio-undetected", respectively.

The sources in the radio-undetected sample might fall into this sample for three main reasons. They might be truly weak radio emitters, for example, sources like 4U~1957+115 which, despite extensive radio observations, has never been observed to reach a hard \xray state and thus produce a jet~\citep{Russell2011}. Others might instead simply not have been observed with the right radio telescope at the right time, always a possibility given the observational bias of our sample. Finally, they might not actually be black-hole \xray binaries, but another type of \xray transient from which radio emission would not be expected. In fact, while the radio-detected sample includes 59 objects from classes A and B and only 7 class C objects (see Section~\ref{sec:the_cat} for the definition of the classes), the numbers for the radio-undetected sample are 14 and 20 for A+B and C, respectively. This implies that a much larger fraction of the radio-undetected sources might simply not be black-hole \xray binaries, but some other type of \xray transient.

For some sources, the jet can be spatially resolved into either a compact jet or moving blobs. In those cases, the direction in which the jet is propagating can be determined. If the jets have been resolved in the radio, we also collect their reported position angle. This is an important quantity when linking gamma-ray sources to \xray binaries because if the emission is produced by interactions between the jet and the ISM, one would expect the emission region to be offset in the direction of the jets, as exemplified by the cases of \ssftt, \maxi and \cyg.

Our source sample, which now includes information on the properties of the sources in radio, optical, \xray and gamma-ray bands is thus the most complete attempt to date to characterize the multi-wavelength behaviour of the black-hole \xray binary population.

\section{Does gamma-ray luminosity correlate with any other property?}
\label{sec:the_corr}
We now compare the properties of our black-hole (candidate) \xray binary sample to the constraints on their gamma-ray luminosity. We focus on the luminosity above 100~TeV because in all cases it is the less model-dependent estimate (see Section~\ref{sec:gamma-ray}.) Appendix~\ref{app:above1TeV} presents the equivalent scatter plots using gamma-ray constraints above 1~TeV.  While the number of sources is much higher thanks to the \hess observations of the inner Galaxy, the systematic uncertainty introduced by the assumed soft spectrum is deemed too high to draw meaningful conclusions. Those results are thus only included for completeness and reference.

We note that we do not expect to find any significant correlation with the resulting dataset, composed of (at best) five measurements and eight upper limits of the gamma-ray luminosity above 100~TeV. However, the presence or absence of strong correlation might hint  to what are the requirements for an \xray binary system to become an extreme particle accelerator, a conclusion which could be used, for example, to motivate further gamma-ray observations.
\begin{figure*}[b]
    \centering
    \begin{subfigure}{0.33\textwidth}
        \centering
        \includegraphics[width=0.9\textwidth]{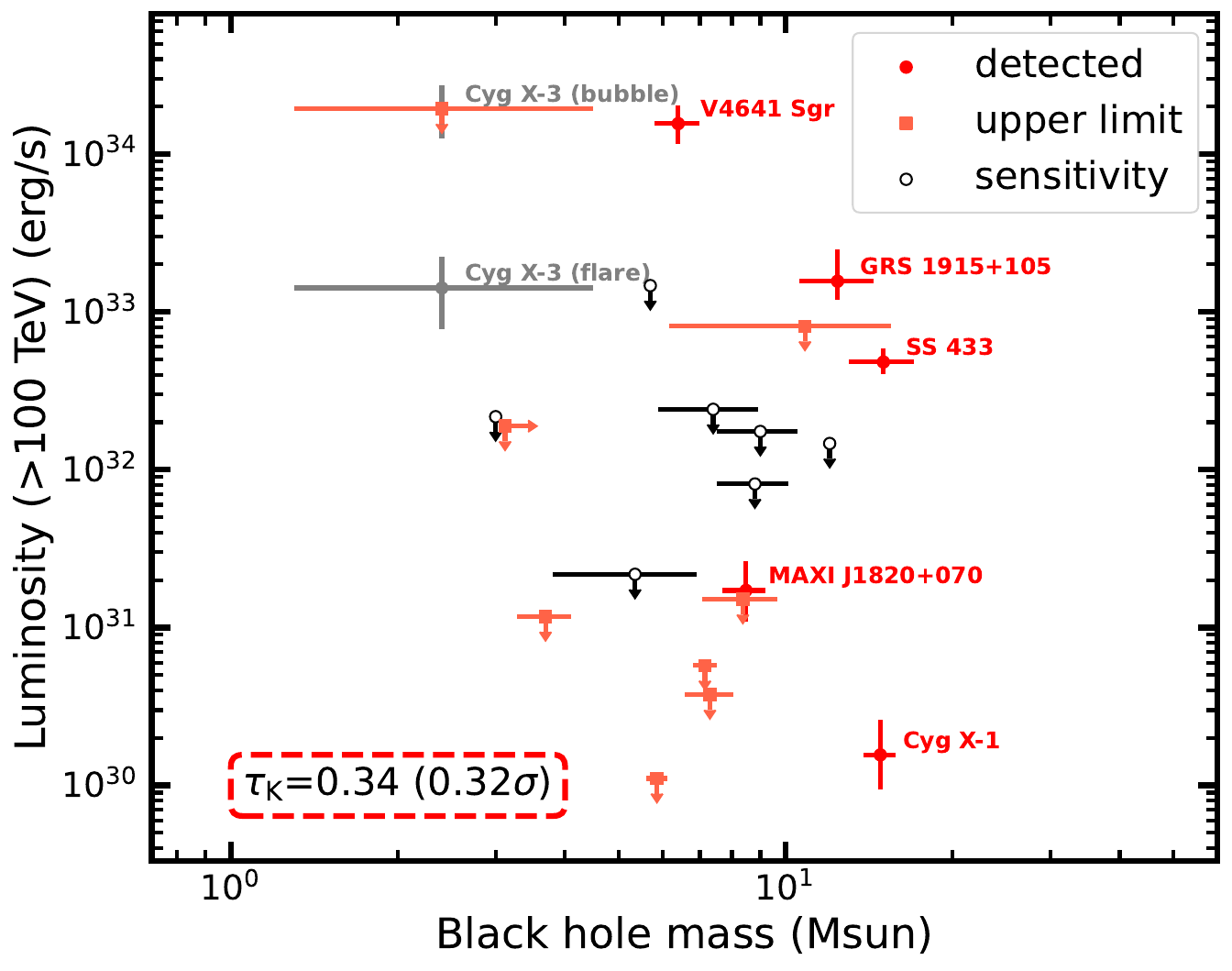} 
    \end{subfigure}
    \begin{subfigure}{0.33\textwidth}
        \centering
		\includegraphics[width=0.9\textwidth]{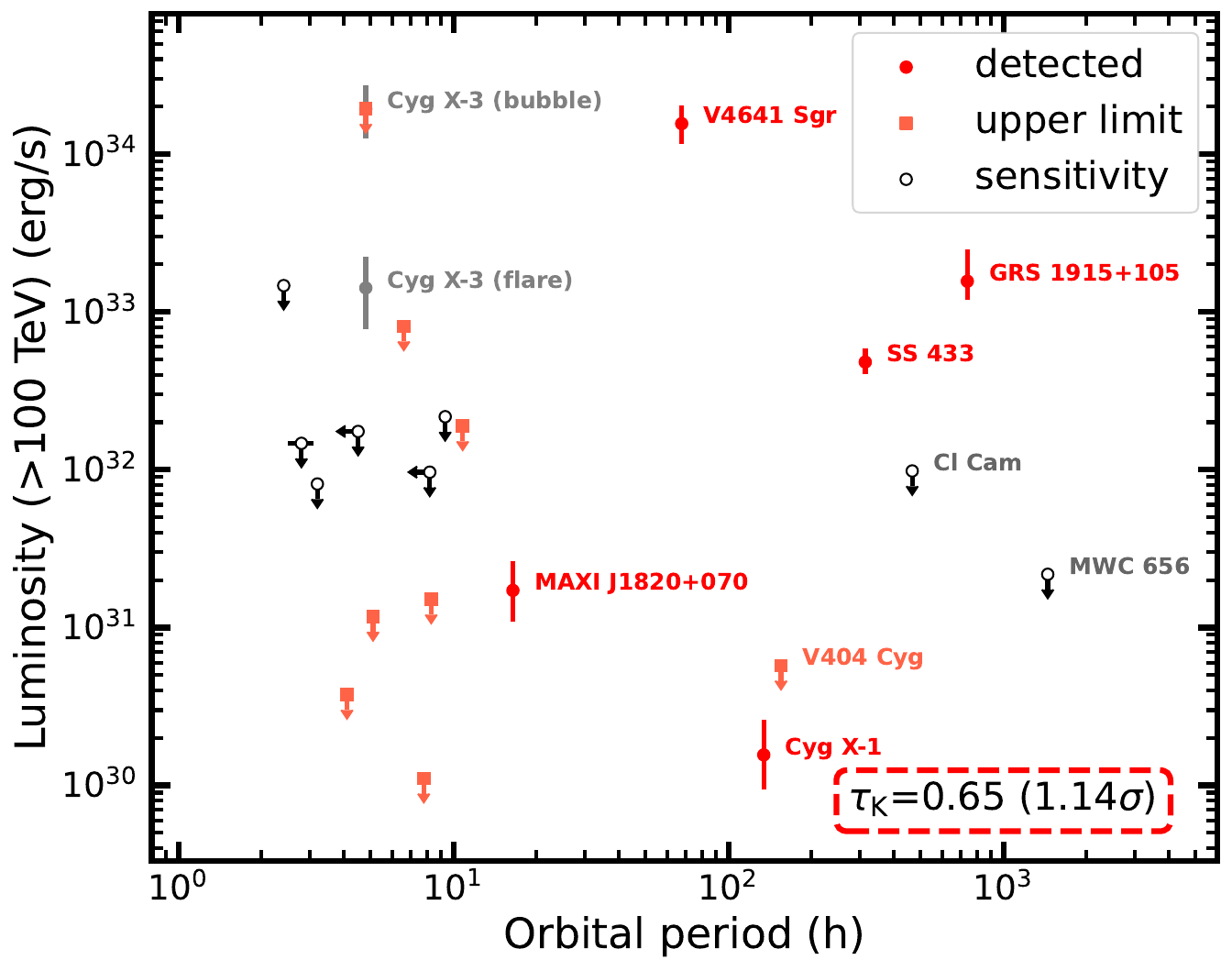} 
    \end{subfigure}
    \begin{subfigure}{0.33\textwidth}
        \centering
        \includegraphics[width=0.9\textwidth]{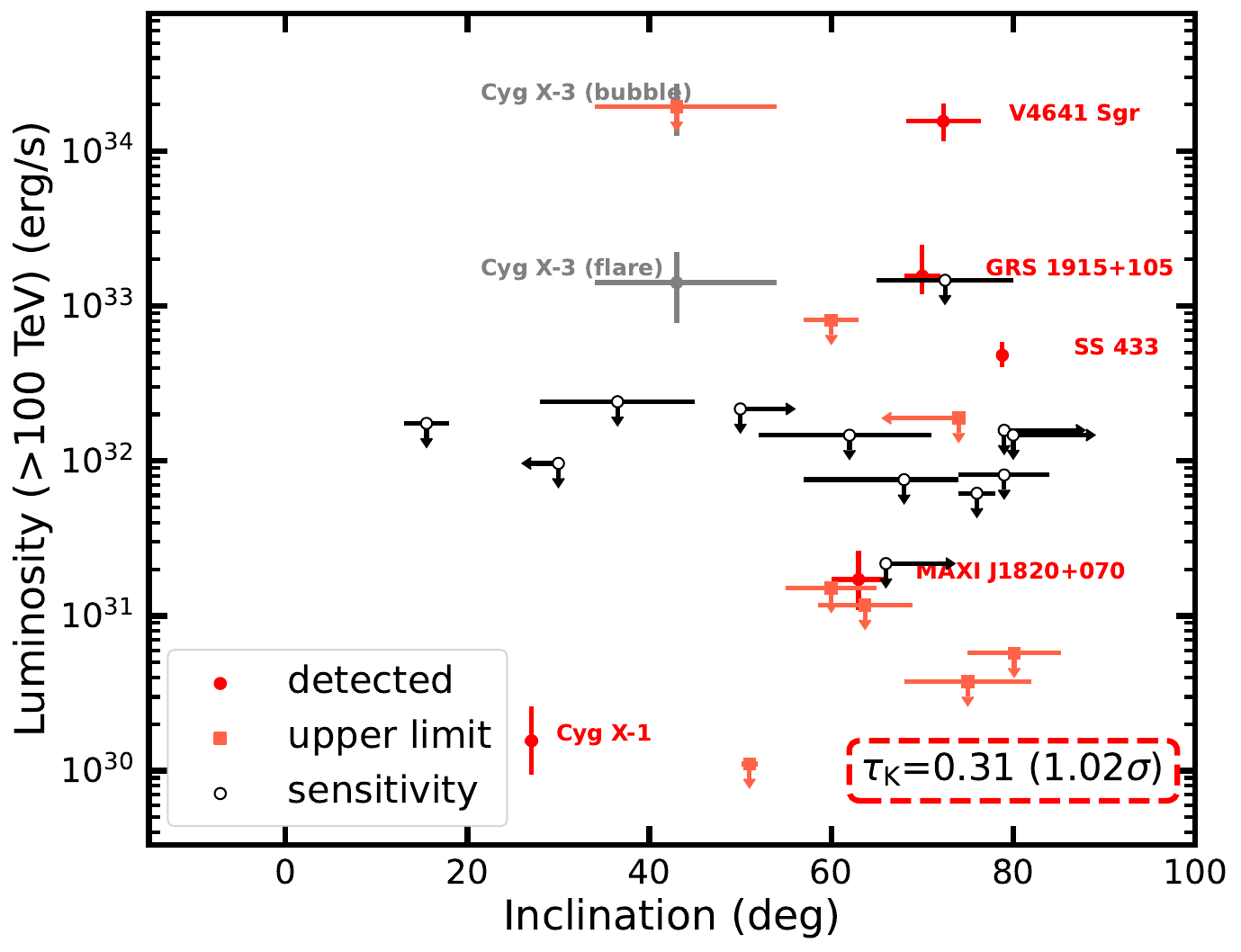} 
    \end{subfigure}
    \caption{\label{fig:arch} Comparisons between the gamma-ray luminosity above 100~TeV and the black hole mass (left), orbital period (middle) and system inclination (right). Only sources for which a constraint on their gamma-ray emission above 100~TeV exists are included. Red circles indicate the detected systems, light red squares indicate those for which upper limits are reported and empty black circles those for which the constraint is derived from a sensitivity curve. When deriving the correlation estimates, we use the gamma-ray flux from the Cygnus bubble as an upper limit for \cygt.}
\end{figure*}

\subsection{Correlation estimates}
Our goal is to determine the presence or absence of a correlation for a dataset which includes upper limits (i.e. censored data). Because the classic Pearson or Spearman tests cannot properly handle censored data, we use instead the Kendall rank correlation coefficient (or $\tau_{K}$-test), first introduced by~\cite{Kendall1938} and expanded to censored data by~\cite{Akritas1996}. It is a non-parametric test for monotonic correlation which measures the rank dependence of two datasets. The $\tau_{K}$ coefficient ranges between -1 (perfect monotonic decreasing correlation) and 1 (perfect monotonic increasing correlation), with a value of zero indicating the lack of a monotonic association. The extension to censored data simply uses the upper (or lower)  limit to determine the order of the sample pair. If the order cannot be determined, for example because both values are censored or because there is a tie, the pair is excluded. We determine the statistical significance of the $\tau_{K}$ coefficient value by performing a permutation test, in which one of the two arrays being compared is permuted 4000 times, with a new value of $\tau_{K}$ being computed for each permutation. The p-value is then derived by counting how many times the permuted $\tau_{K}$ had an absolute value larger the one derived from the real arrays. We take into account the uncertainty on each measurement by sampling the (often asymmetrical) distribution defined by the uncertainty 500 times, and then deriving the correlation estimate as the average between all realizations.

We only consider measured upper limits and not limits derived using sensitivity curves when deriving correlations. We use the flux from the Cygnus bubble as an upper limit when considering \cygt, but note that this is equivalent to excluding it, since the resulting upper limit is higher than all the other gamma-ray luminosities considered. Consequently, when pairing it with any other data point, the rank order cannot be determined and the comparison is excluded from the estimator calculation. To circumvent this, we also derived the correlation estimates using the flaring luminosity as a lower limit. However, we note that this is only done for reference, as the transient emission might have a different origin than the persistent and extended emission observed in other systems.

\subsection{System properties}
We examine the possible relation between the gamma-ray output of a system and its architecture.
It is already from the description in Section~\ref{sec:gamma-ray} that whether a system is classified as a HMXB or LMXB appears uncorrelated with its gamma-ray emission: \grs and \ssftt are similarly luminous  above 100~TeV but host a $0.5\mathrm{M}_{\odot}$ and $>10\mathrm{M}_{\odot}$ companion, respectively. \vsgr is more than an order of magnitude more luminous, yet its companion sits in the middle at $\sim 3\mathrm{M}_{\odot}$. A similar situation is true for the compact object mass, as shown in the left panel of Figure~\ref{fig:arch}. We find a value of the Kendall correlation coefficient of $\tau_{K}=0.34$ (0.3$\sigma$), meaning neither strong nor significant correlation. Including the luminosity of the flares from \cygt as a lower limit has no impact on this conclusion.

\subsubsection{Orbital period}
There are a few physical reasons why the period of a black hole \xray binary and its high energy particle output might be related. The output energy of an outburst should be proportional to the mass in the accretion disk~\citep[e.g.][]{Yu2009} for low-mass \xray binaries, and in general, a longer orbital period suggests a larger maximum radius for the accretion disk. Consequently, many models predict a correlation between the amounts of energy liberated by the binary system and the period, with longer period systems being expected to liberate more energy~\citep[see e.g.][]{Frank2002}. There are observational indications that this is indeed the case for LMXB, although the evidence is not yet conclusive~\citep[e.g.][]{Wu2010}. When comparing the gamma-ray luminosity above 100~TeV and the orbital period, we find a value of the Kendall correlation coefficient of $\tau_{K}=0.65$ (1.14$\sigma$), meaning a strong but not statistically significant  positive correlation. We note that the HMXB \cygt has the shortest orbital period among the detected sources, and when including the luminosity of the flares from \cygt as a lower limit, we find $\tau_{K}=0.49$ (1.30$\sigma$), which indicates a weaker and still not statistically significant correlation.

\subsubsection{Inclination}
The difference in gamma-ray emission between the different systems might be explained by different viewing angles. Low values of the inclination angle might favour detectability in the gamma-ray band, as emission arising from the jet would benefit from Doppler boosting. Conversely, systems for which the line of sight is perpendicular to the jets might result in more extended emission aligned with the jets (as e.g. the case of \ssftt), which in turn, facilitates the association of the gamma-ray emission with the \xray binary - often not straightforward in crowded, Galactic environments. When comparing the reported inclination estimates to the gamma-ray output, we find a value of the Kendall correlation coefficient of $\tau_{K}=0.33$ (1.09$\sigma$), indicating a neither strong nor significant correlation. Including the luminosity of the flares from \cygt as a lower limit has no impact on this conclusion.
\begin{figure*}
    \centering
    \begin{subfigure}{0.33\textwidth}
        \centering
        \includegraphics[width=0.9\textwidth]{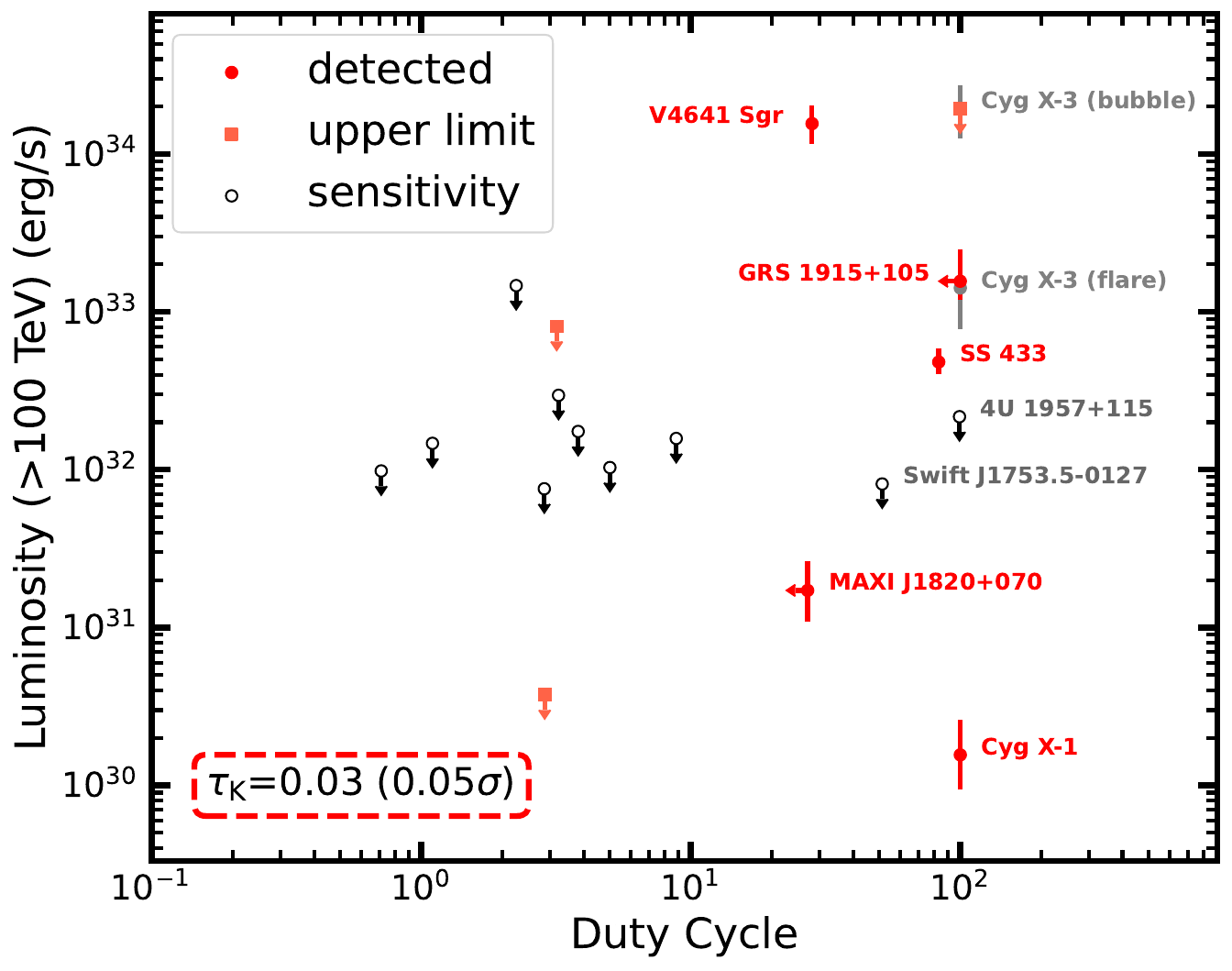} 
    \end{subfigure}
    \begin{subfigure}{0.33\textwidth}
        \centering
		\includegraphics[width=0.9\textwidth]{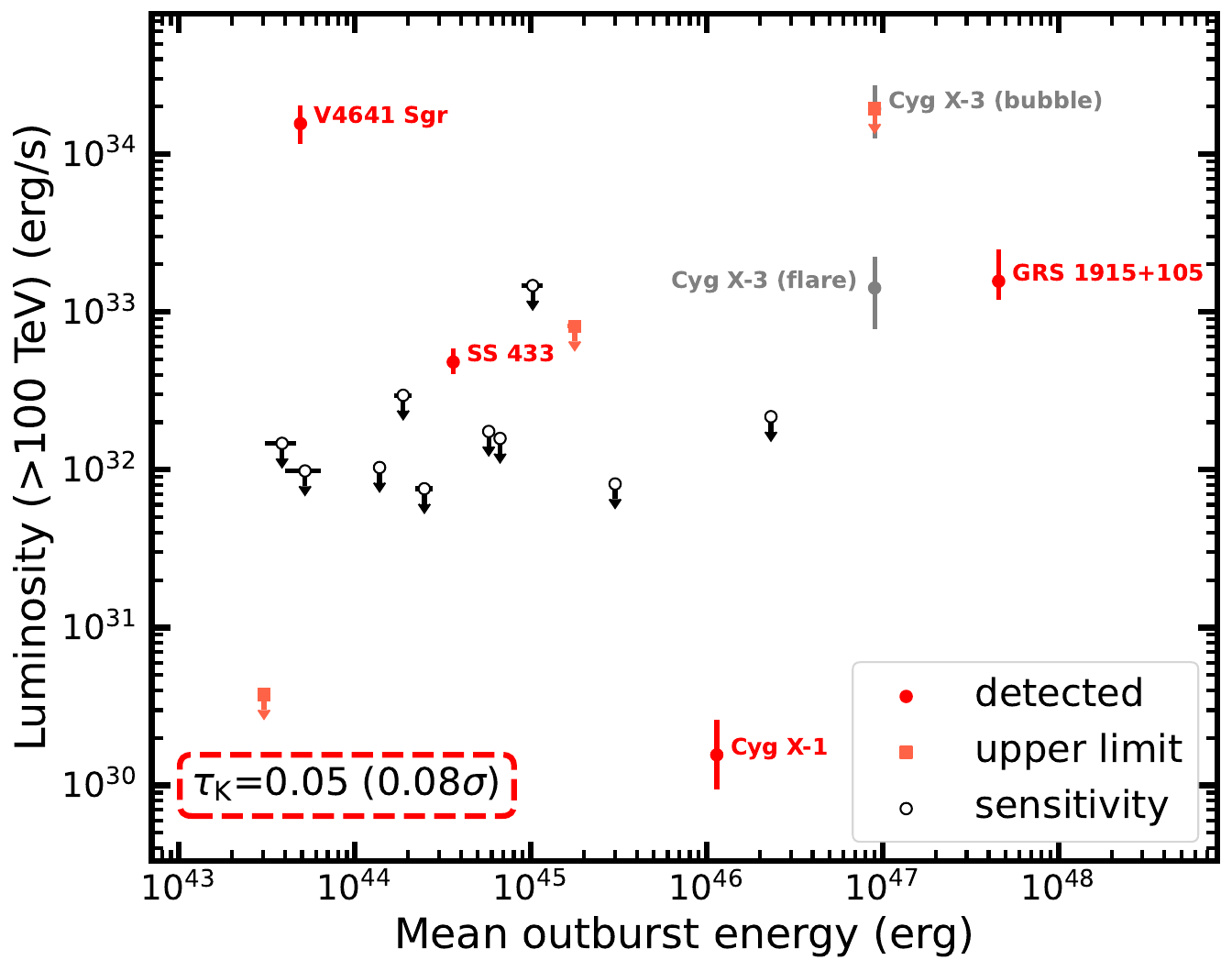} 
    \end{subfigure}
    \begin{subfigure}{0.33\textwidth}
        \centering
		\includegraphics[width=0.9\textwidth]{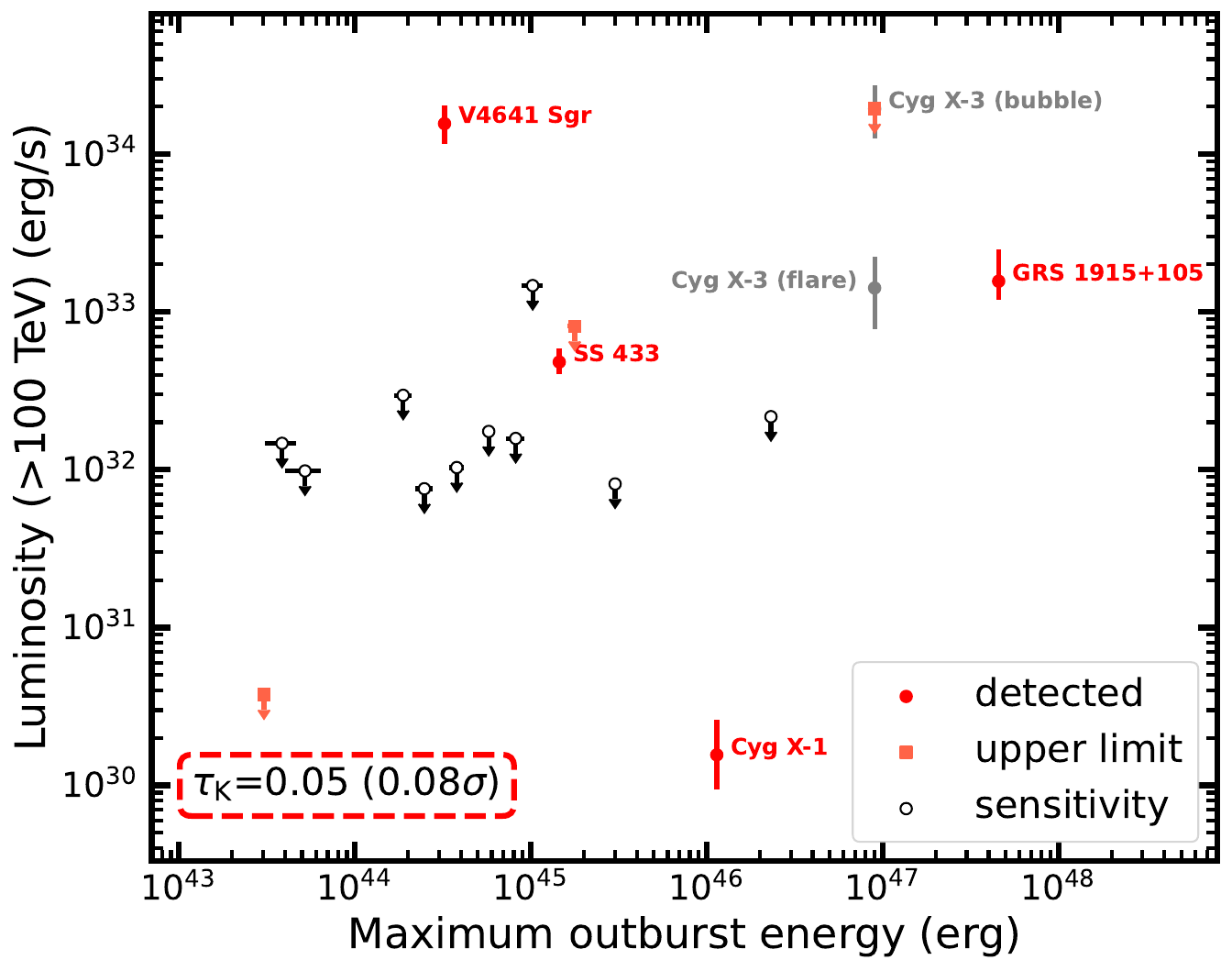} 
    \end{subfigure}
    \begin{subfigure}{0.33\textwidth}
        \centering
        \includegraphics[width=0.9\textwidth]{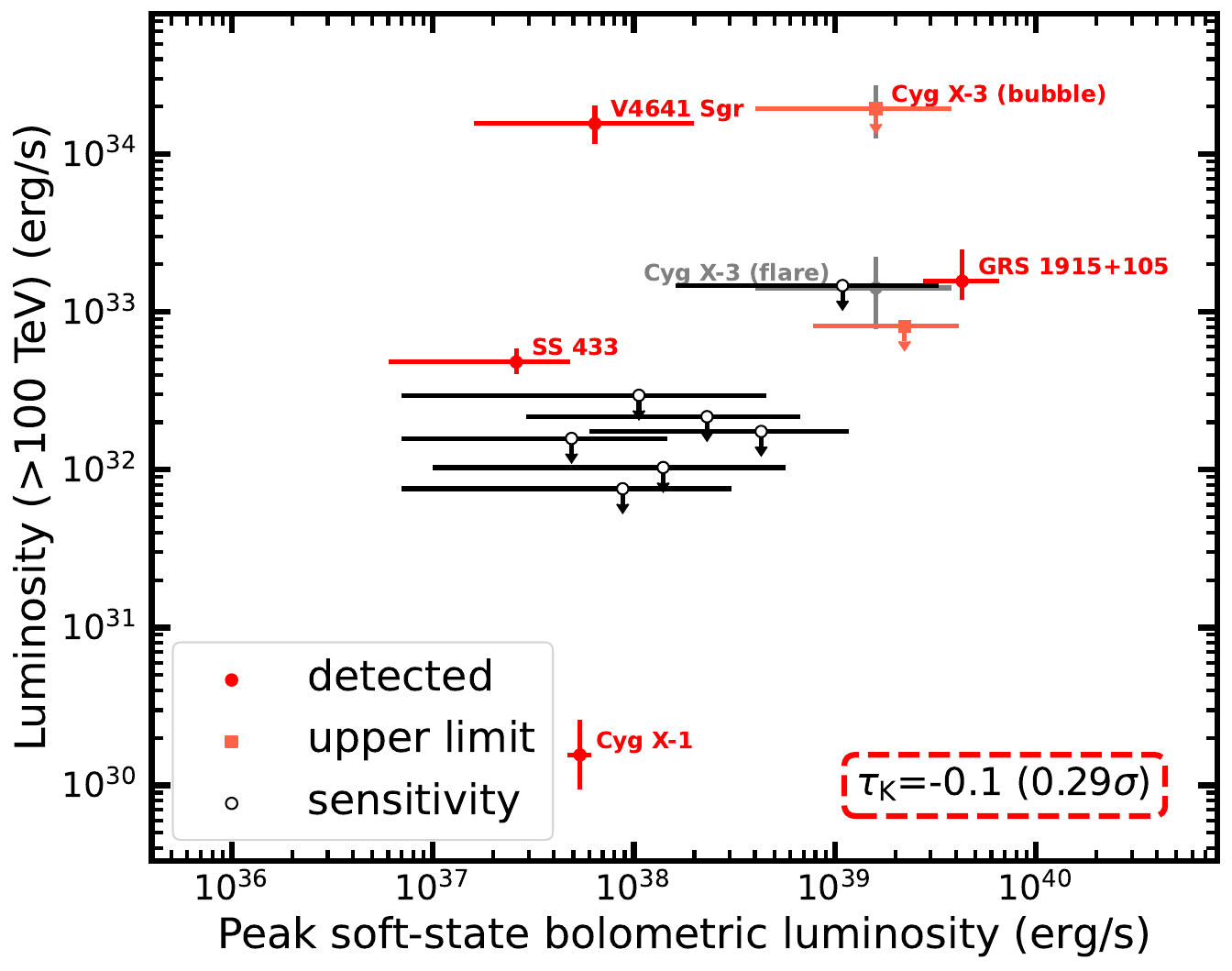} 
    \end{subfigure}
    \begin{subfigure}{0.33\textwidth}
        \centering
		\includegraphics[width=0.9\textwidth]{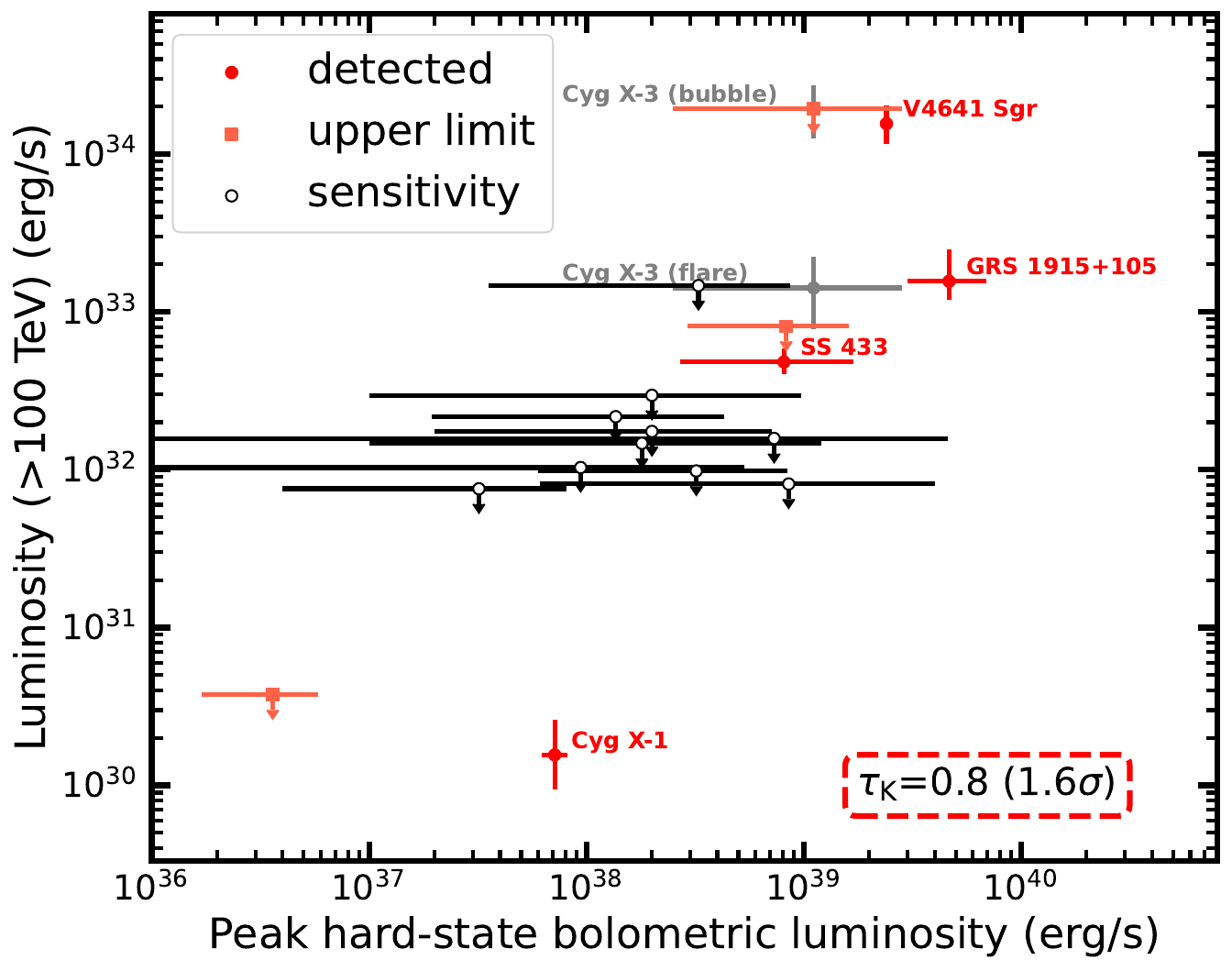} 
    \end{subfigure}
    \begin{subfigure}{0.33\textwidth}
        \centering
		\includegraphics[width=0.9\textwidth]{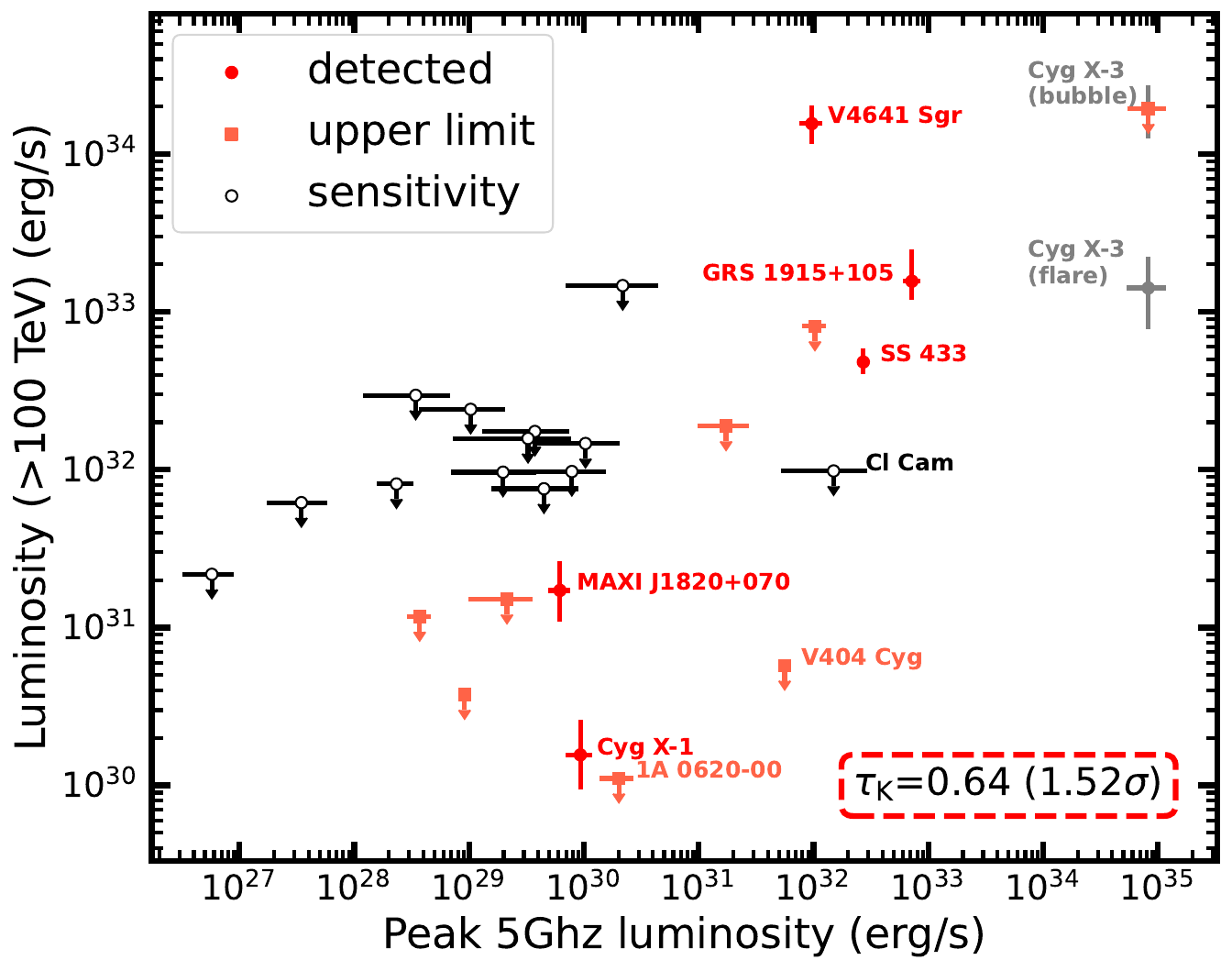} 
    \end{subfigure}
    \caption{\label{fig:power} Same as Figure~\ref{fig:arch}, but comparing the gamma-ray luminosity above 100~TeV to (first row, left to right) the duty cycle, mean and maximum total energy liberated during one outburst, as well as (second row, left to right) the maximum \xray luminosity measured in the soft and hard spectral states, and maximum radio luminosity at 5~GHz.}
\end{figure*}

\subsection{Power}
In many cases, the timescales of the processes responsible for the observed pc-scale gamma-ray emission must be of the order of at least hundreds of years, most likely tens of thousands~\citep[e.g.][]{HESSCollaboration2024, Wan2026}. It is thus not evident how to compare the existing estimates for the power output from black-hole binaries which cover, at best, a few decades to their gamma-ray output. Any such comparison must contend with the obvious caveat that the behaviour of any given source might simply have been different for a large fraction of time in the past. In an attempt to mitigate this as much as possible, we consider three representative proxies for the power output of the sources: their duty cycle (fraction of its lifetime that a transient source has spent in outburst), their total energy released during outbursts, and their instantaneous peak luminosity during said outbursts. Figure~\ref{fig:power} shows these results.

To ensure that any obtained correlation is not induced by the distance to the sources, we also compute flux-flux correlations whenever possible. In all cases, we find consistent results to the luminosity comparisons presented below.

\subsubsection{Duty cycle}
The duty cycle is a crucial parameter to characterize the power output of a source over long timescales. We use the estimations provided in the WATCHDOG catalogue~\citep{Tetarenko2016}, which unfortunately means that \maxi and V404~Cyg are missing from our sample. We can include \maxi by adopting as outburst starting time when it was first discovered~\citep[March 11th 2018,][]{Kawamuro2018} and end when it was reported to return to quiescence~\citep[June 11th 2023,][]{Baglio2023}, which amounts to a total of 1918 days. Taking the same total time interval as in the WATCHDOG catalogue for consistency translates to an upper limit on the duty cycle of 27\%. 

While we do not find a strong or significant correlation between gamma-ray luminosity and duty cycle, with $\tau_{K}=0.03$ (0.05$\sigma$), we note (a posteriori) that every system detected in the gamma-ray band has a duty cycle consistent with $>$20\%. In fact the majority of the detected systems have a duty cycle estimated to be close to 100\%. There are two systems with similarly high duty cycles which are visible by LHAASO but for which only upper limits have been reported: 4U~1957+115 (see Section~\ref{subsec:4u}) and Swift~J1753.5-0127 (one of the faintest radio emitters in our sample, see Figure~\ref{fig:radlum}).  Including the luminosity of the flares from \cygt as a lower limit has no impact on this conclusion.

\subsubsection{Total energy radiated during an outburst}
The temporal evolution of an \xray binary outburst can vary dramatically between different outbursts and sources~\citep[see e.g. ][]{Homan2005, Tremou2026}. The WATCHDOG catalogue~\citep{Tetarenko2016} compiled the outburst history for its 77 sources, including detailed \xray light-curves. They derived the total energy radiated in each outburst by integrating the \xray bolometric (0.001–1000~keV) luminosity of each outburst over its duration. We compare the gamma-ray luminosity to both the mean and maximum total energy released during outbursts, as reported in the WATCHDOG catalogue. In both cases, we find neither a strong nor significant correlation, with $\tau_{K}=0.05$ (0.08$\sigma$).  We note that the maximum and mean radiated energy are strongly correlated, which is why the values of their respective $\tau_{K}$ estimates match up to two significant digits. Including the luminosity of the flares from \cygt as a lower limit has no impact on this conclusion.
\subsubsection{Instantaneous luminosity}
We now consider the maximum luminosity recorded from  each of the sources. In the \xray range, the peak luminosity in the soft state is usually understood to reflect the instantaneous dissipation rate of accretion energy, and thus should be proportional to the rate of mass inflow from the disk. In the hard state, emission could also arise from plasma within the jet, in which dissipation of internal energy might not be instantaneous 

The same is true about the radio emission in the hard state, which is produced by plasma within the jet, and thus the peak radio luminosity is proportional to the average rate at which relativistic electrons are injected into the jet and the length of time for which they are injected~\citep{Fender2001b}. Additionally, the radio emission might suffer from significant self-absorption, which means that the observed peak flux might reflect a transition between the optically thin and thick regimes, and not a real change in the electron injection~\citep{Fender2019, Cowie2026a}.

\vspace{-0.3cm}
\paragraph{Peak \xray luminosity in the soft state:} We compare to the peak \xray bolometric luminosity during the soft \xray spectrum phase of an outburst as collected by the WATCHDOG catalogue.
We find $\tau_{K}=-0.1$ (0.28$\sigma$), meaning neither strong nor significant correlation. Including the luminosity of the flares from \cygt as a lower limit has no impact on this conclusion.
\paragraph{Peak \xray luminosity in the hard state.} We compare the gamma-ray luminosity to the peak \xray bolometric luminosity during the hard \xray spectral state of an outburst as collected by the WATCHDOG catalogue. We find a strong but not statistically significant correlation $\tau_{K}=0.81$ (1.62$\sigma$). A high value of $\tau_{K}$ with a significance below 3$\sigma$ means that while the effect size is large, the observed correlation might be the result of a statistical fluctuation, due to the low number of measurements of source luminosity above 100~TeV.  Including the luminosity of the flares from \cygt as a lower limit has no impact on this conclusion.

\paragraph{Peak radio luminosity:} Finally, we compared the gamma-ray luminosity above 100~TeV with the radio luminosities we collected for all sources in our sample (see Section~\ref{sec:jets}). Similarly to the hard-state luminosity, we find a strong but not statistically significant correlation $\tau_{K}=0.64$ (1.53$\sigma$) between the radio brightness of the sources and their gamma-ray output. That a correlation is found between the gamma-ray luminosity and both hard-state \xray and radio luminosities is not surprising given the well-established relation between the hard-state \xray luminosity and radio luminosity of black-holes~\citep[i.e. the Fundamental Plane, see][]{Merloni2003, Falcke2004, Crook-Mansour2026} and the fact that many (although not all) of the radio measurements in our sample were taken during the hard \xray state. Including the luminosity of the flares from \cygt as a lower limit has no impact on this conclusion.

\section{Cross-matching with gamma-ray catalogues}
\label{sec:the_cat}

The majority of gamma-ray sources are unassociated, meaning they do not have a clear counterpart in other wavelengths~\citep{HESSCollaboration2018, Cao2024a}. Due to their highly transient behaviour in other wavelengths, the majority of efforts to reveal their theorized gamma-ray emission has focused on short-duration observations during or around periods of brightening or spectral hardening in other wavelengths~\citep[e.g.][]{Atoyan1999a,Romero2003, Bosch-Ramon2006, Dubus2010}. Emission was thus expected to be point-like, an assumption which biased previous findings from gamma-ray telescope arrays~\citep[e.g.][]{H.E.S.S.Collaboration2018b,Abe2022}. Similarly, most earlier efforts to catalogue gamma-ray sources did not consider \xray binaries as a possible counterpart for spatially extended Galactic sources. Consequently, it is possible that previously unassociated sources can be attributed to \xray binaries, now that a clearer picture of the expected spectrum and morphology is arising. A complete study of the gamma-ray emission in the regions around the \xray binaries in our sample would require access to the data from the different gamma-ray instruments and is thus out of the scope of this paper. We instead consider a simple quantity: the distribution of angular distances between our sample of \xray binaries and unassociated gamma-ray sources. 

\subsection{Synthetic \xray binary population and strategy}

\begin{figure}
	\centering
		\includegraphics[width=0.98\linewidth]{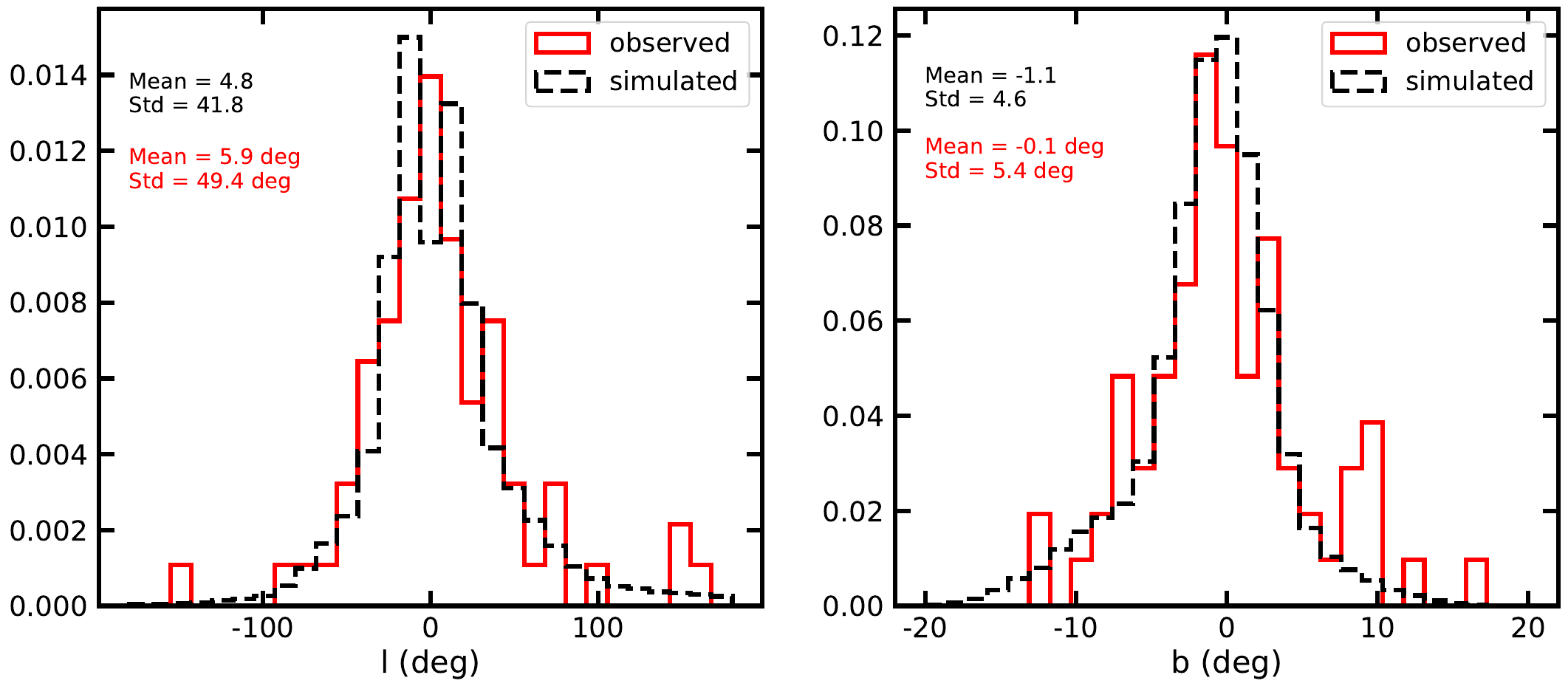} 
	\caption{\label{fig:sim} Distribution of Galactic latitude (left), longitude (right) for the real sample of \xray binaries (red lines) and the simulated one (black dashed lines). Sources located within a radius of 4$\degree$ of the Galactic centre have been excluded.}
\end{figure}

\begin{figure*}
	\centering
		\includegraphics[width=0.98\linewidth]{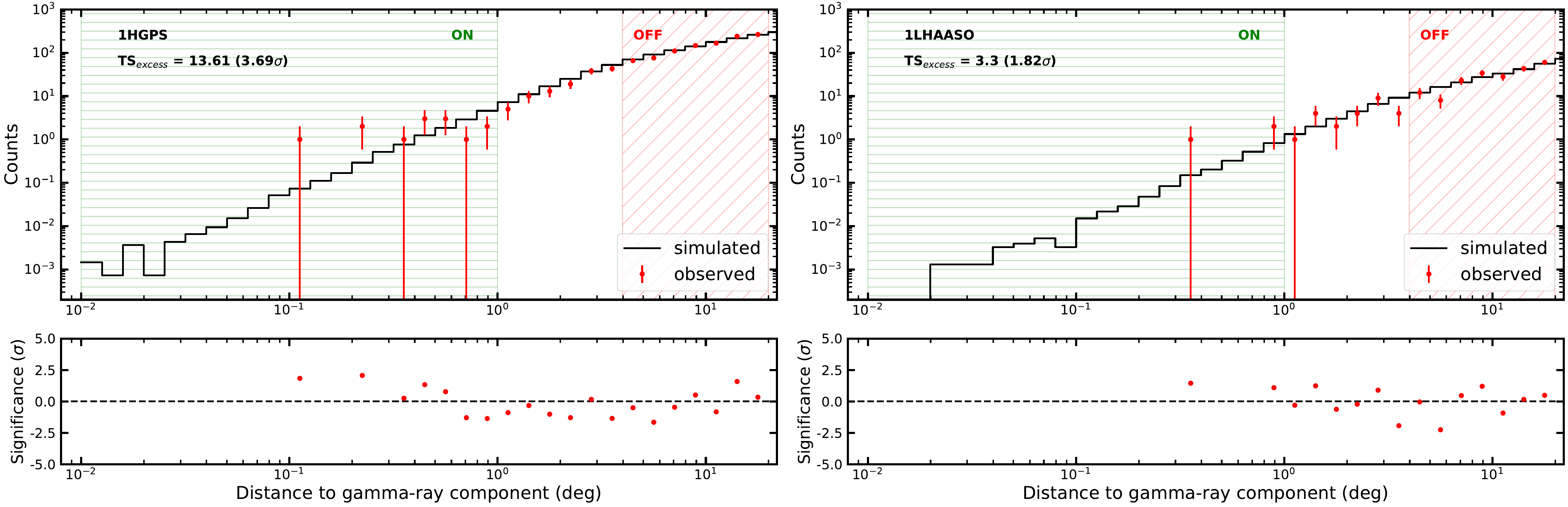} 
	\caption{\label{fig:cat} Distribution of simulated (black lines) and observed (red points) distances between \xray binaries in our sample and gamma-ray catalogue components without association for the 1HGPS catalogue (left) and 1LHAASO catalogue (right). In both panels, the region used to normalize the expected distribution (off) and derive the deviation (on) are indicated with a red diagonal and green horizontal hatch, respectively. In both panels, the residual between the observed and simulated distribution is provided in units of significance.}
\end{figure*}
We first need to determine our expectations for the distribution of distances between unassociated gamma-ray sources and a synthetic population which follows the same  Galactic coordinate distribution as our real \xray binary sample. To do this, we use a Conditional Tabular Generative Adversarial Network~\citep[CTGANs, ][]{Xu2019}. A CTGAN is a machine learning method in which a generator and a discriminator network compete to create increasingly realistic data, specifically designed to model conditional relationships between features in tabular data. In basic terms, the generator is trained to generate synthetic data which resembles the input data, while the discriminator is trained to distinguish between real and synthetic data, thus evaluating the performance of the generator. The goal of the model optimization is to create synthetic data so realistic that the discriminator is unable to distinguish it from the real input data.

We compare our \xray binary sample with unassociated gamma-ray sources from the 1HGPS~\citep{HESSCollaboration2018} and 1LHAASO~\citep{Cao2024a} catalogues. We include only objects within Galactic latitude $|b|<20\degree$ because there are no unassociated gamma-ray sources in our catalogues outside this range, and it avoids biases in the  synthetic sample from high $b$ outliers in the \xray binary distribution. 

When comparing to the gamma-ray catalogues, we exclude synthetic and real sources within a radius of 4$\degree$ of the Galactic Center because it is a crowded region hosting both large-scale diffuse emission and unassociated sources~\citep[see e.g.][]{HESSCollaboration2018a,Albert2024a} which should be subject of dedicated studies and are thus not well characterized by the catalogues. Figures~\ref{fig:sim} and~\ref{fig:sim2d} show the resulting distributions of Galactic longitude and latitude for the synthetic and real populations. There are 75 black-hole (candidate) \xray binaries in the real sample and 90774 in the synthetic sample.

We derive the distribution of distances between unassociated gamma-ray sources and our two samples. We define unassociated as not having a (non-gamma-ray) counterpart in any of the catalogues considered. We also exclude the sources in Table~\ref{tab:detected}, which already have a gamma-ray counterpart.

Because the synthetic \xray binary population has a much larger number of sources than the real sample, we need to normalize the resulting expected distribution so that we can compare between the two. We define the offset range between 4$\degree$ and 20$\degree$ as an "off" region 
in which we expect the distribution of distances for both synthetic and real samples to match, and normalize the synthetic distribution in that range. We then define an "on" region as the range between offsets 0$\degree$ and 1$\degree$ (twice the largest known offset between an \xray binary and the gamma-ray centroid) and compare the synthetic and real distributions in that range.  To do so, we derive the significance of the deviation of the number of observed sources within the "on" region to the expectation derived from the synthetic population taking into account the uncertainty on the background\footnote{\url{https://docs.gammapy.org/2.1/api/gammapy.stats.WStatCountsStatistic.html\#gammapy.stats.WStatCountsStatistic}}.

\begin{figure*}[b]
	\centering
		\includegraphics[width=0.98\linewidth]{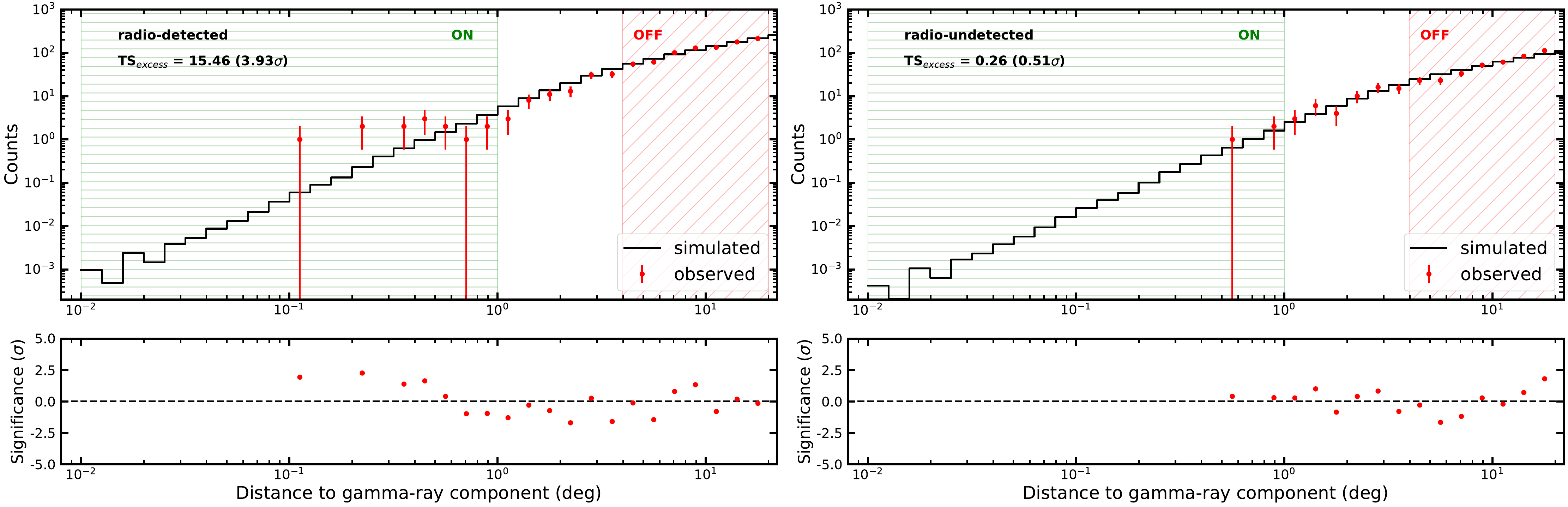} 
	\caption{\label{fig:radiocat} Symbols and colours follow those of Figure~\ref{fig:cat} but for the radio-detected (left) and radio-undetected (right) samples.}
\end{figure*}

\subsection{The different gamma-ray catalogues}
We compare our real and synthetic \xray binary samples to both the 1HGPS and 1LHAASO catalogues separately. They are chosen because of the differences in their sky coverage, namely the $\sim (-20\degree,80\degree)$ and $\sim (-83\degree,37\degree)$ declination ranges for LHAASO and \hess, respectively. In Galactic coordinates this translates (roughly) to coverage for Galactic longitudes of $\sim (0\degree,220\degree)$ and  $\sim (65\degree,-180\degree)$, respectively. 

To determine whether a source in the 1HGPS catalogue has a counterpart, we rely on the \texttt{HGPS\_Associations} table~\citep{HESSCollaboration2018}. We add a handful of associations missed by the catalogue and by now considered firm, such as HESS~J1747-248 with Terzan 5~\citep{H.E.S.S.Collaboration2011}, HESS~J1646-458 with Westerlund~1~\citep{Aharonian2022}, HESS~J1857+026 with PSR~J1856+0245~\citep{MAGICCollaboration2014} and HESS~J1713-381 with CTB 37B~\citep{Aharonian2008}. For the 1LHAASO catalogue we rely on the associations provided in~\citet{Cao2024a}.

We note that while every single one of the TeV-detected sources in Table~\ref{tab:detected} has been detected by LHAASO, there are not yet any \xray binaries detected only by \hess. The lack of \hess-only detections leads to the unlikely conclusion that while one half of the Galaxy hosts a handful of TeV-emitting \xray binaries, the other half does not. Consequently, if there is a correlation between the location of unassociated gamma-ray sources and \xray binaries, we would expect it to be stronger when comparing to the 1HGPS catalogue, as LHAASO has already identified several \xray binaries within their field of view (which we do not consider in our exercise, as mentioned earlier).

Figure~\ref{fig:cat} shows the distributions of distance between unassociated gamma-ray sources and \xray binaries (real and synthetic) for both the 1HGPS and 1LHAASO catalogues. When comparing to the 1HGPS, we find a significant deviation from the expectation of the synthetic sample with test-statistic ($TS$) of 13.6, which translates to over 3.7$\sigma$ deviation. The closest ($<0.5\degree$) separations are due to MAXI~J1631-479 (located 0.12$\degree$ away from a 1HGPS source component), 4U~1630-472 (located 0.22$\degree$ away from a 1HGPS source component), GRO~J1655-40 (located 0.25$\degree$ away from a 1HGPS source component) and MAXI~J1848-015 (located 0.4$\degree$ away from a 1HGPS source component). When comparing to the 1LHAASO catalogue, the significance is lower, with $TS=3.3$ (1.8$\sigma$). The only object with a separation smaller than 0.5$\degree$ is 4U~1957+115, already introduced in Section~\ref{subsec:4u}. 

Note that these results are not sufficient to claim an association between any particular \xray binary and their closest gamma-ray components - a detailed study of each region would be needed to make that claim. Additionally, our comparison does not take into account whether there is already another \xray binary located closer to the gamma-ray component, effectively allowing double-counting (in both the real and synthetic samples).

\subsection{Radio-detected and radio-undetected systems}
Around two thirds (N=49) of the \xray binaries which fall within the criteria of our comparison have been detected in the radio band (see Section~\ref{sec:jets}), whereas the other third (N=21) have no reported radio detection. We refer to these two sub-samples as radio-detected and radio-undetected, respectively. We perform the same comparison as before, but this time distinguishing between the two \xray binary samples and comparing against both gamma-ray catalogues at once. Figure~\ref{fig:radiocat} shows the distributions of distance between unassociated gamma-ray sources and \xray binaries (real and synthetic) for both the radio-detected and radio-undetected cases. When comparing to the radio-detected sample, we find a significant deviation between the simulated and observed distributions with $TS=15.46$ (3.9$\sigma$). For those with no reported radio detection case, the deviation becomes not statistically significant, with $TS=0.26$ (0.51$\sigma$). We note, however, that the fraction of class "C" objects, that is, those for which the evidence of a black-hole binary is weakest, is much higher in the radio-undetected sample (66\%) than in the radio-detected sample (13\%). We repeated the comparison using only class A+B sources, and found an excess at low distances which deviates by $TS=16.8$ ($\sim 4.1 \sigma$) from the expectation for the radio-detected sample, whereas none of the 7 black-hole (candidate) \xray binaries without a reported radio detection fall within 1$\degree$ of an unassociated gamma-ray source.

\subsection{Taking into account the jet direction}
There are 23 sources in our sample for which the jet position angle has been constrained (see Table~\ref{tab:sources}). All five of the sources detected in the gamma-ray band have a relatively well constrained jet position angle. An additional 4 systems with reported jet position angles are located within a 4$\degree$ radius of the Galactic centre and thus excluded from our comparison.  We used the remaining 14 sources to perform an additional comparison with gamma-ray catalogues, which takes into account the jet direction. A jet position angle was assigned to every synthetic \xray binary by sampling from a uniform distribution ranging between 0 and 180$\degree$. We then counted the number of unidentified gamma-ray sources which lie less than 2$\degree$ and within a cone of 15$\degree$ opening angle in the direction of the jet (on both sides) from the position of an \xray binary (real or simulated).  When comparing to the 1HGPS catalogue, we find 2 such coincidences in the real sample: GRO~J1655-40 and EXO~1846-031 (0.25$\degree$ and 1.14$\degree$ away from the gamma-ray source, respectively) and 11013 in the synthetic one. Ignoring the cone requirement, there are 11 occurrences of an \xray binary in the real sample (with reported jet position angle) being within 2$\degree$ of an unassociated gamma-ray source. In the synthetic sample that number is 66222. Taking the ratios results in an expectation of 1.8 such coincidences, which is statistically consistent with the observed 2. For the LHAASO catalogue there are no coincidences.

\subsection{The Gaia dormant black holes}
Because of their low number, we do not include the Gaia dormant black holes in our systematic comparison with gamma-ray catalogues, and instead perform a check by hand. 
Gaia BH1 is located outside the Galactic plane~\citep{ElBadry2022}, at $b\sim 18 \degree$, in a region of the sky not covered by the 1HGPS and more than 10$\degree$ away from the nearest 1LHAASO catalogue source. Gaia BH2~\citep{Tanikawa2023} is only visible by \hess but located more than 3$\degree$ away from the nearest 1HGPS catalogue source. Finally, Gaia BH3~\citep{GaiaColl2024} is visible by both but located 4$\degree$  and 2.7$\degree$ from the nearest 1HGPS and 1LHAASO sources, respectively. We thus find no evidence of an association between the unidentified gamma-ray sources in the considered catalogues and the three dormant black holes identified so far.

\section{Conclusions}
\label{sec:the_conclusions}
By combining the different results presented in the paper, we arrive at our main conclusion: radio-detected black-hole \xray binaries with a high duty cycle are more likely to be detected as gamma-ray sources than radio-undetected objects or those with low duty cycles. The latter can be understood trivially as a requirement on the sustained power necessary to produce enough gamma-rays to be detected by the current generation of instruments. The first is less obvious and indicates that the particles responsible for the emission would be accelerated as a consequence of jet activity, either within the jets themselves or in the interaction of the jets with their surrounding medium. The fact that the observed gamma-ray morphology is, in most cases, offset from the binary in a direction consistent with that of jets supports this conclusion. 

We find a strong but not yet significant - as expected due to the low number of sources detected so far - correlation  between the gamma-ray luminosity above 100~TeV and observational proxies for the jet power, suggesting a link between the power available in the jets and the power carried by accelerated particles. We interpret this link to suggest that particle acceleration is either taking place within the jets or triggered by interaction between the jets and other components in the system. The lack of such correlation with the \xray luminosity in the soft state, which is dominated by the accretion disk and not the jet, supports this picture. Recently~\citet{2025ApJ...989L..25W} posited that outflows in highly accreting \xray binaries re natural super-PeVatron candidates, linking outflow power and maximum attainable particle energy. While here we do not test this hypothesis directly here, our results support the overall picture presented in that work.

We stress, however, that even though the measured correlations imply a strong effect, their significance is low, meaning that the available data is not sufficient to firmly establish it. Applying the same correlation estimate to a dummy dataset consistent of five perfectly linearly correlated points with uncertainties in the order of 10\% yields, as expected, a strong correlation ($\tau_{K}\sim 1$) but only 2$\sigma$ significance. In this dummy, perfectly correlated dataset, a sample of at least $\sim$9 entries and $\sim$15 entries is required to reach 3 and 5$\sigma$ significance, respectively. Consequently, increasing the number of objects detected in the gamma-ray band will be crucial to establish the observed correlation. Beyond additional detections, a more detailed characterization of the gamma-ray emission above 100~TeV, is also critical to make the jet connection suggested by our results firm. In particular, detailed, energy-resolved studies of the gamma-ray morphology are crucial to establish the location of particle acceleration in \xray binary systems and thus, provide a complementary constraint to the theoretical picture.

As detailed in Section~\ref{sec:gamma-ray}, most of the systems detected in the gamma-ray band display extended and/or offset emission, in spatial scales which, even assuming purely ballistic transport at the speed of light, translate to at least hundreds of years. In some cases, providing the necessary power to explain the gamma-ray emission while keeping the average fraction of total jet power accounted for by particles relatively low ($<10\%$) requires particles to accumulate for thousands of years~\citep[e.g.][]{HESSCollaboration2024, Acharyya2026, Wan2026}. These timescales are obviously much longer than the few decades (at best) that \xray binaries have been monitored in the \xray and radio bands. This highlights a contradiction in our sample: the gamma-ray luminosities reflect the average behaviour of the sources on timescales much longer than the few decades for which \xray binary monitoring exists at any other wavelength. Furthermore, we find a correlation between the gamma-ray luminosities and peak radio/\xray luminosities but not with time-averaged quantities like the total energy radiated during an outburst, leading to the somewhat counter-intuitive conclusion that the maximum instantaneous power released seems to be more relevant to the $>100$~TeV gamma-ray output than how long the powerful outburst is sustained for. We interpret the obtained correlations to imply that the maximum luminosity might reflect a representative quantity of the capabilities of the source, but note that effects like obscuration or viewing angle can significantly impact these values. 

The evidence for "dark jets" in a number of black-hole \xray binaries~\citep[e.g.][]{Gallo2005, Motta2025, Cowie2026} suggests that the electromagnetic output of these systems might only be a lower limit on their actual power output. In such a scenario, the maximum luminosity would act as a proxy for the amount of power a source can possibly produce, while most of the power required to produce the PeV particles is carried instead in a radiatively inefficient manner.

We find no correlation between the gamma-ray luminosity and quantities like the compact object mass, the companion mass or the inclination angle, although note that many of these quantities are subject to relatively large observational uncertainties.

Additionally, when comparing to gamma-ray catalogues, we find significant ($\sim$3.7$\sigma$) evidence that some unassociated gamma-ray sources in the Southern sky are associated with \xray binaries. This is, in a sense, an obvious statement since LHAASO has detected 5 sources in the range of Galactic longitudes between -180 and 0$\degree$, and there is no reason to believe that the range between 0 and 180$\degree$ would be different. 

Perhaps more intriguingly, we find that the association is only significant when considering the population of radio-detected \xray binaries, indicating again a connection between gamma-ray emission and jet activity. However, we note that out of the 21 sources in the sample with no reported radio detection, we were able to find a report of observations of these sources for only half of them. While it is possible that some of the remaining sources have been observed with a radio telescope and simply not reported, this lack of coverage highlights an important observational bias in our sample. Additionally, the sample of objects without a reported radio detection includes a much larger fraction of objects for which the identification as a black-hole \xray binary is based on relatively weak evidence, meaning they might actually be some other kind of X-ray transient. In that case, they might not be detected in the radio band because they are simply not expected to ever produce jets. Consequently, the increased significance obtained with the radio-detected sample could also be the result of a cleaner black-hole binary sample which includes less "noise" from other types of sources. Either case is, however, consistent with the conclusion that detection of radio emission (as proxy for jet activity) and gamma-ray emission are tentatively linked.

We find no gamma-ray sources (unassociated or otherwise) in the vicinity of the Gaia dormant black holes, although note that there are only three such objects known so far, and that they lie at least a few degrees away from the Galactic plane, where exposure is much lower for the 1HGPS catalogue. The fourth Gaia data release\footnote{\url{https://www.cosmos.esa.int/web/gaia/dr4}} is expected to reveal the positions of many more dormant black holes, which should be investigated as possible gamma-ray sources.

Our results disprove the hypothesis that recent flaring activity is correlated with a gamma-ray detection, proposed by~\citet{LhaasoCollaboration2025}. By definition, every single one of the \xray binaries in our sample has flared recently (otherwise, they would not have been catalogued as a candidate black-hole \xray binary) and yet, a large fraction of them remain undetected in the gamma-ray band. We note that LHAASO has detected gamma-ray emission from approximately a third of the radio-detected class A+B sources within its field of view, which, under a naive scaling assumption indicates that there should be over 10 other radio-detected systems within reach of current and future generation gamma-ray telescopes. In particular, instruments like the proposed Southern Wide-Field Gamma-ray Observatory~\citep[SWGO, ][]{SWGOCollaboration2025} and the upcoming Cherenkov Telescope Array Observatory~\citep[CTA, ][]{ctagps} will provide unprecedented coverage of the southern sky, and should be able to (perhaps more than) double the number of black-hole \xray binaries detected in the TeV and PeV range.

\begin{acknowledgements}
We thank the anonymous referee for their valuable comments and suggestions. LON thanks Andrzej Zdziarski for pointing out a typo in Table A.1. LON acknowledges support from the Max Planck Society. RF, SM, JCM and LON acknowledge support from European Research Council Synergy Grant ‘Blackholistic’ (grant agreement no. 101071643). FJC was supported by STFC grant ST/Y509474/1. RF further acknowledges support from the UKRI and the Hintze family charitable foundation. JCM further acknowledges financial support from the Rhodes Trust.
\end{acknowledgements}

\bibliographystyle{aa}
\bibliography{uQ}

@Article{Chauhan2021,
  author        = {Chauhan, J. and Miller-Jones, J. C. A. and Raja, W. and Allison, J. R. and Jacob, P. F. L. and Anderson, G. E. and Carotenuto, F. and Corbel, S. and Fender, R. and Hotan, A. and Whiting, M. and Woudt, P. A. and Koribalski, B. and Mahony, E.},
  journal       = {\mnras},
  title         = {Measuring the distance to the black hole candidate X-ray binary MAXI J1348-630 using H I absorption},
  year          = {2021},
  month         = feb,
  number        = {1},
  pages         = {L60-L64},
  volume        = {501},
  archiveprefix = {arXiv},
  doi           = {10.1093/mnrasl/slaa195},
  eprint        = {2009.14419},
  primaryclass  = {astro-ph.HE},
  url           = {https://ui.adsabs.harvard.edu/abs/2021MNRAS.501L..60C},
}

@Article{Charles2019,
  author        = {Charles, Phil and Matthews, James H. and Buckley, David A. H. and Gandhi, Poshak and Kotze, Enrico and Paice, John},
  journal       = {\mnras},
  title         = {Hot, dense He II outflows during the 2017 outburst of the X-ray transient Swift J1357.2-0933},
  year          = {2019},
  month         = oct,
  number        = {1},
  pages         = {L47-L52},
  volume        = {489},
  archiveprefix = {arXiv},
  doi           = {10.1093/mnrasl/slz120},
  eprint        = {1908.00320},
  primaryclass  = {astro-ph.HE},
  url           = {https://ui.adsabs.harvard.edu/abs/2019MNRAS.489L..47C},
}

@Article{Atri2020,
  author        = {Atri, P. and Miller-Jones, J. C. A. and Bahramian, A. and Plotkin, R. M. and Deller, A. T. and Jonker, P. G. and Maccarone, T. J. and Sivakoff, G. R. and Soria, R. and Altamirano, D. and Belloni, T. and Fender, R. and Koerding, E. and Maitra, D. and Markoff, S. and Migliari, S. and Russell, D. and Russell, T. and Sarazin, C. L. and Tetarenko, A. J. and Tudose, V.},
  journal       = {\mnras},
  title         = {A radio parallax to the black hole X-ray binary MAXI J1820+070},
  year          = {2020},
  month         = mar,
  number        = {1},
  pages         = {L81-L86},
  volume        = {493},
  archiveprefix = {arXiv},
  doi           = {10.1093/mnrasl/slaa010},
  eprint        = {1912.04525},
  primaryclass  = {astro-ph.HE},
  url           = {https://ui.adsabs.harvard.edu/abs/2020MNRAS.493L..81A},
}

@Article{Gandhi2019,
  author        = {Gandhi, Poshak and Rao, Anjali and Johnson, Michael A. C. and Paice, John A. and Maccarone, Thomas J.},
  journal       = {\mnras},
  title         = {Gaia Data Release 2 distances and peculiar velocities for Galactic black hole transients},
  year          = {2019},
  month         = may,
  number        = {2},
  pages         = {2642-2655},
  volume        = {485},
  archiveprefix = {arXiv},
  doi           = {10.1093/mnras/stz438},
  eprint        = {1804.11349},
  primaryclass  = {astro-ph.HE},
  url           = {https://ui.adsabs.harvard.edu/abs/2019MNRAS.485.2642G},
}

@Article{Shaposhnikov2010,
  author        = {Shaposhnikov, Nikolai and Markwardt, Craig and Swank, Jean and Krimm, Hans},
  journal       = {\apj},
  title         = {Discovery and Monitoring of a New Black Hole Candidate XTE J1752-223 with RXTE: Rms Spectrum Evolution, Black Hole Mass, and the Source Distance},
  year          = {2010},
  month         = nov,
  number        = {2},
  pages         = {1817-1824},
  volume        = {723},
  archiveprefix = {arXiv},
  doi           = {10.1088/0004-637X/723/2/1817},
  eprint        = {1008.0597},
  primaryclass  = {astro-ph.HE},
  url           = {https://ui.adsabs.harvard.edu/abs/2010ApJ...723.1817S},
}

@Article{Ratti2012,
  author        = {Ratti, E. M. and Jonker, P. G. and Miller-Jones, J. C. A. and Torres, M. A. P. and Homan, J. and Markoff, S. and Tomsick, J. A. and Kaaret, P. and Wijnands, R. and Gallo, E. and {\"O}zel, F. and Steeghs, D. T. H. and Fender, R. P.},
  journal       = {\mnras},
  title         = {The black hole candidate XTE J1752-223 towards and in quiescence: optical and simultaneous X-ray-radio observations},
  year          = {2012},
  month         = jul,
  number        = {3},
  pages         = {2656-2667},
  volume        = {423},
  archiveprefix = {arXiv},
  doi           = {10.1111/j.1365-2966.2012.21071.x},
  eprint        = {1204.2735},
  primaryclass  = {astro-ph.HE},
  url           = {https://ui.adsabs.harvard.edu/abs/2012MNRAS.423.2656R},
}

@Article{Reid2023,
  author        = {Reid, M. J. and Miller-Jones, J. C. A.},
  journal       = {\apj},
  title         = {On the Distances to the X-Ray Binaries Cygnus X-3 and GRS 1915+105},
  year          = {2023},
  month         = dec,
  number        = {2},
  pages         = {85},
  volume        = {959},
  archiveprefix = {arXiv},
  doi           = {10.3847/1538-4357/acfe0c},
  eid           = {85},
  eprint        = {2309.15027},
  primaryclass  = {astro-ph.HE},
  url           = {https://ui.adsabs.harvard.edu/abs/2023ApJ...959...85R},
}

@Article{CorralSantana2011,
  author        = {Corral-Santana, J. M. and Casares, J. and Shahbaz, T. and Zurita, C. and Mart{\'\i}nez-Pais, I. G. and Rodr{\'\i}guez-Gil, P.},
  journal       = {\mnras},
  title         = {Evidence for a black hole in the X-ray transient XTE J1859+226},
  year          = {2011},
  month         = may,
  number        = {1},
  pages         = {L15-L19},
  volume        = {413},
  archiveprefix = {arXiv},
  doi           = {10.1111/j.1745-3933.2011.01022.x},
  eprint        = {1102.0654},
  primaryclass  = {astro-ph.SR},
  url           = {https://ui.adsabs.harvard.edu/abs/2011MNRAS.413L..15C},
}

@Article{Yamaoka2025,
  author        = {Yamaoka, Kazutaka and Kawaguchi, Toshihiro and McCollough, Michael L. and Farinelli, Ruben and Trushkin, Sergei},
  journal       = {\pasj},
  title         = {X-ray spectral and timing properties of the black hole binary XTE J1859+226 and their relation to jets},
  year          = {2025},
  month         = apr,
  number        = {2},
  pages         = {237-259},
  volume        = {77},
  archiveprefix = {arXiv},
  doi           = {10.1093/pasj/psae113},
  eprint        = {2412.02977},
  primaryclass  = {astro-ph.HE},
  url           = {https://ui.adsabs.harvard.edu/abs/2025PASJ...77..237Y},
}

@Article{Sala2007,
  author        = {Sala, G. and Greiner, J. and Ajello, M. and Bottacini, E. and Haberl, F.},
  journal       = {\aap},
  title         = {XMM-Newton and INTEGRAL observations of the black hole candidate <ASTROBJ>XTE J1817-330</ASTROBJ>},
  year          = {2007},
  month         = oct,
  number        = {2},
  pages         = {561-568},
  volume        = {473},
  archiveprefix = {arXiv},
  doi           = {10.1051/0004-6361:20077360},
  eprint        = {0707.4155},
  primaryclass  = {astro-ph},
  url           = {https://ui.adsabs.harvard.edu/abs/2007A&A...473..561S},
}

@Article{MataSanchez2022,
  author        = {Mata S{\'a}nchez, D. and Mu{\~n}oz-Darias, T. and C{\'u}neo, V. A. and Armas Padilla, M. and S{\'a}nchez-Sierras, J. and Panizo-Espinar, G. and Casares, J. and Corral-Santana, J. M. and Torres, M. A. P.},
  journal       = {\apjl},
  title         = {Hard-state Optical Wind during the Discovery Outburst of the Black Hole X-Ray Dipper MAXI J1803-298},
  year          = {2022},
  month         = feb,
  number        = {2},
  pages         = {L10},
  volume        = {926},
  archiveprefix = {arXiv},
  doi           = {10.3847/2041-8213/ac502f},
  eid           = {L10},
  eprint        = {2201.09896},
  primaryclass  = {astro-ph.HE},
  url           = {https://ui.adsabs.harvard.edu/abs/2022ApJ...926L..10M},
}

@Article{ArmasPadilla2019,
  author        = {Armas Padilla, M. and Mu{\~n}oz-Darias, T. and S{\'a}nchez-Sierras, J. and De Marco, B. and Jim{\'e}nez-Ibarra, F. and Casares, J. and Corral-Santana, J. M. and Torres, M. A. P.},
  journal       = {\mnras},
  title         = {Multiwavelength spectroscopy of the black hole candidate MAXI J1813-095 during its discovery outburst},
  year          = {2019},
  month         = jun,
  number        = {4},
  pages         = {5235-5243},
  volume        = {485},
  archiveprefix = {arXiv},
  doi           = {10.1093/mnras/stz737},
  eprint        = {1903.04498},
  primaryclass  = {astro-ph.HE},
  url           = {https://ui.adsabs.harvard.edu/abs/2019MNRAS.485.5235A},
}

@Article{Jana2021,
  author        = {Jana, Arghajit and Jaisawal, Gaurava K. and Naik, Sachindra and Kumari, Neeraj and Chhotaray, Birendra and Altamirano, D. and Remillard, R. A. and Gendreau, Keith C.},
  journal       = {\mnras},
  title         = {NICER observations of the black hole candidate MAXI J0637-430 during the 2019-2020 outburst},
  year          = {2021},
  month         = jul,
  number        = {4},
  pages         = {4793-4805},
  volume        = {504},
  archiveprefix = {arXiv},
  doi           = {10.1093/mnras/stab1231},
  eprint        = {2104.13005},
  primaryclass  = {astro-ph.HE},
  url           = {https://ui.adsabs.harvard.edu/abs/2021MNRAS.504.4793J},
}

@Article{Peng2024,
  author        = {Peng, Jing-Qiang and Zhang, Shu and Shui, Qing-Cang and Chen, Yu-Peng and Zhang, Shuang-Nan and Kong, Ling-Da and Santangelo, A. and Yu, Zhuo-Li and Ji, Long and Wang, Peng-Ju and Chang, Zhi and Li, Jian and Li, Zhao-sheng},
  journal       = {\apjl},
  title         = {Insight-HXMT, NICER, and NuSTAR Views to the Newly Discovered Black Hole X-Ray Binary Swift J151857.0{\textendash}572147},
  year          = {2024},
  month         = sep,
  number        = {1},
  pages         = {L7},
  volume        = {973},
  archiveprefix = {arXiv},
  doi           = {10.3847/2041-8213/ad74ec},
  eid           = {L7},
  eprint        = {2503.03093},
  primaryclass  = {astro-ph.HE},
  url           = {https://ui.adsabs.harvard.edu/abs/2024ApJ...973L...7P},
}

@Article{Rout2023,
  author        = {Rout, Sandeep K. and Vadawale, Santosh and Gar{\'c}ia, Javier and Connors, Riley},
  journal       = {\apj},
  title         = {Revisiting the Galactic X-Ray Binary MAXI J1631-479: Implications for High Inclination and a Massive Black Hole},
  year          = {2023},
  month         = feb,
  number        = {1},
  pages         = {68},
  volume        = {944},
  archiveprefix = {arXiv},
  doi           = {10.3847/1538-4357/acaaa4},
  eid           = {68},
  eprint        = {2212.05293},
  primaryclass  = {astro-ph.HE},
  url           = {https://ui.adsabs.harvard.edu/abs/2023ApJ...944...68R},
}

@Article{Bozzo2016,
  author        = {Bozzo, E. and Pjanka, P. and Romano, P. and Papitto, A. and Ferrigno, C. and Motta, S. and Zdziarski, A. A. and Pintore, F. and Di Salvo, T. and Burderi, L. and Lazzati, D. and Ponti, G. and Pavan, L.},
  journal       = {\aap},
  title         = {IGR J17451-3022: A dipping and eclipsing low mass X-ray binary},
  year          = {2016},
  month         = may,
  pages         = {A42},
  volume        = {589},
  archiveprefix = {arXiv},
  doi           = {10.1051/0004-6361/201527501},
  eid           = {A42},
  eprint        = {1603.03353},
  primaryclass  = {astro-ph.HE},
  url           = {https://ui.adsabs.harvard.edu/abs/2016A&A...589A..42B},
}

@Article{Paizis2015,
  author        = {Paizis, A. and Nowak, M. A. and Rodriguez, J. and Segreto, A. and Chaty, S. and Rau, A. and Chenevez, J. and Del Santo, M. and Greiner, J. and Schmidl, S.},
  journal       = {\apj},
  title         = {Investigating the Nature of IGR J17454-2919 Using X-Ray and Near-infrared Observations},
  year          = {2015},
  month         = jul,
  number        = {1},
  pages         = {34},
  volume        = {808},
  archiveprefix = {arXiv},
  doi           = {10.1088/0004-637X/808/1/34},
  eid           = {34},
  eprint        = {1506.01205},
  primaryclass  = {astro-ph.HE},
  url           = {https://ui.adsabs.harvard.edu/abs/2015ApJ...808...34P},
}

@Article{Tetarenko2016,
  author        = {Tetarenko, B. E. and Sivakoff, G. R. and Heinke, C. O. and Gladstone, J. C.},
  journal       = {\apjs},
  title         = {WATCHDOG: A Comprehensive All-sky Database of Galactic Black Hole X-ray Binaries},
  year          = {2016},
  month         = feb,
  number        = {2},
  pages         = {15},
  volume        = {222},
  archiveprefix = {arXiv},
  doi           = {10.3847/0067-0049/222/2/15},
  eid           = {15},
  eprint        = {1512.00778},
  primaryclass  = {astro-ph.HE},
  url           = {https://ui.adsabs.harvard.edu/abs/2016ApJS..222...15T},
}

@Article{CorralSantana2016,
  author        = {Corral-Santana, J. M. and Casares, J. and Mu{\~n}oz-Darias, T. and Bauer, F. E. and Mart{\'\i}nez-Pais, I. G. and Russell, D. M.},
  journal       = {\aap},
  title         = {BlackCAT: A catalogue of stellar-mass black holes in X-ray transients},
  year          = {2016},
  month         = mar,
  pages         = {A61},
  volume        = {587},
  archiveprefix = {arXiv},
  doi           = {10.1051/0004-6361/201527130},
  eid           = {A61},
  eprint        = {1510.08869},
  primaryclass  = {astro-ph.HE},
  url           = {https://ui.adsabs.harvard.edu/abs/2016A&A...587A..61C},
}

@Article{Fender1999,
  author        = {Fender, R. P. and Garrington, S. T. and McKay, D. J. and Muxlow, T. W. B. and Pooley, G. G. and Spencer, R. E. and Stirling, A. M. and Waltman, E. B.},
  journal       = {\mnras},
  title         = {MERLIN observations of relativistic ejections from GRS 1915+105},
  year          = {1999},
  month         = apr,
  number        = {4},
  pages         = {865-876},
  volume        = {304},
  archiveprefix = {arXiv},
  doi           = {10.1046/j.1365-8711.1999.02364.x},
  eprint        = {astro-ph/9812150},
  primaryclass  = {astro-ph},
  url           = {https://ui.adsabs.harvard.edu/abs/1999MNRAS.304..865F},
}

@Article{Chaty2001,
  author        = {Chaty, S. and Rodr{\'\i}guez, L. F. and Mirabel, I. F. and Geballe, T. R. and Fuchs, Y. and Claret, A. and Cesarsky, C. J. and Cesarsky, D.},
  journal       = {\aap},
  title         = {A search for possible interactions between ejections from GRS 1915+105 and the surrounding interstellar medium},
  year          = {2001},
  month         = feb,
  pages         = {1035-1046},
  volume        = {366},
  archiveprefix = {arXiv},
  doi           = {10.1051/0004-6361:20000266},
  eprint        = {astro-ph/0011297},
  primaryclass  = {astro-ph},
  url           = {https://ui.adsabs.harvard.edu/abs/2001A&A...366.1035C},
}

@Article{Mirabel1994,
  author   = {Mirabel, I. F. and Rodr{\'\i}guez, L. F.},
  journal  = {\nat},
  title    = {A superluminal source in the Galaxy},
  year     = {1994},
  month    = sep,
  number   = {6492},
  pages    = {46-48},
  volume   = {371},
  doi      = {10.1038/371046a0},
  url      = {https://ui.adsabs.harvard.edu/abs/1994Natur.371...46M},
}

@Article{Tetarenko2018,
  author        = {Tetarenko, A. J. and Freeman, P. and Rosolowsky, E. W. and Miller-Jones, J. C. A. and Sivakoff, G. R.},
  journal       = {\mnras},
  title         = {Mapping jet-ISM interactions in X-ray binaries with ALMA: a GRS 1915+105 case study},
  year          = {2018},
  month         = mar,
  number        = {1},
  pages         = {448-468},
  volume        = {475},
  archiveprefix = {arXiv},
  doi           = {10.1093/mnras/stx3151},
  eprint        = {1712.00432},
  primaryclass  = {astro-ph.HE},
  url           = {https://ui.adsabs.harvard.edu/abs/2018MNRAS.475..448T},
}

@Article{Zdziarski2014,
  author        = {Zdziarski, Andrzej A.},
  journal       = {\mnras},
  title         = {The jet kinetic power, distance and inclination of GRS 1915+105},
  year          = {2014},
  month         = oct,
  number        = {2},
  pages         = {1113-1118},
  volume        = {444},
  archiveprefix = {arXiv},
  doi           = {10.1093/mnras/stu1525},
  eprint        = {1406.1908},
  primaryclass  = {astro-ph.HE},
  url           = {https://ui.adsabs.harvard.edu/abs/2014MNRAS.444.1113Z},
}

@Article{Fender2000,
  author        = {Fender, R. P. and Pooley, G. G.},
  journal       = {\mnras},
  title         = {Giant repeated ejections from GRS 1915+105},
  year          = {2000},
  month         = oct,
  number        = {1},
  pages         = {L1-L5},
  volume        = {318},
  archiveprefix = {arXiv},
  doi           = {10.1046/j.1365-8711.2000.03847.x},
  eprint        = {astro-ph/0006278},
  primaryclass  = {astro-ph},
  url           = {https://ui.adsabs.harvard.edu/abs/2000MNRAS.318L...1F},
}

@Article{2025A&A...696A.222M,
  author        = {Motta, S. E. and Atri, P. and Matthews, James H. and van den Eijnden, Jakob and Fender, Rob P. and Miller-Jones, James C. A. and Heywood, Ian and Woudt, Patrick},
  journal       = {\aap},
  title         = {MeerKAT discovers a jet-driven bow shock near GRS 1915+105: How an invisible large-scale jet sculpts a microquasar's environment},
  year          = {2025},
  month         = apr,
  pages         = {A222},
  volume        = {696},
  archiveprefix = {arXiv},
  doi           = {10.1051/0004-6361/202452838},
  eid           = {A222},
  eprint        = {2504.17425},
  primaryclass  = {astro-ph.HE},
  url           = {https://ui.adsabs.harvard.edu/abs/2025A&A...696A.222M},
}

@Article{Hjellming1995,
  author   = {Hjellming, R. M. and Rupen, M. P.},
  journal  = {\nat},
  title    = {Episodic ejection of relativistic jets by the X-ray transient GRO J1655 - 40},
  year     = {1995},
  month    = jun,
  number   = {6531},
  pages    = {464-468},
  volume   = {375},
  doi      = {10.1038/375464a0},
  url      = {https://ui.adsabs.harvard.edu/abs/1995Natur.375..464H},
}

@Article{Tingay1995,
  author   = {Tingay, S. J. and Jauncey, D. L. and Preston, R. A. and Reynolds, J. E. and Meier, D. L. and Murphy, D. W. and Tzioumis, A. K. and McKay, D. J. and Kesteven, M. J. and Lovell, J. E. J. and Campbell-Wilson, D. and Ellingsen, S. P. and Gough, R. and Hunstead, R. W. and Jonos, D. L. and McCulloch, P. M. and Migenes, V. and Quick, J. and Sinclair, M. W. and Smits, D.},
  journal  = {\nat},
  title    = {Relativistic motion in a nearby bright X-ray source},
  year     = {1995},
  month    = mar,
  number   = {6518},
  pages    = {141-143},
  volume   = {374},
  doi      = {10.1038/374141a0},
  url      = {https://ui.adsabs.harvard.edu/abs/1995Natur.374..141T},
}

@Article{Migliari2007,
  author        = {Migliari, S. and Tomsick, J. A. and Markoff, S. and Kalemci, E. and Bailyn, C. D. and Buxton, M. and Corbel, S. and Fender, R. P. and Kaaret, P.},
  journal       = {\apj},
  title         = {Tracing the Jet Contribution to the Mid-IR over the 2005 Outburst of GRO J1655-40 via Broadband Spectral Modeling},
  year          = {2007},
  month         = nov,
  number        = {1},
  pages         = {610-623},
  volume        = {670},
  archiveprefix = {arXiv},
  doi           = {10.1086/522023},
  eprint        = {0707.4500},
  primaryclass  = {astro-ph},
  url           = {https://ui.adsabs.harvard.edu/abs/2007ApJ...670..610M},
}

@Article{Greene2001,
  author        = {Greene, Jenny and Bailyn, Charles D. and Orosz, Jerome A.},
  journal       = {\apj},
  title         = {Optical and Infrared Photometry of the Microquasar GRO J1655-40 in Quiescence},
  year          = {2001},
  month         = jun,
  number        = {2},
  pages         = {1290-1297},
  volume        = {554},
  archiveprefix = {arXiv},
  comment       = {mass and inclination},
  doi           = {10.1086/321411},
  eprint        = {astro-ph/0101337},
  primaryclass  = {astro-ph},
  url           = {https://ui.adsabs.harvard.edu/abs/2001ApJ...554.1290G},
}

@Article{Gallo2004,
  author        = {Gallo, E. and Corbel, S. and Fender, R. P. and Maccarone, T. J. and Tzioumis, A. K.},
  journal       = {\mnras},
  title         = {A transient large-scale relativistic radio jet from GX 339-4},
  year          = {2004},
  month         = jan,
  number        = {3},
  pages         = {L52-L56},
  volume        = {347},
  archiveprefix = {arXiv},
  doi           = {10.1111/j.1365-2966.2004.07435.x},
  eprint        = {astro-ph/0311452},
  primaryclass  = {astro-ph},
  url           = {https://ui.adsabs.harvard.edu/abs/2004MNRAS.347L..52G},
}

@Article{Corbel2000,
  author        = {Corbel, S. and Fender, R. P. and Tzioumis, A. K. and Nowak, M. and McIntyre, V. and Durouchoux, P. and Sood, R.},
  journal       = {\aap},
  title         = {Coupling of the X-ray and radio emission in the black hole candidate and compact jet source GX 339-4},
  year          = {2000},
  month         = jul,
  pages         = {251-268},
  volume        = {359},
  archiveprefix = {arXiv},
  doi           = {10.48550/arXiv.astro-ph/0003460},
  eprint        = {astro-ph/0003460},
  primaryclass  = {astro-ph},
  url           = {https://ui.adsabs.harvard.edu/abs/2000A&A...359..251C},
}

@Article{Corbel2002,
  author        = {Corbel, S. and Fender, R. P.},
  journal       = {\apjl},
  title         = {Near-Infrared Synchrotron Emission from the Compact Jet of GX 339-4},
  year          = {2002},
  month         = jul,
  number        = {1},
  pages         = {L35-L39},
  volume        = {573},
  archiveprefix = {arXiv},
  doi           = {10.1086/341870},
  eprint        = {astro-ph/0205402},
  primaryclass  = {astro-ph},
  url           = {https://ui.adsabs.harvard.edu/abs/2002ApJ...573L..35C},
}

@Article{Casella2010,
  author        = {Casella, P. and Maccarone, T. J. and O'Brien, K. and Fender, R. P. and Russell, D. M. and van der Klis, M. and Pe'Er, A. and Maitra, D. and Altamirano, D. and Belloni, T. and Kanbach, G. and Klein-Wolt, M. and Mason, E. and Soleri, P. and Stefanescu, A. and Wiersema, K. and Wijnands, R.},
  journal       = {\mnras},
  title         = {Fast infrared variability from a relativistic jet in GX 339-4},
  year          = {2010},
  month         = may,
  number        = {1},
  pages         = {L21-L25},
  volume        = {404},
  archiveprefix = {arXiv},
  doi           = {10.1111/j.1745-3933.2010.00826.x},
  eprint        = {1002.1233},
  primaryclass  = {astro-ph.HE},
  url           = {https://ui.adsabs.harvard.edu/abs/2010MNRAS.404L..21C},
}

@Article{Zdziarski2019,
  author        = {Zdziarski, Andrzej A. and Zi{\'o}{\l}kowski, Janusz and Miko{\l}ajewska, Joanna},
  journal       = {\mnras},
  title         = {The X-ray binary GX 339-4/V821 Ara: the distance, inclination, evolutionary status, and mass transfer},
  year          = {2019},
  month         = sep,
  number        = {1},
  pages         = {1026-1034},
  volume        = {488},
  archiveprefix = {arXiv},
  doi           = {10.1093/mnras/stz1787},
  eprint        = {1904.07803},
  primaryclass  = {astro-ph.SR},
  url           = {https://ui.adsabs.harvard.edu/abs/2019MNRAS.488.1026Z},
}

@Article{Rodriguez2003,
  author        = {Rodriguez, J. and Corbel, S. and Tomsick, J. A.},
  journal       = {\apj},
  title         = {Spectral Evolution of the Microquasar XTE J1550-564 over Its Entire 2000 Outburst},
  year          = {2003},
  month         = oct,
  number        = {2},
  pages         = {1032-1038},
  volume        = {595},
  archiveprefix = {arXiv},
  doi           = {10.1086/377478},
  eprint        = {astro-ph/0306227},
  primaryclass  = {astro-ph},
  url           = {https://ui.adsabs.harvard.edu/abs/2003ApJ...595.1032R},
}

@InProceedings{Hannikainen2001,
  author        = {Hannikainen, Diana and Wu, Kinwah and Campbell-Wilson, Duncan and Hunstead, Richard and Lovell, Jim and McIntyre, Vince and Reynolds, John and Soria, Roberto and Tzioumis, Tasso},
  booktitle     = {Exploring the Gamma-Ray Universe},
  title         = {Radio emission from the X-ray transient XTE J1550-564},
  year          = {2001},
  editor        = {{Gimenez}, A. and {Reglero}, V. and {Winkler}, C.},
  month         = sep,
  pages         = {291-294},
  series        = {ESA Special Publication},
  volume        = {459},
  archiveprefix = {arXiv},
  doi           = {10.48550/arXiv.astro-ph/0102070},
  eprint        = {astro-ph/0102070},
  primaryclass  = {astro-ph},
  url           = {https://ui.adsabs.harvard.edu/abs/2001ESASP.459..291H},
}

@Article{Kaaret2003,
  author        = {Kaaret, P. and Corbel, S. and Tomsick, J. A. and Fender, R. and Miller, J. M. and Orosz, J. A. and Tzioumis, A. K. and Wijnands, R.},
  journal       = {\apj},
  title         = {X-Ray Emission from the Jets of XTE J1550-564},
  year          = {2003},
  month         = jan,
  number        = {2},
  pages         = {945-953},
  volume        = {582},
  archiveprefix = {arXiv},
  doi           = {10.1086/344540},
  eprint        = {astro-ph/0210401},
  primaryclass  = {astro-ph},
  url           = {https://ui.adsabs.harvard.edu/abs/2003ApJ...582..945K},
}

@Article{2011ApJ...730...75O,
  author        = {Orosz, Jerome A. and Steiner, James F. and McClintock, Jeffrey E. and Torres, Manuel A. P. and Remillard, Ronald A. and Bailyn, Charles D. and Miller, Jon M.},
  journal       = {\apj},
  title         = {An Improved Dynamical Model for the Microquasar XTE J1550-564},
  year          = {2011},
  month         = apr,
  number        = {2},
  pages         = {75},
  volume        = {730},
  archiveprefix = {arXiv},
  doi           = {10.1088/0004-637X/730/2/75},
  eid           = {75},
  eprint        = {1101.2499},
  primaryclass  = {astro-ph.SR},
  url           = {https://ui.adsabs.harvard.edu/abs/2011ApJ...730...75O},
}

@Article{Hannikainen2009,
  author        = {Hannikainen, D. C. and Hunstead, R. W. and Wu, K. and McIntyre, V. and Lovell, J. E. J. and Campbell-Wilson, D. and McCollough, M. L. and Reynolds, J. and Tzioumis, A. K.},
  journal       = {\mnras},
  title         = {Revisiting the relativistic ejection event in XTE J1550-564 during the 1998 outburst},
  year          = {2009},
  month         = jul,
  number        = {2},
  pages         = {569-576},
  volume        = {397},
  archiveprefix = {arXiv},
  doi           = {10.1111/j.1365-2966.2009.14997.x},
  eprint        = {0904.4849},
  primaryclass  = {astro-ph.HE},
  url           = {https://ui.adsabs.harvard.edu/abs/2009MNRAS.397..569H},
}

@Article{Motta2014,
  author        = {Motta, S. E. and Belloni, T. M. and Stella, L. and Mu{\~n}oz-Darias, T. and Fender, R.},
  journal       = {\mnras},
  title         = {Precise mass and spin measurements for a stellar-mass black hole through X-ray timing: the case of GRO J1655-40},
  year          = {2014},
  month         = jan,
  number        = {3},
  pages         = {2554-2565},
  volume        = {437},
  archiveprefix = {arXiv},
  doi           = {10.1093/mnras/stt2068},
  eprint        = {1309.3652},
  primaryclass  = {astro-ph.HE},
  url           = {https://ui.adsabs.harvard.edu/abs/2014MNRAS.437.2554M},
}

@Article{Marti2017,
  author   = {Mart{\'\i}, Josep and Luque-Escamilla, Pedro L. and Bosch-Ramon, Valent{\'\i} and Paredes, Josep M.},
  journal  = {Nature Communications},
  title    = {A galactic microquasar mimicking winged radio galaxies},
  year     = {2017},
  month    = nov,
  pages    = {1757},
  volume   = {8},
  doi      = {10.1038/s41467-017-01976-5},
  eid      = {1757},
  url      = {https://ui.adsabs.harvard.edu/abs/2017NatCo...8.1757M},
}

@Article{Mariani2025,
  author        = {Mariani, I. and Motta, S. E. and Atri, P. and Matthews, J. H. and Fender, R. P. and Mart{\'\i}, J. and Luque-Escamilla, P. L. and Heywood, I.},
  journal       = {\aap},
  title         = {A MeerKAT view of the parsec-scale jets in the black-hole X-ray binary GRS 1758─258},
  year          = {2025},
  month         = dec,
  pages         = {A239},
  volume        = {704},
  archiveprefix = {arXiv},
  doi           = {10.1051/0004-6361/202556290},
  eid           = {A239},
  eprint        = {2509.10275},
  primaryclass  = {astro-ph.HE},
  url           = {https://ui.adsabs.harvard.edu/abs/2025A&A...704A.239M},
}

@Article{Casares2023,
  author        = {Casares, J. and Yanes-Rizo, I. V. and Torres, M. A. P. and Abbott, T. M. C. and Armas Padilla, M. and Charles, P. A. and C{\'u}neo, V. A. and Mu{\~n}oz-Darias, T. and Jonker, P. G. and Maguire, K.},
  journal       = {\mnras},
  title         = {The orbital period, black hole mass, and distance to the X-ray transient GRS 1716-249 ( =N Oph 93)},
  year          = {2023},
  month         = dec,
  number        = {4},
  pages         = {5209-5219},
  volume        = {526},
  archiveprefix = {arXiv},
  doi           = {10.1093/mnras/stad3068},
  eprint        = {2310.11561},
  primaryclass  = {astro-ph.HE},
  url           = {https://ui.adsabs.harvard.edu/abs/2023MNRAS.526.5209C},
}

@Article{Wood2023,
  author        = {Wood, C. M. and Miller-Jones, J. C. A. and Bahramian, A. and Tingay, S. J. and Russell, T. D. and Tetarenko, A. J. and Altamirano, D. and Belloni, T. and Carotenuto, F. and Ceccobello, C. and Corbel, S. and Espinasse, M. and Fender, R. P. and K{\"o}rding, E. and Migliari, S. and Russell, D. M. and Sarazin, C. L. and Sivakoff, G. R. and Soria, R. and Tudose, V.},
  journal       = {\mnras},
  title         = {Time-dependent visibility modelling of a relativistic jet in the X-ray binary MAXI J1803-298},
  year          = {2023},
  month         = jun,
  number        = {1},
  pages         = {70-89},
  volume        = {522},
  archiveprefix = {arXiv},
  doi           = {10.1093/mnras/stad939},
  eprint        = {2303.15648},
  primaryclass  = {astro-ph.HE},
  url           = {https://ui.adsabs.harvard.edu/abs/2023MNRAS.522...70W},
}

@Article{2016A&A...596A..46M,
  author        = {Mart{\'\i}, Josep and Luque-Escamilla, Pedro L. and Mu{\~n}oz-Arjonilla, {\'A}lvaro J.},
  journal       = {\aap},
  title         = {Optical spectroscopy of the microquasar GRS 1758-258: a possible intermediate mass system?},
  year          = {2016},
  month         = nov,
  pages         = {A46},
  volume        = {596},
  archiveprefix = {arXiv},
  doi           = {10.1051/0004-6361/201629630},
  eid           = {A46},
  eprint        = {1611.03683},
  primaryclass  = {astro-ph.HE},
  url           = {https://ui.adsabs.harvard.edu/abs/2016A&A...596A..46M},
}

@Article{Brocksopp2013,
  author        = {Brocksopp, C. and Corbel, S. and Tzioumis, A. and Broderick, J. W. and Rodriguez, J. and Yang, J. and Fender, R. P. and Paragi, Z.},
  journal       = {\mnras},
  title         = {XTE J1752-223 in outburst: a persistent radio jet, dramatic flaring, multiple ejections and linear polarization},
  year          = {2013},
  month         = jun,
  number        = {2},
  pages         = {931-943},
  volume        = {432},
  archiveprefix = {arXiv},
  doi           = {10.1093/mnras/stt493},
  eprint        = {1303.6702},
  primaryclass  = {astro-ph.HE},
  url           = {https://ui.adsabs.harvard.edu/abs/2013MNRAS.432..931B},
}

@Article{Garcia2018,
  author        = {Garc{\'\i}a, Javier A. and Steiner, James F. and Grinberg, Victoria and Dauser, Thomas and Connors, Riley M. T. and McClintock, Jeffrey E. and Remillard, Ronald A. and Wilms, J{\"o}rn and Harrison, Fiona A. and Tomsick, John A.},
  journal       = {\apj},
  title         = {Reflection Spectroscopy of the Black Hole Binary XTE J1752-223 in Its Long-stable Hard State},
  year          = {2018},
  month         = sep,
  number        = {1},
  pages         = {25},
  volume        = {864},
  archiveprefix = {arXiv},
  doi           = {10.3847/1538-4357/aad231},
  eid           = {25},
  eprint        = {1807.01949},
  primaryclass  = {astro-ph.HE},
  url           = {https://ui.adsabs.harvard.edu/abs/2018ApJ...864...25G},
}

@Article{Hjellming1998,
  author   = {Hjellming, R. M. and Rupen, M. P. and Ghigo, F. and Waltman, E. B. and Mioduszewski, A. J. and Fender, R. P. and Stappers, B. W. and Wieringa, M. and Wark, R. and Green, D. W. E.},
  journal  = {\iaucirc},
  title    = {XTE J1748-288},
  year     = {1998},
  month    = jun,
  pages    = {1},
  volume   = {6937},
  url      = {https://ui.adsabs.harvard.edu/abs/1998IAUC.6937....1H},
}

@Article{Saikia2022,
  author        = {Saikia, Payaswini and Russell, David M. and Baglio, M. C. and Bramich, D. M. and Casella, Piergiorgio and Diaz Trigo, Maria and Gandhi, Poshak and Jiang, Jiachen and Maccarone, Thomas and Soria, Roberto and Al Noori, Hind and Al Yazeedi, Aisha and Alabarta, Kevin and Belloni, Tomaso and Bel, Marion Cadolle and Ceccobello, Chiara and Corbel, St{\'e}phane and Fender, Rob and Gallo, Elena and Homan, Jeroen and Koljonen, Karri and Lewis, Fraser and Markoff, Sera B. and Miller-Jones, James C. A. and Rodriguez, Jerome and Russell, Thomas D. and Shahbaz, Tariq and Sivakoff, Gregory R. and Testa, Vincenzo and Tetarenko, Alexandra J.},
  journal       = {\apj},
  title         = {A Multiwavelength Study of GRS 1716-249 in Outburst: Constraints on Its System Parameters},
  year          = {2022},
  month         = jun,
  number        = {1},
  pages         = {38},
  volume        = {932},
  archiveprefix = {arXiv},
  doi           = {10.3847/1538-4357/ac6ce1},
  eid           = {38},
  eprint        = {2205.04452},
  primaryclass  = {astro-ph.HE},
  url           = {https://ui.adsabs.harvard.edu/abs/2022ApJ...932...38S},
}

@Article{2007MNRAS.378.1111B,
  author        = {Brocksopp, C. and Miller-Jones, J. C. A. and Fender, R. P. and Stappers, B. W.},
  journal       = {\mnras},
  title         = {A highly polarized radio jet during the 1998 outburst of the black hole transient XTE J1748-288},
  year          = {2007},
  month         = jul,
  number        = {3},
  pages         = {1111-1117},
  volume        = {378},
  archiveprefix = {arXiv},
  doi           = {10.1111/j.1365-2966.2007.11846.x},
  eprint        = {0705.1125},
  primaryclass  = {astro-ph},
  url           = {https://ui.adsabs.harvard.edu/abs/2007MNRAS.378.1111B},
}

@Article{Paizis2007,
  author        = {Paizis, A. and Nowak, M. A. and Chaty, S. and Rodriguez, J. and Courvoisier, T. J.-L. and Del Santo, M. and Ebisawa, K. and Farinelli, R. and Ubertini, P. and Wilms, J.},
  journal       = {\apjl},
  title         = {Unveiling the Nature of IGR J17497-2821 Using X-Ray and Near-Infrared Observations},
  year          = {2007},
  month         = mar,
  number        = {2},
  pages         = {L109-L112},
  volume        = {657},
  archiveprefix = {arXiv},
  doi           = {10.1086/513313},
  eprint        = {astro-ph/0611344},
  primaryclass  = {astro-ph},
  url           = {https://ui.adsabs.harvard.edu/abs/2007ApJ...657L.109P},
}

@Article{Torres2021,
  author        = {Torres, M. A. P. and Jonker, P. G. and Casares, J. and Miller-Jones, J. C. A. and Steeghs, D.},
  journal       = {\mnras},
  title         = {Delimiting the black hole mass in the X-ray transient MAXI J1659-152 with H{\ensuremath{\alpha}} spectroscopy},
  year          = {2021},
  month         = feb,
  number        = {2},
  pages         = {2174-2181},
  volume        = {501},
  archiveprefix = {arXiv},
  doi           = {10.1093/mnras/staa3786},
  eprint        = {2011.02383},
  primaryclass  = {astro-ph.HE},
  url           = {https://ui.adsabs.harvard.edu/abs/2021MNRAS.501.2174T},
}

@Article{Xie2020,
  author        = {Xie, Fu-Guo and Yan, Zhen and Wu, Zhongzu},
  journal       = {\apj},
  title         = {Radio/X-Ray Correlation in the Mini-outbursts of Black Hole X-Ray Transient GRS 1739-278},
  year          = {2020},
  month         = mar,
  number        = {1},
  pages         = {31},
  volume        = {891},
  archiveprefix = {arXiv},
  doi           = {10.3847/1538-4357/ab711f},
  eid           = {31},
  eprint        = {1911.06447},
  primaryclass  = {astro-ph.HE},
  url           = {https://ui.adsabs.harvard.edu/abs/2020ApJ...891...31X},
}

@Article{Jonker2012,
  author        = {Jonker, P. G. and Miller-Jones, J. C. A. and Homan, J. and Tomsick, J. and Fender, R. P. and Kaaret, P. and Markoff, S. and Gallo, E.},
  journal       = {\mnras},
  title         = {The black hole candidate MAXI J1659-152 in and towards quiescence in X-ray and radio},
  year          = {2012},
  month         = jul,
  number        = {4},
  pages         = {3308-3315},
  volume        = {423},
  archiveprefix = {arXiv},
  doi           = {10.1111/j.1365-2966.2012.21116.x},
  eprint        = {1204.4832},
  primaryclass  = {astro-ph.HE},
  url           = {https://ui.adsabs.harvard.edu/abs/2012MNRAS.423.3308J},
}

@Article{Corbel2014,
  author   = {Corbel, St{\'e}phane and Tomsick, John A. and Tzioumis, Tasso},
  journal  = {The Astronomer's Telegram},
  title    = {A radio counterpart to the X-ray transient MAXI J1828-249 detected with the ATCA},
  year     = {2014},
  month    = feb,
  pages    = {1},
  volume   = {5911},
  url      = {https://ui.adsabs.harvard.edu/abs/2014ATel.5911....1C},
}

@Article{Yang2011,
  author        = {Yang, J. and Paragi, Z. and Corbel, S. and Gurvits, L. I. and Campbell, R. M. and Brocksopp, C.},
  journal       = {\mnras},
  title         = {Transient relativistic ejections and stationary core in XTE J1752-223},
  year          = {2011},
  month         = nov,
  number        = {1},
  pages         = {L25-L29},
  volume        = {418},
  archiveprefix = {arXiv},
  doi           = {10.1111/j.1745-3933.2011.01136.x},
  eprint        = {1108.3492},
  primaryclass  = {astro-ph.HE},
  url           = {https://ui.adsabs.harvard.edu/abs/2011MNRAS.418L..25Y},
}

@Article{MataSanchez2025,
  author        = {Mata S{\'a}nchez, D. and Torres, M. A. P. and Casares, J. and Mu{\~n}oz-Darias, T. and Armas Padilla, M. and Yanes-Rizo, I. V.},
  journal       = {\aap},
  title         = {Dynamical confirmation of a black hole in the X-ray transient Swift J1727.8‑1613},
  year          = {2025},
  month         = jan,
  pages         = {A129},
  volume        = {693},
  archiveprefix = {arXiv},
  doi           = {10.1051/0004-6361/202451960},
  eid           = {A129},
  eprint        = {2408.13310},
  primaryclass  = {astro-ph.HE},
  url           = {https://ui.adsabs.harvard.edu/abs/2025A&A...693A.129M},
}

@Article{DelSanto2023,
  author        = {Del Santo, M. and Pinto, C. and Marino, A. and D'A{\`\i}, A. and Petrucci, P.-O. and Malzac, J. and Ferreira, J. and Pintore, F. and Motta, S. E. and Russell, T. D. and Segreto, A. and Sanna, A.},
  journal       = {\mnras},
  title         = {An ultrafast outflow in the black hole candidate MAXI J1810-222?},
  year          = {2023},
  month         = jul,
  number        = {1},
  pages         = {L15-L20},
  volume        = {523},
  archiveprefix = {arXiv},
  doi           = {10.1093/mnrasl/slad048},
  eprint        = {2304.08514},
  primaryclass  = {astro-ph.HE},
  url           = {https://ui.adsabs.harvard.edu/abs/2023MNRAS.523L..15D},
}

@Article{2022MNRAS.513.6196R,
  author        = {Russell, T. D. and Del Santo, M. and Marino, A. and Segreto, A. and Motta, S. E. and Bahramian, A. and Corbel, S. and D'A{\`\i}, A. and Di Salvo, T. and Miller-Jones, J. C. A. and Pinto, C. and Pintore, F. and Tzioumis, A.},
  journal       = {\mnras},
  title         = {Investigating the nature and properties of MAXI J1810-222 with radio and X-ray observations},
  year          = {2022},
  month         = jul,
  number        = {4},
  pages         = {6196-6209},
  volume        = {513},
  archiveprefix = {arXiv},
  doi           = {10.1093/mnras/stac1332},
  eprint        = {2205.05721},
  primaryclass  = {astro-ph.HE},
  url           = {https://ui.adsabs.harvard.edu/abs/2022MNRAS.513.6196R},
}

@Article{Russell2013,
  author        = {Russell, D. M. and Russell, T. D. and Miller-Jones, J. C. A. and O'Brien, K. and Soria, R. and Sivakoff, G. R. and Slaven-Blair, T. and Lewis, F. and Markoff, S. and Homan, J. and Altamirano, D. and Curran, P. A. and Rupen, M. P. and Belloni, T. M. and Cadolle Bel, M. and Casella, P. and Corbel, S. and Dhawan, V. and Fender, R. P. and Gallo, E. and Gandhi, P. and Heinz, S. and K{\"o}rding, E. G. and Krimm, H. A. and Maitra, D. and Migliari, S. and Remillard, R. A. and Sarazin, C. L. and Shahbaz, T. and Tudose, V.},
  journal       = {\apjl},
  title         = {An Evolving Compact Jet in the Black Hole X-Ray Binary MAXI J1836-194},
  year          = {2013},
  month         = may,
  number        = {2},
  pages         = {L35},
  volume        = {768},
  archiveprefix = {arXiv},
  doi           = {10.1088/2041-8205/768/2/L35},
  eid           = {L35},
  eprint        = {1304.3510},
  primaryclass  = {astro-ph.HE},
  url           = {https://ui.adsabs.harvard.edu/abs/2013ApJ...768L..35R},
}

@Article{Bassi2019,
  author        = {Bassi, T. and Del Santo, M. and D'A{\i}, A. and Motta, S. E. and Malzac, J. and Segreto, A. and Miller-Jones, J. C. A. and Atri, P. and Plotkin, R. M. and Belloni, T. M. and Mineo, T. and Tzioumis, A. K.},
  journal       = {\mnras},
  title         = {The long outburst of the black hole transient GRS 1716-249 observed in the X-ray and radio band},
  year          = {2019},
  month         = jan,
  number        = {2},
  pages         = {1587-1601},
  volume        = {482},
  archiveprefix = {arXiv},
  doi           = {10.1093/mnras/sty2739},
  eprint        = {1810.03914},
  primaryclass  = {astro-ph.HE},
  url           = {https://ui.adsabs.harvard.edu/abs/2019MNRAS.482.1587B},
}

@Article{Soria2011,
  author        = {Soria, Roberto and Broderick, Jess W. and Hao, Jingfang and Hannikainen, Diana C. and Mehdipour, Missagh and Pottschmidt, Katja and Zhang, Shuang-Nan},
  journal       = {\mnras},
  title         = {Accretion states of the Galactic microquasar GRS 1758-258},
  year          = {2011},
  month         = jul,
  number        = {1},
  pages         = {410-424},
  volume        = {415},
  archiveprefix = {arXiv},
  doi           = {10.1111/j.1365-2966.2011.18714.x},
  eprint        = {1103.3009},
  primaryclass  = {astro-ph.HE},
  url           = {https://ui.adsabs.harvard.edu/abs/2011MNRAS.415..410S},
}

@Article{CadolleBel2009,
  author        = {Cadolle Bel, M. and Prat, L. and Rodriguez, J. and Rib{\'o}, M. and Barrag{\'a}n, L. and D'Avanzo, P. and Hannikainen, D. C. and Kuulkers, E. and Campana, S. and Mold{\'o}n, J. and Chaty, S. and Zurita-Heras, J. and Goldwurm, A. and Goldoni, P.},
  journal       = {\aap},
  title         = {Detailed radio to soft {\ensuremath{\gamma}}-ray studies of the 2005 outburst of the new X-ray transient XTE J1818-245},
  year          = {2009},
  month         = jul,
  number        = {1},
  pages         = {1-13},
  volume        = {501},
  archiveprefix = {arXiv},
  doi           = {10.1051/0004-6361/200810744},
  eprint        = {0903.4714},
  primaryclass  = {astro-ph.HE},
  url           = {https://ui.adsabs.harvard.edu/abs/2009A&A...501....1C},
}

@Article{Russell2015,
  author        = {Russell, T. D. and Miller-Jones, J. C. A. and Curran, P. A. and Soria, R. and Altamirano, D. and Corbel, S. and Coriat, M. and Moin, A. and Russell, D. M. and Sivakoff, G. R. and Slaven-Blair, T. J. and Belloni, T. M. and Fender, R. P. and Heinz, S. and Jonker, P. G. and Krimm, H. A. and K{\"o}rding, E. G. and Maitra, D. and Markoff, S. and Middleton, M. and Migliari, S. and Remillard, R. A. and Rupen, M. P. and Sarazin, C. L. and Tetarenko, A. J. and Torres, M. A. P. and Tudose, V. and Tzioumis, A. K.},
  journal       = {\mnras},
  title         = {Radio monitoring of the hard state jets in the 2011 outburst of MAXI J1836-194},
  year          = {2015},
  month         = jun,
  number        = {2},
  pages         = {1745-1759},
  volume        = {450},
  archiveprefix = {arXiv},
  doi           = {10.1093/mnras/stv723},
  eprint        = {1503.08634},
  primaryclass  = {astro-ph.HE},
  url           = {https://ui.adsabs.harvard.edu/abs/2015MNRAS.450.1745R},
}

@Article{Russell2018,
  author   = {Russell, T. D. and Miller-Jones, J. C. A. and Sivakoff, G. R. and Tetarenko, A. J. and JACPOT XRB Collaboration},
  journal  = {The Astronomer's Telegram},
  title    = {ATCA radio detection of the new X-ray transient MAXI J1813-095 as a candidate radio-quiet black hole X-ray binary},
  year     = {2018},
  month    = feb,
  pages    = {1},
  volume   = {11356},
  url      = {https://ui.adsabs.harvard.edu/abs/2018ATel11356....1R},
}

@Article{Zhang2022,
  author        = {Zhang, X. and Yu, W. and Motta, S. E. and Fender, R. and Woudt, P. and Miller-Jones, J. C. A. and Sivakoff, G. R.},
  journal       = {\mnras},
  title         = {MeerKAT radio detection of the Galactic black hole candidate Swift J1842.5-1124 during its 2020 outburst},
  year          = {2022},
  month         = feb,
  number        = {1},
  pages         = {1258-1263},
  volume        = {510},
  archiveprefix = {arXiv},
  doi           = {10.1093/mnras/stab3463},
  eprint        = {2112.02202},
  primaryclass  = {astro-ph.HE},
  url           = {https://ui.adsabs.harvard.edu/abs/2022MNRAS.510.1258Z},
}

@Article{CorralSantana2025,
  author        = {Corral-Santana, J. M. and Rodr{\'\i}guez-Gil, P. and Torres, M. A. P. and Casares, J. and Jonker, P. G. and Perdomo Garc{\'\i}a, A. and Trelawny, D. T. and Carballo-Bello, J. A. and Charles, P. A. and Mata S{\'a}nchez, D. and Mu{\~n}oz-Darias, T. and Ringwald, F. A. and Mart{\'\i}nez-Pais, I. G. and Corradi, R. L. M. and Saikia, P. and Russell, D. M.},
  journal       = {\aap},
  title         = {Characterising the short-orbital period X-ray transient Swift J1910.2─0546},
  year          = {2025},
  month         = oct,
  pages         = {A225},
  volume        = {702},
  archiveprefix = {arXiv},
  doi           = {10.1051/0004-6361/202556452},
  eid           = {A225},
  eprint        = {2508.16775},
  primaryclass  = {astro-ph.HE},
  url           = {https://ui.adsabs.harvard.edu/abs/2025A&A...702A.225C},
}

@Article{King2012,
  author   = {King, A. L. and Miller, J. M. and Degenaar, N. and Reynolds, M. and Reis, R.},
  journal  = {The Astronomer's Telegram},
  title    = {New radio detection of MAXI J1910-057 in hard-state transition},
  year     = {2012},
  month    = aug,
  pages    = {1},
  volume   = {4295},
  url      = {https://ui.adsabs.harvard.edu/abs/2012ATel.4295....1K},
}

@Article{Saikia2023,
  author        = {Saikia, Payaswini and Russell, David M. and Pirbhoy, Saarah F. and Baglio, M. C. and Bramich, D. M. and Alabarta, Kevin and Lewis, Fraser and Charles, Phil},
  journal       = {\mnras},
  title         = {Clockwise evolution in the hardness-intensity diagram of the black hole X-ray binary Swift J1910.2-0546},
  year          = {2023},
  month         = sep,
  number        = {3},
  pages         = {4543-4553},
  volume        = {524},
  archiveprefix = {arXiv},
  doi           = {10.1093/mnras/stad2044},
  eprint        = {2307.08407},
  primaryclass  = {astro-ph.HE},
  url           = {https://ui.adsabs.harvard.edu/abs/2023MNRAS.524.4543S},
}

@Article{Williams2022,
  author        = {Williams, D. R. A. and Motta, S. E. and Fender, R. and Miller-Jones, J. C. A. and Neilsen, J. and Allison, J. R. and Bright, J. and Heywood, I. and Jacob, P. F. L. and Rhodes, L. and Tremou, E. and Woudt, P. A. and Eijnden, J. van den and Carotenuto, F. and Green, D. A. and Titterington, D. and van der Horst, A. J. and Saikia, P.},
  journal       = {\mnras},
  title         = {Radio observations of the Black Hole X-ray Binary EXO 1846-031 re-awakening from a 34-year slumber},
  year          = {2022},
  month         = dec,
  number        = {2},
  pages         = {2801-2817},
  volume        = {517},
  archiveprefix = {arXiv},
  doi           = {10.1093/mnras/stac2700},
  eprint        = {2209.10228},
  primaryclass  = {astro-ph.HE},
  url           = {https://ui.adsabs.harvard.edu/abs/2022MNRAS.517.2801W},
}

@Article{Bahramian2023,
  author        = {Bahramian, A. and Tremou, E. and Tetarenko, A. J. and Miller-Jones, J. C. A. and Fender, R. P. and Corbel, S. and Williams, D. R. A. and Strader, J. and Carotenuto, F. and Salinas, R. and Kennea, J. A. and Motta, S. E. and Woudt, P. A. and Matthews, J. H. and Russell, T. D.},
  journal       = {\apjl},
  title         = {MAXI J1848-015: The First Detection of Relativistically Moving Outflows from a Globular Cluster X-Ray Binary},
  year          = {2023},
  month         = may,
  number        = {1},
  pages         = {L7},
  volume        = {948},
  archiveprefix = {arXiv},
  doi           = {10.3847/2041-8213/accde1},
  eid           = {L7},
  eprint        = {2305.03764},
  primaryclass  = {astro-ph.HE},
  url           = {https://ui.adsabs.harvard.edu/abs/2023ApJ...948L...7B},
}

@Article{Bright2025,
  author        = {Bright, Joe S. and Fender, Rob and Russell, David M. and Motta, Sara E. and Man, Ethan and van den Eijnden, Jakob and Alabarta, Kevin and Crook-Mansour, Justine and Baglio, Maria C. and Green, David A. and Heywood, Ian and Lewis, Fraser and Saikia, Payaswini and Scott, Paul F. and Titterington, David J.},
  journal       = {\mnras},
  title         = {The accretion{\textendash}ejection connection in the black hole X-ray binary MAXI J1820+070},
  year          = {2025},
  month         = aug,
  number        = {2},
  pages         = {1851-1865},
  volume        = {541},
  archiveprefix = {arXiv},
  doi           = {10.1093/mnras/staf1098},
  eprint        = {2507.11303},
  primaryclass  = {astro-ph.HE},
  url           = {https://ui.adsabs.harvard.edu/abs/2025MNRAS.541.1851B},
}

@Article{Poutanen2022,
  author        = {Poutanen, Juri and Veledina, Alexandra and Berdyugin, Andrei V. and Berdyugina, Svetlana V. and Jermak, Helen and Jonker, Peter G. and Kajava, Jari J. E. and Kosenkov, Ilia A. and Kravtsov, Vadim and Piirola, Vilppu and Shrestha, Manisha and Perez Torres, Manuel A. and Tsygankov, Sergey S.},
  journal       = {Science},
  title         = {Black hole spin{\textendash}orbit misalignment in the x-ray binary MAXI J1820+070},
  year          = {2022},
  month         = feb,
  number        = {6583},
  pages         = {874-876},
  volume        = {375},
  archiveprefix = {arXiv},
  doi           = {10.1126/science.abl4679},
  eprint        = {2109.07511},
  primaryclass  = {astro-ph.HE},
  url           = {https://ui.adsabs.harvard.edu/abs/2022Sci...375..874P},
}

@Article{Blundell2011,
  author        = {Blundell, Katherine M. and Schmidtobreick, Linda and Trushkin, Sergei},
  journal       = {\mnras},
  title         = {SS433's accretion disc, wind and jets: before, during and after a major flare},
  year          = {2011},
  month         = nov,
  number        = {4},
  pages         = {2401-2410},
  volume        = {417},
  archiveprefix = {arXiv},
  doi           = {10.1111/j.1365-2966.2011.18785.x},
  eprint        = {1104.2917},
  primaryclass  = {astro-ph.GA},
  url           = {https://ui.adsabs.harvard.edu/abs/2011MNRAS.417.2401B},
}

@Article{Pandey2006,
  author        = {Pandey, M. and Manchanda, R. K. and Rao, A. P. and Durouchoux, P. and Ishwara-Chandra},
  journal       = {\aap},
  title         = {GMRT observations of the field of INTEGRAL X-ray sources - I},
  year          = {2006},
  month         = feb,
  number        = {2},
  pages         = {471-483},
  volume        = {446},
  archiveprefix = {arXiv},
  doi           = {10.1051/0004-6361:20042317},
  eprint        = {astro-ph/0509645},
  primaryclass  = {astro-ph},
  url           = {https://ui.adsabs.harvard.edu/abs/2006A&A...446..471P},
}

@Article{Rushton2017,
  author        = {Rushton, A. P. and Miller-Jones, J. C. A. and Curran, P. A. and Sivakoff, G. R. and Rupen, M. P. and Paragi, Z. and Spencer, R. E. and Yang, J. and Altamirano, D. and Belloni, T. and Fender, R. P. and Krimm, H. A. and Maitra, D. and Migliari, S. and Russell, D. M. and Russell, T. D. and Soria, R. and Tudose, V.},
  journal       = {\mnras},
  title         = {Resolved, expanding jets in the Galactic black hole candidate XTE J1908+094},
  year          = {2017},
  month         = jul,
  number        = {3},
  pages         = {2788-2802},
  volume        = {468},
  archiveprefix = {arXiv},
  doi           = {10.1093/mnras/stx526},
  eprint        = {1703.02110},
  primaryclass  = {astro-ph.HE},
  url           = {https://ui.adsabs.harvard.edu/abs/2017MNRAS.468.2788R},
}

@Article{Chaty2006,
  author        = {Chaty, S. and Mignani, R. P. and Israel, G. L.},
  journal       = {\mnras},
  title         = {A closer look at the X-ray transient XTE J1908+094: identification of two new near-infrared candidate counterparts},
  year          = {2006},
  month         = feb,
  number        = {4},
  pages         = {1387-1391},
  volume        = {365},
  archiveprefix = {arXiv},
  doi           = {10.1111/j.1365-2966.2005.09838.x},
  eprint        = {astro-ph/0511560},
  primaryclass  = {astro-ph},
  url           = {https://ui.adsabs.harvard.edu/abs/2006MNRAS.365.1387C},
}

@Article{Torres2020,
  author        = {Torres, M. A. P. and Casares, J. and Jim{\'e}nez-Ibarra, F. and {\'A}lvarez-Hern{\'a}ndez, A. and Mu{\~n}oz-Darias, T. and Armas Padilla, M. and Jonker, P. G. and Heida, M.},
  journal       = {\apjl},
  title         = {The Binary Mass Ratio in the Black Hole Transient MAXI J1820+070},
  year          = {2020},
  month         = apr,
  number        = {2},
  pages         = {L37},
  volume        = {893},
  archiveprefix = {arXiv},
  doi           = {10.3847/2041-8213/ab863a},
  eid           = {L37},
  eprint        = {2003.02360},
  primaryclass  = {astro-ph.HE},
  url           = {https://ui.adsabs.harvard.edu/abs/2020ApJ...893L..37T},
}

@Article{Russell2011,
  author        = {Russell, D. M. and Miller-Jones, J. C. A. and Maccarone, T. J. and Yang, Y. J. and Fender, R. P. and Lewis, F.},
  journal       = {\apjl},
  title         = {Testing the Jet Quenching Paradigm with an Ultradeep Observation of a Steadily Soft State Black Hole},
  year          = {2011},
  month         = sep,
  number        = {1},
  pages         = {L19},
  volume        = {739},
  archiveprefix = {arXiv},
  doi           = {10.1088/2041-8205/739/1/L19},
  eid           = {L19},
  eprint        = {1106.0723},
  primaryclass  = {astro-ph.HE},
  url           = {https://ui.adsabs.harvard.edu/abs/2011ApJ...739L..19R},
}

@Article{Brocksopp2002,
  author        = {Brocksopp, C. and Fender, R. P. and McCollough, M. and Pooley, G. G. and Rupen, M. P. and Hjellming, R. M. and de la Force, C. J. and Spencer, R. E. and Muxlow, T. W. B. and Garrington, S. T. and Trushkin, S.},
  journal       = {\mnras},
  title         = {Initial low/hard state, multiple jet ejections and X-ray/radio correlations during the outburst of XTE J1859+226},
  year          = {2002},
  month         = apr,
  number        = {3},
  pages         = {765-775},
  volume        = {331},
  archiveprefix = {arXiv},
  doi           = {10.1046/j.1365-8711.2002.05230.x},
  eprint        = {astro-ph/0112137},
  primaryclass  = {astro-ph},
  url           = {https://ui.adsabs.harvard.edu/abs/2002MNRAS.331..765B},
}

@Article{Fender2023,
  author        = {Fender, R. P. and Mooley, K. P. and Motta, S. E. and Bright, J. S. and Williams, D. R. A. and Rushton, A. P. and Beswick, R. J. and Miller-Jones, J. C. A. and Kimura, M. and Isogai, K. and Kato, T.},
  journal       = {\mnras},
  title         = {Comprehensive coverage of particle acceleration and kinetic feedback from the stellar mass black hole V404 Cygni},
  year          = {2023},
  month         = jan,
  number        = {1},
  pages         = {1243-1259},
  volume        = {518},
  archiveprefix = {arXiv},
  doi           = {10.1093/mnras/stac1836},
  eprint        = {2206.09831},
  primaryclass  = {astro-ph.HE},
  url           = {https://ui.adsabs.harvard.edu/abs/2023MNRAS.518.1243F},
}

@Article{MillerJones2019,
  author        = {Miller-Jones, James C. A. and Tetarenko, Alexandra J. and Sivakoff, Gregory R. and Middleton, Matthew J. and Altamirano, Diego and Anderson, Gemma E. and Belloni, Tomaso M. and Fender, Rob P. and Jonker, Peter G. and K{\"o}rding, Elmar G. and Krimm, Hans A. and Maitra, Dipankar and Markoff, Sera and Migliari, Simone and Mooley, Kunal P. and Rupen, Michael P. and Russell, David M. and Russell, Thomas D. and Sarazin, Craig L. and Soria, Roberto and Tudose, Valeriu},
  journal       = {\nat},
  title         = {A rapidly changing jet orientation in the stellar-mass black-hole system V404 Cygni},
  year          = {2019},
  month         = apr,
  number        = {7756},
  pages         = {374-377},
  volume        = {569},
  archiveprefix = {arXiv},
  doi           = {10.1038/s41586-019-1152-0},
  eprint        = {1906.05400},
  primaryclass  = {astro-ph.HE},
  url           = {https://ui.adsabs.harvard.edu/abs/2019Natur.569..374M},
}

@Article{Dzib2015,
  author        = {Dzib, S. A. and Massi, M. and Jaron, F.},
  journal       = {\aap},
  title         = {Radio emission from the Be/black hole binary MWC 656},
  year          = {2015},
  month         = aug,
  pages         = {L6},
  volume        = {580},
  archiveprefix = {arXiv},
  doi           = {10.1051/0004-6361/201526867},
  eid           = {L6},
  eprint        = {1507.04488},
  primaryclass  = {astro-ph.HE},
  url           = {https://ui.adsabs.harvard.edu/abs/2015A&A...580L...6D},
}

@Article{Bartlett2013,
  author        = {Bartlett, E. S. and Clark, J. S. and Coe, M. J. and Garcia, M. R. and Uttley, P.},
  journal       = {\mnras},
  title         = {Timing and spectral analysis of the unusual X-ray transient XTE J0421+560/CI Camelopardalis},
  year          = {2013},
  month         = feb,
  number        = {2},
  pages         = {1213-1220},
  volume        = {429},
  archiveprefix = {arXiv},
  doi           = {10.1093/mnras/sts411},
  eprint        = {1211.3400},
  primaryclass  = {astro-ph.HE},
  url           = {https://ui.adsabs.harvard.edu/abs/2013MNRAS.429.1213B},
}

@Article{Clark2000,
  author   = {Clark, J. S. and Miroshnichenko, A. S. and Larionov, V. M. and Lyuty, V. M. and Hynes, R. I. and Pooley, G. G. and Coe, M. J. and McCollough, M. and Dieters, S. and Efimov, Yu. S. and Fabregat, J. and Goranskii, V. P. and Haswell, C. A. and Metlova, N. V. and Robinson, E. L. and Roche, P. and Shenavrin, V. I. and Welsh, W. F.},
  journal  = {\aap},
  title    = {Photometric observations of the radio bright B[e]/X-ray binary CI Cam},
  year     = {2000},
  month    = apr,
  pages    = {50-62},
  volume   = {356},
  url      = {https://ui.adsabs.harvard.edu/abs/2000A&A...356...50C},
}

@Article{Yadlapalli2021,
  author        = {Yadlapalli, Nitika and Ravi, Vikram and Yao, Yuhan and Kulkarni, S. R. and Brisken, Walter},
  journal       = {\apjl},
  title         = {VLBA Discovery of a Resolved Source in the Candidate Black Hole X-Ray Binary AT2019wey},
  year          = {2021},
  month         = mar,
  number        = {2},
  pages         = {L27},
  volume        = {909},
  archiveprefix = {arXiv},
  doi           = {10.3847/2041-8213/abea19},
  eid           = {L27},
  eprint        = {2012.13426},
  primaryclass  = {astro-ph.HE},
  url           = {https://ui.adsabs.harvard.edu/abs/2021ApJ...909L..27Y},
}

@Article{Chatterjee2019,
  author        = {Chatterjee, Debjit and Debnath, Dipak and Jana, Arghajit and Chakrabarti, Sandip K.},
  journal       = {\apss},
  title         = {Properties of the black hole candidate XTE J1118+480 with the TCAF solution during its jet activity induced 2000 outburst},
  year          = {2019},
  month         = jan,
  number        = {1},
  pages         = {14},
  volume        = {364},
  archiveprefix = {arXiv},
  doi           = {10.1007/s10509-019-3495-2},
  eid           = {14},
  eprint        = {1803.09438},
  primaryclass  = {astro-ph.HE},
  url           = {https://ui.adsabs.harvard.edu/abs/2019Ap&SS.364...14C},
}

@Article{Fender2001,
  author        = {Fender, R. P. and Hjellming, R. M. and Tilanus, R. P. J. and Pooley, G. G. and Deane, J. R. and Ogley, R. N. and Spencer, R. E.},
  journal       = {\mnras},
  title         = {Spectral evidence for a powerful compact jet from XTE J1118+480},
  year          = {2001},
  month         = apr,
  number        = {2},
  pages         = {L23-L27},
  volume        = {322},
  archiveprefix = {arXiv},
  doi           = {10.1046/j.1365-8711.2001.04362.x},
  eprint        = {astro-ph/0101346},
  primaryclass  = {astro-ph},
  url           = {https://ui.adsabs.harvard.edu/abs/2001MNRAS.322L..23F},
}

@Article{Shrader1994,
  author   = {Shrader, C. R. and Wagner, R. Mark and Hjellming, R. M. and Han, X. H. and Starrfield, S. G.},
  journal  = {\apj},
  title    = {The Early Ultraviolet, Optical, and Radio Evolution of the Soft X-Ray Transient GRO J0422+32},
  year     = {1994},
  month    = oct,
  pages    = {698},
  volume   = {434},
  doi      = {10.1086/174771},
  url      = {https://ui.adsabs.harvard.edu/abs/1994ApJ...434..698S},
}

@Article{Fender2001a,
  author        = {Fender, R. P.},
  journal       = {\mnras},
  title         = {Powerful jets from black hole X-ray binaries in low/hard X-ray states},
  year          = {2001},
  month         = mar,
  number        = {1},
  pages         = {31-42},
  volume        = {322},
  archiveprefix = {arXiv},
  doi           = {10.1046/j.1365-8711.2001.04080.x},
  eprint        = {astro-ph/0008447},
  primaryclass  = {astro-ph},
  url           = {https://ui.adsabs.harvard.edu/abs/2001MNRAS.322...31F},
}

@Article{Harrison2007,
  author        = {Harrison, Thomas E. and Howell, Steve B. and Szkody, Paula and Cordova, France A.},
  journal       = {\aj},
  title         = {The Nature of the Secondary Star in the Black Hole X-Ray Transient V616 Mon (=A0620-00)},
  year          = {2007},
  month         = jan,
  number        = {1},
  pages         = {162-168},
  volume        = {133},
  archiveprefix = {arXiv},
  doi           = {10.1086/509572},
  eprint        = {astro-ph/0609535},
  primaryclass  = {astro-ph},
  url           = {https://ui.adsabs.harvard.edu/abs/2007AJ....133..162H},
}

@Article{Grunsven2017,
  author        = {van Grunsven, Theo F. J. and Jonker, Peter G. and Verbunt, Frank W. M. and Robinson, Edward L.},
  journal       = {\mnras},
  title         = {The mass of the black hole in 1A 0620-00, revisiting the ellipsoidal light curve modelling},
  year          = {2017},
  month         = dec,
  number        = {2},
  pages         = {1907-1914},
  volume        = {472},
  archiveprefix = {arXiv},
  doi           = {10.1093/mnras/stx2071},
  eprint        = {1708.08209},
  primaryclass  = {astro-ph.HE},
  url           = {https://ui.adsabs.harvard.edu/abs/2017MNRAS.472.1907V},
}

@Article{Gallo2006,
  author        = {Gallo, E. and Fender, R. P. and Miller-Jones, J. C. A. and Merloni, A. and Jonker, P. G. and Heinz, S. and Maccarone, T. J. and van der Klis, M.},
  journal       = {\mnras},
  title         = {A radio-emitting outflow in the quiescent state of A0620-00: implications for modelling low-luminosity black hole binaries},
  year          = {2006},
  month         = aug,
  number        = {3},
  pages         = {1351-1360},
  volume        = {370},
  archiveprefix = {arXiv},
  doi           = {10.1111/j.1365-2966.2006.10560.x},
  eprint        = {astro-ph/0605376},
  primaryclass  = {astro-ph},
  url           = {https://ui.adsabs.harvard.edu/abs/2006MNRAS.370.1351G},
}

@Article{2022MNRAS.515.3105S,
  author        = {Soria, Roberto and Ma, Ruican and Tao, Lian and Zhang, Shuang-Nan},
  journal       = {\mnras},
  title         = {A 2-hr binary period for the black hole transient MAXI J0637-430},
  year          = {2022},
  month         = sep,
  number        = {2},
  pages         = {3105-3112},
  volume        = {515},
  archiveprefix = {arXiv},
  doi           = {10.1093/mnras/stac1896},
  eprint        = {2207.01631},
  primaryclass  = {astro-ph.HE},
  url           = {https://ui.adsabs.harvard.edu/abs/2022MNRAS.515.3105S},
}

@Article{2019ApJ...883..198R,
  author        = {Russell, T. D. and Tetarenko, A. J. and Miller-Jones, J. C. A. and Sivakoff, G. R. and Parikh, A. S. and Rapisarda, S. and Wijnands, R. and Corbel, S. and Tremou, E. and Altamirano, D. and Baglio, M. C. and Ceccobello, C. and Degenaar, N. and van den Eijnden, J. and Fender, R. and Heywood, I. and Krimm, H. A. and Lucchini, M. and Markoff, S. and Russell, D. M. and Soria, R. and Woudt, P. A.},
  journal       = {\apj},
  title         = {Disk-Jet Coupling in the 2017/2018 Outburst of the Galactic Black Hole Candidate X-Ray Binary MAXI J1535-571},
  year          = {2019},
  month         = oct,
  number        = {2},
  pages         = {198},
  volume        = {883},
  archiveprefix = {arXiv},
  doi           = {10.3847/1538-4357/ab3d36},
  eid           = {198},
  eprint        = {1906.00998},
  primaryclass  = {astro-ph.HE},
  url           = {https://ui.adsabs.harvard.edu/abs/2019ApJ...883..198R},
}

@Article{2019ATel13275....1R,
  author   = {Russell, T. D. and Miller-Jones, J. C. A. and Sivakoff, G. R. and Tetarenko, A. J.},
  journal  = {The Astronomer's Telegram},
  title    = {ATCA radio detection of the new X-ray transient MAXI J0637-430},
  year     = {2019},
  month    = nov,
  pages    = {1},
  volume   = {13275},
  url      = {https://ui.adsabs.harvard.edu/abs/2019ATel13275....1R},
}

@Article{Tetarenko2021,
  author        = {Tetarenko, B. E. and Shaw, A. W. and Manrow, E. R. and Charles, P. A. and Miller, J. M. and Russell, T. D. and Tetarenko, A. J.},
  journal       = {\mnras},
  title         = {Using optical spectroscopy to map the geometry and structure of the irradiated accretion discs in low-mass X-ray binaries: the pilot study of MAXI J0637-430},
  year          = {2021},
  month         = mar,
  number        = {3},
  pages         = {3406-3420},
  volume        = {501},
  archiveprefix = {arXiv},
  doi           = {10.1093/mnras/staa3861},
  eprint        = {2011.13414},
  primaryclass  = {astro-ph.HE},
  url           = {https://ui.adsabs.harvard.edu/abs/2021MNRAS.501.3406T},
}

@Article{Orosz2014,
  author        = {Orosz, Jerome A. and Steiner, James F. and McClintock, Jeffrey E. and Buxton, Michelle M. and Bailyn, Charles D. and Steeghs, Danny and Guberman, Alec and Torres, Manuel A. P.},
  journal       = {\apj},
  title         = {The Mass of the Black Hole in LMC X-3},
  year          = {2014},
  month         = oct,
  number        = {2},
  pages         = {154},
  volume        = {794},
  archiveprefix = {arXiv},
  doi           = {10.1088/0004-637X/794/2/154},
  eid           = {154},
  eprint        = {1402.0085},
  primaryclass  = {astro-ph.SR},
  url           = {https://ui.adsabs.harvard.edu/abs/2014ApJ...794..154O},
}

@Article{Naze2025,
  author        = {Naz{\'e}, Ya{\"e}l and Rauw, Gregor},
  journal       = {\aap},
  title         = {Another one (BH+OB pair) bites the dust},
  year          = {2025},
  month         = apr,
  pages         = {A84},
  volume        = {696},
  archiveprefix = {arXiv},
  doi           = {10.1051/0004-6361/202453493},
  eid           = {A84},
  eprint        = {2503.08190},
  primaryclass  = {astro-ph.SR},
  url           = {https://ui.adsabs.harvard.edu/abs/2025A&A...696A..84N},
}

@Article{Plotkin2021,
  author        = {Plotkin, R. M. and Bahramian, A. and Miller-Jones, J. C. A. and Reynolds, M. T. and Atri, P. and Maccarone, T. J. and Shaw, A. W. and Gandhi, P.},
  journal       = {\mnras},
  title         = {Towards a larger sample of radio jets from quiescent black hole X-ray binaries},
  year          = {2021},
  month         = may,
  number        = {3},
  pages         = {3784-3795},
  volume        = {503},
  archiveprefix = {arXiv},
  doi           = {10.1093/mnras/stab644},
  eprint        = {2103.02178},
  primaryclass  = {astro-ph.HE},
  url           = {https://ui.adsabs.harvard.edu/abs/2021MNRAS.503.3784P},
}

@Article{Carotenuto2021,
  author        = {Carotenuto, F. and Corbel, S. and Tremou, E. and Russell, T. D. and Tzioumis, A. and Fender, R. P. and Woudt, P. A. and Motta, S. E. and Miller-Jones, J. C. A. and Chauhan, J. and Tetarenko, A. J. and Sivakoff, G. R. and Heywood, I. and Horesh, A. and van der Horst, A. J. and Koerding, E. and Mooley, K. P.},
  journal       = {\mnras},
  title         = {The black hole transient MAXI J1348-630: evolution of the compact and transient jets during its 2019/2020 outburst},
  year          = {2021},
  month         = jun,
  number        = {1},
  pages         = {444-468},
  volume        = {504},
  archiveprefix = {arXiv},
  doi           = {10.1093/mnras/stab864},
  eprint        = {2103.12190},
  primaryclass  = {astro-ph.HE},
  url           = {https://ui.adsabs.harvard.edu/abs/2021MNRAS.504..444C},
}

@Article{Brocksopp2001,
  author        = {Brocksopp, C. and Jonker, P. G. and Fender, R. P. and Groot, P. J. and van der Klis, M. and Tingay, S. J.},
  journal       = {\mnras},
  title         = {The 1997 hard-state outburst of the X-ray transient GS 1354-64/BW Cir},
  year          = {2001},
  month         = may,
  number        = {2},
  pages         = {517-528},
  volume        = {323},
  archiveprefix = {arXiv},
  doi           = {10.1046/j.1365-8711.2001.04193.x},
  eprint        = {astro-ph/0011145},
  primaryclass  = {astro-ph},
  url           = {https://ui.adsabs.harvard.edu/abs/2001MNRAS.323..517B},
}

@Article{Koljonen2016,
  author        = {Koljonen, K. I. I. and Russell, D. M. and Corral-Santana, J. M. and Armas Padilla, M. and Mu{\~n}oz-Darias, T. and Lewis, F. and Coriat, M. and Bauer, F. E.},
  journal       = {\mnras},
  title         = {A `high-hard' outburst of the black hole X-ray binary GS 1354-64},
  year          = {2016},
  month         = jul,
  number        = {1},
  pages         = {942-955},
  volume        = {460},
  archiveprefix = {arXiv},
  doi           = {10.1093/mnras/stw1007},
  eprint        = {1602.06414},
  primaryclass  = {astro-ph.HE},
  url           = {https://ui.adsabs.harvard.edu/abs/2016MNRAS.460..942K},
}

@Article{YanesRizo2024,
  author        = {Yanes-Rizo, I. V. and Torres, M. A. P. and Casares, J. and Monelli, M. and Jonker, P. G. and Abbot, T. and Armas Padilla, M. and Mu{\~n}oz-Darias, T.},
  journal       = {\mnras},
  title         = {Evidence for a black hole in the historical X-ray transient A 1524-61 (= KY TrA)},
  year          = {2024},
  month         = jan,
  number        = {3},
  pages         = {5949-5955},
  volume        = {527},
  archiveprefix = {arXiv},
  doi           = {10.1093/mnras/stad3522},
  eprint        = {2311.08480},
  primaryclass  = {astro-ph.HE},
  url           = {https://ui.adsabs.harvard.edu/abs/2024MNRAS.527.5949Y},
}

@Article{Carotenuto2024a,
  author   = {Carotenuto, F. and Russell, T. D.},
  journal  = {The Astronomer's Telegram},
  title    = {ATCA detection of an extremely bright radio flare from Swift J151857.0-572147},
  year     = {2024},
  month    = mar,
  pages    = {1},
  volume   = {16518},
  url      = {https://ui.adsabs.harvard.edu/abs/2024ATel16518....1C},
}

@Article{Mondal2024,
  author        = {Mondal, Santanu and Suribhatla, S. Pujitha and Chatterjee, Kaushik and Singh, Chandra B. and Chatterjee, Rwitika},
  journal       = {\apj},
  title         = {The First Detection of X-Ray Polarization in a Newly Discovered Galactic Transient Swift J151857.0-572147},
  year          = {2024},
  month         = nov,
  number        = {2},
  pages         = {257},
  volume        = {975},
  archiveprefix = {arXiv},
  doi           = {10.3847/1538-4357/ad7d92},
  eid           = {257},
  eprint        = {2404.09643},
  primaryclass  = {astro-ph.HE},
  url           = {https://ui.adsabs.harvard.edu/abs/2024ApJ...975..257M},
}

@Article{MillerJones2011,
  author   = {Miller-Jones, J. C. A. and Tzioumis, A. K. and Jonker, P. G. and Sivakoff, G. R. and Maccarone, T. J. and Nelemans, G.},
  journal  = {The Astronomer's Telegram},
  title    = {ATCA radio observations of MAXI J1543-564},
  year     = {2011},
  month    = may,
  pages    = {1},
  volume   = {3364},
  url      = {https://ui.adsabs.harvard.edu/abs/2011ATel.3364....1M},
}

@Article{2017MNRAS.472..141M,
  author        = {Migliori, Giulia and Corbel, S. and Tomsick, J. A. and Kaaret, P. and Fender, R. P. and Tzioumis, A. K. and Coriat, M. and Orosz, J. A.},
  journal       = {\mnras},
  title         = {Evolving morphology of the large-scale relativistic jets from XTE J1550-564},
  year          = {2017},
  month         = nov,
  number        = {1},
  pages         = {141-165},
  volume        = {472},
  archiveprefix = {arXiv},
  doi           = {10.1093/mnras/stx1864},
  eprint        = {1707.06876},
  primaryclass  = {astro-ph.HE},
  url           = {https://ui.adsabs.harvard.edu/abs/2017MNRAS.472..141M},
}

@Article{Russell2018a,
  author        = {Russell, David M. and Qasim, Ahlam Al and Bernardini, Federico and Plotkin, Richard M. and Lewis, Fraser and Koljonen, Karri I. I. and Yang, Yi-Jung},
  journal       = {\apj},
  title         = {Optical Precursors to Black Hole X-Ray Binary Outbursts: An Evolving Synchrotron Jet Spectrum in Swift J1357.2-0933},
  year          = {2018},
  month         = jan,
  number        = {2},
  pages         = {90},
  volume        = {852},
  archiveprefix = {arXiv},
  doi           = {10.3847/1538-4357/aa9d8c},
  eid           = {90},
  eprint        = {1707.05814},
  primaryclass  = {astro-ph.HE},
  url           = {https://ui.adsabs.harvard.edu/abs/2018ApJ...852...90R},
}

@Article{2015MNRAS.454.2199M,
  author        = {Mata S{\'a}nchez, D. and Mu{\~n}oz-Darias, T. and Casares, J. and Corral-Santana, J. M. and Shahbaz, T.},
  journal       = {\mnras},
  title         = {Swift J1357.2-0933: a massive black hole in the Galactic thick disc},
  year          = {2015},
  month         = dec,
  number        = {2},
  pages         = {2199-2204},
  volume        = {454},
  archiveprefix = {arXiv},
  doi           = {10.1093/mnras/stv2111},
  eprint        = {1509.05412},
  primaryclass  = {astro-ph.HE},
  url           = {https://ui.adsabs.harvard.edu/abs/2015MNRAS.454.2199M},
}

@Article{Plotkin2016,
  author        = {Plotkin, Richard M. and Gallo, Elena and Jonker, Peter G. and Miller-Jones, James C. A. and Homan, Jeroen and Mu{\~n}oz-Darias, Teo and Markoff, Sera and Armas Padilla, Montserrat and Fender, Rob and Rushton, Anthony P. and Russell, David M. and Torres, Manuel A. P.},
  journal       = {\mnras},
  title         = {A clean sightline to quiescence: multiwavelength observations of the high Galactic latitude black hole X-ray binary Swift J1357.2-0933},
  year          = {2016},
  month         = mar,
  number        = {3},
  pages         = {2707-2716},
  volume        = {456},
  archiveprefix = {arXiv},
  doi           = {10.1093/mnras/stv2861},
  eprint        = {1512.01941},
  primaryclass  = {astro-ph.HE},
  url           = {https://ui.adsabs.harvard.edu/abs/2016MNRAS.456.2707P},
}

@Article{Zhang2025,
  author        = {Zhang, X. and Yu, W. and Carotenuto, F. and Motta, S. E. and Fender, R. and Miller-Jones, J. C. A. and Russell, T. D. and Bahramian, A. and Woudt, P. and Hughes, A. K. and Sivakoff, G. R.},
  journal       = {\mnras},
  title         = {Peculiar radio-bright behaviour of the Galactic black hole transient 4U 1543-47 in the 2021-2023 outburst},
  year          = {2025},
  month         = mar,
  number        = {1},
  pages         = {L43-L49},
  volume        = {538},
  archiveprefix = {arXiv},
  doi           = {10.1093/mnrasl/slaf008},
  eprint        = {2501.04917},
  primaryclass  = {astro-ph.HE},
  url           = {https://ui.adsabs.harvard.edu/abs/2025MNRAS.538L..43Z},
}

@Article{Zhang2026,
  author        = {Zhang, Zuobin and Jiang, Jiachen and Carotenuto, Francesco and Liu, Honghui and Bambi, Cosimo and Fender, Rob P. and Young, Andrew J. and van den Eijnden, Jakob and Reynolds, Christopher S. and Fabian, Andrew C. and Girard, Julien N. and Neilsen, Joey and Steiner, James F. and Tomsick, John A. and Corbel, St{\'e}phane and Hughes, Andrew K.},
  journal       = {Nature Astronomy},
  title         = {Evidence of mutually exclusive outflow forms from a black hole X-ray binary},
  year          = {2026},
  month         = jan,
  doi           = {10.1038/s41550-025-02753-x},
  primaryclass  = {astro-ph.HE},
  url           = {https://ui.adsabs.harvard.edu/abs/2026NatAs.tmp....9Z},
}

@Article{Curran2011,
  author        = {Curran, P. A. and Chaty, S. and Zurita Heras, J. A.},
  journal       = {\aap},
  title         = {Discovery and identification of infrared counterpart candidates of four Galactic centre low mass X-ray binaries},
  year          = {2011},
  month         = sep,
  pages         = {A3},
  volume        = {533},
  archiveprefix = {arXiv},
  doi           = {10.1051/0004-6361/201117208},
  eid           = {A3},
  eprint        = {1107.2045},
  primaryclass  = {astro-ph.HE},
  url           = {https://ui.adsabs.harvard.edu/abs/2011A&A...533A...3C},
}

@Article{Markwardt2008,
  author   = {Markwardt, C. B. and Pereira, D. and Swank, J. H.},
  journal  = {The Astronomer's Telegram},
  title    = {RXTE Discovers Faint Transient XTE J1637-498},
  year     = {2008},
  month    = sep,
  pages    = {1},
  volume   = {1699},
  url      = {https://ui.adsabs.harvard.edu/abs/2008ATel.1699....1M},
}

@Article{Russell2019,
  author   = {Russell, T. D. and van den Eijnden, J. and Degenaar, N.},
  journal  = {The Astronomer's Telegram},
  title    = {ATCA radio detection of the new X-ray transient MAXI J1631-478},
  year     = {2019},
  month    = jan,
  pages    = {1},
  volume   = {12396},
  url      = {https://ui.adsabs.harvard.edu/abs/2019ATel12396....1R},
}

@Article{Monageng2021,
  author        = {Monageng, I. M. and Motta, S. E. and Fender, R. and Yu, W. and Woudt, P. A. and Tremou, E. and Miller-Jones, J. C. A. and van der Horst, A. J.},
  journal       = {\mnras},
  title         = {Radio flaring and dual radio loud/quiet behaviour in the new candidate black hole X-ray binary MAXI J1631-472},
  year          = {2021},
  month         = mar,
  number        = {4},
  pages         = {5776-5781},
  volume        = {501},
  archiveprefix = {arXiv},
  doi           = {10.1093/mnras/stab043},
  eprint        = {2101.01569},
  primaryclass  = {astro-ph.HE},
  url           = {https://ui.adsabs.harvard.edu/abs/2021MNRAS.501.5776M},
}

@Article{Corbel2004,
  author        = {Corbel, S. and Fender, R. P. and Tomsick, J. A. and Tzioumis, A. K. and Tingay, S.},
  journal       = {\apj},
  title         = {On the Origin of Radio Emission in the X-Ray States of XTE J1650-500 during the 2001-2002 Outburst},
  year          = {2004},
  month         = dec,
  number        = {2},
  pages         = {1272-1283},
  volume        = {617},
  archiveprefix = {arXiv},
  doi           = {10.1086/425650},
  eprint        = {astro-ph/0409154},
  primaryclass  = {astro-ph},
  url           = {https://ui.adsabs.harvard.edu/abs/2004ApJ...617.1272C},
}

@Article{Calvelo2009,
  author   = {Calvelo, D. E. and Tzioumis, T. and Corbel, S. and Brocksopp, C. and Fender, R. P.},
  journal  = {The Astronomer's Telegram},
  title    = {Radio detection and localization of XTE J1652-453 with the ATCA},
  year     = {2009},
  month    = jul,
  pages    = {1},
  volume   = {2135},
  url      = {https://ui.adsabs.harvard.edu/abs/2009ATel.2135....1C},
}

@Article{Russell2018b,
  author   = {Russell, T. D. and Miller-Jones, J. C. A. and Sivakoff, G. R. and Tetarenko, A. J.},
  journal  = {The Astronomer's Telegram},
  title    = {ATCA radio detection of the new X-ray transient Swift J1658.2-4242},
  year     = {2018},
  month    = feb,
  pages    = {1},
  volume   = {11322},
  url      = {https://ui.adsabs.harvard.edu/abs/2018ATel11322....1R},
}

@Article{Mondal2023,
  author        = {Mondal, Santanu and Jithesh, V.},
  journal       = {\mnras},
  title         = {Spectral and temporal studies of Swift J1658.2-4242 using AstroSat observations with the JeTCAF model},
  year          = {2023},
  month         = jun,
  number        = {2},
  pages         = {2065-2074},
  volume        = {522},
  archiveprefix = {arXiv},
  doi           = {10.1093/mnras/stad1058},
  eprint        = {2304.04422},
  primaryclass  = {astro-ph.HE},
  url           = {https://ui.adsabs.harvard.edu/abs/2023MNRAS.522.2065M},
}

@Article{Brocksopp2005,
  author   = {Brocksopp, Catherine and Corbel, Stephane and Rupen, Michael and Sault, Bob and Tzioumis, Tasso and Dhawan, Vivek and Mioduszewski, Amy and Cimo, Giuseppe and Fender, Rob},
  journal  = {The Astronomer's Telegram},
  title    = {Renewed Radio Emission from GRO J1655-40},
  year     = {2005},
  month    = sep,
  pages    = {1},
  volume   = {612},
  url      = {https://ui.adsabs.harvard.edu/abs/2005ATel..612....1B},
}

@Article{Balakrishnan2023,
  author        = {Balakrishnan, Mayura and Draghis, Paul A. and Miller, Jon M. and Bright, Joe and Fender, Robert and Ng, Mason and Cackett, Edward and Fabian, Andrew and Kuntz, Kip and Miller-Jones, James C. A. and Proga, Daniel and Ray, Paul S. and Raymond, John and Reynolds, Mark and Zoghbi, Abderahmen},
  journal       = {\apj},
  title         = {The Black Hole Candidate Swift J1728.9-3613 and the Supernova Remnant G351.9-0.9},
  year          = {2023},
  month         = apr,
  number        = {1},
  pages         = {38},
  volume        = {947},
  archiveprefix = {arXiv},
  doi           = {10.3847/1538-4357/acc1c9},
  eid           = {38},
  eprint        = {2303.04159},
  primaryclass  = {astro-ph.HE},
  url           = {https://ui.adsabs.harvard.edu/abs/2023ApJ...947...38B},
}

@Article{Heinke2023,
  author        = {Heinke, Craig O.},
  journal       = {\apj},
  title         = {Chance Coincidences between Black Hole Low-mass X-Ray Binaries and Supernova Remnants},
  year          = {2023},
  month         = sep,
  number        = {1},
  pages         = {8},
  volume        = {955},
  archiveprefix = {arXiv},
  doi           = {10.3847/1538-4357/acefb6},
  eid           = {8},
  eprint        = {2308.06678},
  primaryclass  = {astro-ph.HE},
  url           = {https://ui.adsabs.harvard.edu/abs/2023ApJ...955....8H},
}

@Article{Onori2021,
  author        = {Onori, F. and Fiocchi, M. and Masetti, N. and Rojas, A. F. and Bazzano, A. and Bassani, L. and Bird, A. J.},
  journal       = {\mnras},
  title         = {Multiwavelength observations of the Galactic X-ray binaries IGR J20155+3827 and Swift J1713.4-4219},
  year          = {2021},
  month         = may,
  number        = {1},
  pages         = {472-483},
  volume        = {503},
  archiveprefix = {arXiv},
  doi           = {10.1093/mnras/stab315},
  eprint        = {2102.01638},
  primaryclass  = {astro-ph.HE},
  url           = {https://ui.adsabs.harvard.edu/abs/2021MNRAS.503..472O},
}

@Article{DelSanto2014,
  author        = {Del Santo, M. and Nucita, A. A. and Lodato, G. and Manni, L. and De Paolis, F. and Farihi, J. and De Cesare, G. and Segreto, A.},
  journal       = {\mnras},
  title         = {The puzzling source IGR J17361-4441 in NGC 6388: a possible planetary tidal disruption event},
  year          = {2014},
  month         = oct,
  number        = {1},
  pages         = {93-101},
  volume        = {444},
  archiveprefix = {arXiv},
  doi           = {10.1093/mnras/stu1436},
  eprint        = {1407.5081},
  primaryclass  = {astro-ph.HE},
  url           = {https://ui.adsabs.harvard.edu/abs/2014MNRAS.444...93D},
}

@Article{Rodriguez2011,
  author        = {Rodriguez, J. and Corbel, S. and Caballero, I. and Tomsick, J. A. and Tzioumis, T. and Paizis, A. and Cadolle Bel, M. and Kuulkers, E.},
  journal       = {\aap},
  title         = {First simultaneous multi-wavelength observations of the black hole candidate IGR J17091-3624. ATCA, INTEGRAL, Swift, and RXTE views of the 2011 outburst},
  year          = {2011},
  month         = sep,
  pages         = {L4},
  volume        = {533},
  archiveprefix = {arXiv},
  doi           = {10.1051/0004-6361/201117511},
  eid           = {L4},
  eprint        = {1108.0666},
  primaryclass  = {astro-ph.HE},
  url           = {https://ui.adsabs.harvard.edu/abs/2011A&A...533L...4R},
}

@Article{Rupen2005,
  author   = {Rupen, M. P. and Mioduszewski, A. J. and Dhawan, V.},
  journal  = {The Astronomer's Telegram},
  title    = {Probable radio counterpart to IGR J17098-3628},
  year     = {2005},
  month    = may,
  pages    = {1},
  volume   = {490},
  url      = {https://ui.adsabs.harvard.edu/abs/2005ATel..490....1R},
}

@Article{Ma2012,
  author        = {Ma, Renyi},
  journal       = {\mnras},
  title         = {Modelling the multiband spectrum of IGR J17177-3656},
  year          = {2012},
  month         = jun,
  number        = {1},
  pages         = {L87-L91},
  volume        = {423},
  archiveprefix = {arXiv},
  doi           = {10.1111/j.1745-3933.2012.01263.x},
  eprint        = {1203.6031},
  primaryclass  = {astro-ph.HE},
  url           = {https://ui.adsabs.harvard.edu/abs/2012MNRAS.423L..87M},
}

@Article{Brocksopp2005a,
  author        = {Brocksopp, C. and Corbel, S. and Fender, R. P. and Rupen, M. and Sault, R. and Tingay, S. J. and Hannikainen, D. and O'Brien, K.},
  journal       = {\mnras},
  title         = {The 2003 radio outburst of a new X-ray transient: XTE J1720-318},
  year          = {2005},
  month         = jan,
  number        = {1},
  pages         = {125-130},
  volume        = {356},
  archiveprefix = {arXiv},
  doi           = {10.1111/j.1365-2966.2004.08427.x},
  eprint        = {astro-ph/0410353},
  primaryclass  = {astro-ph},
  url           = {https://ui.adsabs.harvard.edu/abs/2005MNRAS.356..125B},
}

@InProceedings{Chaty2006a,
  author        = {Chaty, Sylvain},
  booktitle     = {VI Microquasar Workshop: Microquasars and Beyond},
  title         = {Multi-wavelength observations of the microquasar XTE J1720-318: a transition from high-soft to low-hard state},
  year          = {2006},
  month         = jan,
  pages         = {14.1},
  archiveprefix = {arXiv},
  doi           = {10.22323/1.033.0014},
  eid           = {14.1},
  eprint        = {astro-ph/0611558},
  primaryclass  = {astro-ph},
  url           = {https://ui.adsabs.harvard.edu/abs/2006smqw.confE..14C},
}

@Article{ArmasPadilla2011,
  author        = {Armas Padilla, M. and Degenaar, N. and Patruno, A. and Russell, D. M. and Linares, M. and Maccarone, T. J. and Homan, J. and Wijnands, R.},
  journal       = {\mnras},
  title         = {X-ray softening in the new X-ray transient XTE J1719-291 during its 2008 outburst decay},
  year          = {2011},
  month         = oct,
  number        = {1},
  pages         = {659-665},
  volume        = {417},
  archiveprefix = {arXiv},
  doi           = {10.1111/j.1365-2966.2011.19308.x},
  eprint        = {1104.3423},
  primaryclass  = {astro-ph.HE},
  url           = {https://ui.adsabs.harvard.edu/abs/2011MNRAS.417..659A},
}

@Article{Peng2023,
  author        = {Peng, Jing-Qiang and Zhang, Shu and Wang, Peng-Ju and Zhang, Shuang-Nan and Kong, Ling-Da and Chen, Yu-Peng and Shui, Qing-Cang and Ji, Long and Qu, Jin-Lu and Tao, Lian and Ge, Ming-Yu and Ma, Rui-Can and Chang, Zhi and Li, Jian and Li, Zhao-sheng and Yu, Zhuo-Li and Yan, Zhe and Zhang, Peng and Xiao, Yun-Xiang and Zhao, Shu-Jie},
  journal       = {\apj},
  title         = {Back to Business: SLX 1746-331 after 13 Years of Silence},
  year          = {2023},
  month         = oct,
  number        = {2},
  pages         = {96},
  volume        = {955},
  archiveprefix = {arXiv},
  doi           = {10.3847/1538-4357/acf461},
  eid           = {96},
  eprint        = {2503.01218},
  primaryclass  = {astro-ph.HE},
  url           = {https://ui.adsabs.harvard.edu/abs/2023ApJ...955...96P},
}

@Article{Kaaret2006,
  author        = {Kaaret, P. and Corbel, S. and Tomsick, J. A. and Lazendic, J. and Tzioumis, A. K. and Butt, Y. and Wijnands, R.},
  journal       = {\apj},
  title         = {Evolution of the X-Ray Jets from 4U 1755-33},
  year          = {2006},
  month         = apr,
  number        = {1},
  pages         = {410-417},
  volume        = {641},
  archiveprefix = {arXiv},
  doi           = {10.1086/500399},
  eprint        = {astro-ph/0512240},
  primaryclass  = {astro-ph},
  url           = {https://ui.adsabs.harvard.edu/abs/2006ApJ...641..410K},
}

@Article{Angelini2003,
  author        = {Angelini, Lorella and White, Nicholas E.},
  journal       = {\apjl},
  title         = {An XMM-Newton Observation of 4U 1755-33 in Quiescence: Evidence of a Fossil X-Ray Jet},
  year          = {2003},
  month         = mar,
  number        = {1},
  pages         = {L71-L75},
  volume        = {586},
  archiveprefix = {arXiv},
  doi           = {10.1086/374682},
  eprint        = {astro-ph/0302315},
  primaryclass  = {astro-ph},
  url           = {https://ui.adsabs.harvard.edu/abs/2003ApJ...586L..71A},
}

@Article{Tetarenko2016a,
  author   = {Tetarenko, A. and Sivakoff, G. R. and Bahramian, A. and Heinke, C. O. and Miller-Jones, J. C. A. and Maccarone, T. and Degenaar, N. and Wijnands, R.},
  journal  = {The Astronomer's Telegram},
  title    = {VLA observations indicate GRS 1736-297 is either a black hole X-ray binary or accreting millisecond pulsar},
  year     = {2016},
  month    = feb,
  pages    = {1},
  volume   = {8744},
  url      = {https://ui.adsabs.harvard.edu/abs/2016ATel.8744....1T},
}

@Article{Zdziarski2016,
  author        = {Zdziarski, Andrzej A. and Zi{\'o}{\l}kowski, Janusz and Bozzo, Enrico and Pjanka, Patryk},
  journal       = {\aap},
  title         = {IGR J17451-3022: constraints on the nature of the donor star},
  year          = {2016},
  month         = oct,
  pages         = {A52},
  volume        = {595},
  archiveprefix = {arXiv},
  doi           = {10.1051/0004-6361/201628585},
  eid           = {A52},
  eprint        = {1603.07288},
  primaryclass  = {astro-ph.HE},
  url           = {https://ui.adsabs.harvard.edu/abs/2016A&A...595A..52Z},
}

@Article{LuqueEscamilla2015,
  author        = {Luque-Escamilla, Pedro L. and Mart{\'\i}, Josep and Mart{\'\i}nez-Aroza, Jos{\'e}},
  journal       = {\aap},
  title         = {The precessing jets of <ASTROBJ>1E 1740.7-2942</ASTROBJ>},
  year          = {2015},
  month         = dec,
  pages         = {A122},
  volume        = {584},
  archiveprefix = {arXiv},
  doi           = {10.1051/0004-6361/201527238},
  eid           = {A122},
  eprint        = {1511.01425},
  primaryclass  = {astro-ph.HE},
  url           = {https://ui.adsabs.harvard.edu/abs/2015A&A...584A.122L},
}

@Article{Mirabel1992,
  author   = {Mirabel, I. F. and Rodriguez, L. F. and Cordier, B. and Paul, J. and Lebrun, F.},
  journal  = {\nat},
  title    = {A double-sided radio jet from the compact Galactic Centre annihilator 1E1740.7-2942},
  year     = {1992},
  month    = jul,
  number   = {6383},
  pages    = {215-217},
  volume   = {358},
  doi      = {10.1038/358215a0},
  url      = {https://ui.adsabs.harvard.edu/abs/1992Natur.358..215M},
}

@Article{DelSanto2005,
  author    = {Del Santo, M. and Bazzano, A. and Zdziarski, A. A. and Smith, D. M. and Bezayiff, N. and Farinelli, R. and De Cesare, G. and Ubertini, P. and Bird, A. J. and Cadolle Bel, M. and Capitanio, F. and Goldwurm, A. and Malizia, A. and Mirabel, I. F. and Natalucci, L. and Winkler, C.},
  journal   = {Astronomy \&; Astrophysics},
  title     = {1E 1740.7–2942: Temporal and spectral evolution from INTEGRAL and RXTE observations},
  year      = {2005},
  issn      = {1432-0746},
  month     = mar,
  number    = {2},
  pages     = {613--617},
  volume    = {433},
  doi       = {10.1051/0004-6361:20042058},
  publisher = {EDP Sciences},
}

@Article{Gallo2002,
  author        = {Gallo, E. and Fender, R. P.},
  journal       = {\mnras},
  title         = {Chandra imaging spectroscopy of 1E 1740.7-2942},
  year          = {2002},
  month         = dec,
  number        = {3},
  pages         = {869-874},
  volume        = {337},
  archiveprefix = {arXiv},
  doi           = {10.1046/j.1365-8711.2002.05957.x},
  eprint        = {astro-ph/0208296},
  primaryclass  = {astro-ph},
  url           = {https://ui.adsabs.harvard.edu/abs/2002MNRAS.337..869G},
}

@Article{2025arXiv250914465M,
  author        = {Mandel, Shifra and Mori, Kaya and Draghis, Paul A. and Reynolds, Mark and Jin, Chichuan and Parra, Maxime and Levin, Benjamin and Miao, Eric and Grollimund, Noa and Ciurlo, Anna and Granados, Sean A. and Jaisawal, Gaurava K. and Marra, Lorenzo and Bachetti, Matteo and Capitanio, Fiamma and Degenaar, Nathalie and Hailey, Charles J. and Hong, JaeSub and Motta, Sara and Ponti, Gabriele and Shara, Michael M. and Shidatsu, Megumi and Tomsick, John A. and Campbell, Randall and Corbel, St{\'e}phane and Fender, Rob and Ghez, Andrea and Grindlay, Jonathan and Haggard, Daryl and Hosek, Jr., Matthew W. and K{\"o}nig, Ole and Matsunaga, Kai and Miku{\v{s}}incov{\'a}, Romana and Nynka, Melania and Sanger-Johnson, Grace and Stel, Giovanni and Tarana, Antonella and Wijnands, Rudy and Zhang, Shuo},
  journal       = {arXiv e-prints},
  title         = {A multiwavelength study of the new Galactic center black hole candidate MAXI J1744-294},
  year          = {2025},
  month         = sep,
  pages         = {arXiv:2509.14465},
  archiveprefix = {arXiv},
  doi           = {10.48550/arXiv.2509.14465},
  eid           = {arXiv:2509.14465},
  eprint        = {2509.14465},
  primaryclass  = {astro-ph.HE},
  url           = {https://ui.adsabs.harvard.edu/abs/2025arXiv250914465M},
}

@Article{Rupen2006,
  author   = {Rupen, M. P. and Dhawan, V. and Mioduszewski, A. J.},
  journal  = {The Astronomer's Telegram},
  title    = {Possible Radio Counterpart to XTE J1817-330},
  year     = {2006},
  month    = feb,
  pages    = {1},
  volume   = {717},
  url      = {https://ui.adsabs.harvard.edu/abs/2006ATel..717....1R},
}

@Article{Fender2001b,
  author        = {Fender, R. P. and Kuulkers, E.},
  journal       = {\mnras},
  title         = {On the peak radio and X-ray emission from neutron star and black hole candidate X-ray transients},
  year          = {2001},
  month         = jul,
  number        = {4},
  pages         = {923-930},
  volume        = {324},
  archiveprefix = {arXiv},
  doi           = {10.1046/j.1365-8711.2001.04345.x},
  eprint        = {astro-ph/0101155},
  primaryclass  = {astro-ph},
  url           = {https://ui.adsabs.harvard.edu/abs/2001MNRAS.324..923F},
}

@Article{Bower2005,
  author        = {Bower, Geoffrey C. and Roberts, Doug A. and Yusef-Zadeh, Farhad and Backer, Donald C. and Cotton, W. D. and Goss, W. M. and Lang, Cornelia C. and Lithwick, Yoram},
  journal       = {\apj},
  title         = {A Radio Transient 0.1 Parsecs from Sagittarius A*},
  year          = {2005},
  month         = nov,
  number        = {1},
  pages         = {218-227},
  volume        = {633},
  archiveprefix = {arXiv},
  doi           = {10.1086/444587},
  eprint        = {astro-ph/0507221},
  primaryclass  = {astro-ph},
  url           = {https://ui.adsabs.harvard.edu/abs/2005ApJ...633..218B},
}

@Article{Hjellming1988,
  author   = {Hjellming, R. M. and Calovini, T. A. and Han, Xiao Hong and Cordova, F. A.},
  journal  = {\apjl},
  title    = {Transient Radio Emission from the X-Ray Nova ASM 2000+25},
  year     = {1988},
  month    = dec,
  pages    = {L75},
  volume   = {335},
  doi      = {10.1086/185343},
  url      = {https://ui.adsabs.harvard.edu/abs/1988ApJ...335L..75H},
}

@Article{Ball1995,
  author   = {Ball, Lewis and Kesteven, M.~J. and Campbell-Wilson, D. and Turtle, A.~J. and Hjellming, R.~M.},
  journal  = {\mnras},
  title    = {Radio observations of GRS 1124-68 (X-ray Nova MUSCAE 1991)},
  year     = {1995},
  month    = apr,
  number   = {3},
  pages    = {722-730},
  volume   = {273},
  doi      = {10.1093/mnras/273.3.722},
  url      = {https://ui.adsabs.harvard.edu/abs/1995MNRAS.273..722B},
}

@Article{Wu2016,
  author        = {Wu, Jianfeng and Orosz, Jerome A. and McClintock, Jeffrey E. and Hasan, Imran and Bailyn, Charles D. and Gou, Lijun and Chen, Zihan},
  journal       = {\apj},
  title         = {The Mass of the Black Hole in the X-ray Binary Nova Muscae 1991},
  year          = {2016},
  month         = jul,
  number        = {1},
  pages         = {46},
  volume        = {825},
  archiveprefix = {arXiv},
  doi           = {10.3847/0004-637X/825/1/46},
  eid           = {46},
  eprint        = {1601.00616},
  primaryclass  = {astro-ph.HE},
  url           = {https://ui.adsabs.harvard.edu/abs/2016ApJ...825...46W},
}

@Article{Burridge2025,
  author        = {Burridge, Benjamin J. and Miller-Jones, James C.~A. and Bahramian, Arash and Prabu, Steve R. and Streeter, Reagan and Castro Segura, Noel and Corral-Santana, Jes{\'u}s M. and Knigge, Christian and Zdziarski, Andrzej and Mata S{\'a}nchez, Daniel and Tremou, Evangelia and Carotenuto, Francesco and Fender, Rob and Saikia, Payaswini},
  journal       = {\apj},
  title         = {On the Distance to the Black Hole X-Ray Binary Swift J1727.8─1613},
  year          = {2025},
  month         = dec,
  number        = {2},
  pages         = {243},
  volume        = {994},
  archiveprefix = {arXiv},
  doi           = {10.3847/1538-4357/ae11a0},
  eid           = {243},
  eprint        = {2502.06448},
  primaryclass  = {astro-ph.HE},
  url           = {https://ui.adsabs.harvard.edu/abs/2025ApJ...994..243B},
}

@Article{Wood2024,
  author        = {Wood, Callan M. and Miller-Jones, James C.~A. and Bahramian, Arash and Tingay, Steven J. and Prabu, Steve and Russell, Thomas D. and Atri, Pikky and Carotenuto, Francesco and Altamirano, Diego and Motta, Sara E. and Hyland, Lucas and Reynolds, Cormac and Weston, Stuart and Fender, Rob and K{\"o}rding, Elmar and Maitra, Dipankar and Markoff, Sera and Migliari, Simone and Russell, David M. and Sarazin, Craig L. and Sivakoff, Gregory R. and Soria, Roberto and Tetarenko, Alexandra J. and Tudose, Valeriu},
  journal       = {\apjl},
  title         = {Swift J1727.8{\textendash}1613 Has the Largest Resolved Continuous Jet Ever Seen in an X-Ray Binary},
  year          = {2024},
  month         = aug,
  number        = {1},
  pages         = {L9},
  volume        = {971},
  archiveprefix = {arXiv},
  doi           = {10.3847/2041-8213/ad6572},
  eid           = {L9},
  eprint        = {2405.12370},
  primaryclass  = {astro-ph.HE},
  url           = {https://ui.adsabs.harvard.edu/abs/2024ApJ...971L...9W},
}

@Article{Kourmpetis2026,
  author        = {Kourmpetis, Kostas and Riaz, Shafqat and Liu, Honghui and Mirzaev, Temurbek and Bambi, Cosimo and Das, Debtroy and Jiang, Jiachen and Kokkotas, Kostas D. and Santangelo, Andrea},
  journal       = {arXiv e-prints},
  title         = {Modeling X-ray reflection spectra from returning radiation: application to 4U 1630$-$47},
  year          = {2026},
  month         = jan,
  pages         = {arXiv:2601.14860},
  archiveprefix = {arXiv},
  doi           = {10.48550/arXiv.2601.14860},
  eid           = {arXiv:2601.14860},
  eprint        = {2601.14860},
  primaryclass  = {astro-ph.HE},
  url           = {https://ui.adsabs.harvard.edu/abs/2026arXiv260114860K},
}

@Article{Xu2018,
  author        = {Xu, Yanjun and Harrison, Fiona A. and Kennea, Jamie A. and Walton, Dominic J. and Tomsick, John A. and Miller, Jon M. and Barret, Didier and Fabian, Andrew C. and Forster, Karl and F{\"u}rst, Felix and Gandhi, Poshak and Garc{\'\i}a, Javier A.},
  journal       = {\apj},
  title         = {The Hard State of the Highly Absorbed High Inclination Black Hole Binary Candidate Swift J1658.2-4242 Observed by NuSTAR and Swift},
  year          = {2018},
  month         = sep,
  number        = {1},
  pages         = {18},
  volume        = {865},
  archiveprefix = {arXiv},
  doi           = {10.3847/1538-4357/aada03},
  eid           = {18},
  eprint        = {1805.07705},
  primaryclass  = {astro-ph.HE},
  url           = {https://ui.adsabs.harvard.edu/abs/2018ApJ...865...18X},
}

@Article{Yao2021,
  author        = {Yao, Yuhan and Kulkarni, S.~R. and Burdge, Kevin B. and Caiazzo, Ilaria and De, Kishalay and Dong, Dillon and Fremling, C. and Kasliwal, Mansi M. and Kupfer, Thomas and van Roestel, Jan and Sollerman, Jesper and Bagdasaryan, Ashot and Bellm, Eric C. and Cenko, S. Bradley and Drake, Andrew J. and Duev, Dmitry A. and Graham, Matthew J. and Kaye, Stephen and Masci, Frank J. and Miranda, Nicolas and Prince, Thomas A. and Riddle, Reed and Rusholme, Ben and Soumagnac, Maayane T.},
  journal       = {\apj},
  title         = {Multi-wavelength Observations of AT2019wey: a New Candidate Black Hole Low-mass X-ray Binary},
  year          = {2021},
  month         = oct,
  number        = {2},
  pages         = {120},
  volume        = {920},
  archiveprefix = {arXiv},
  doi           = {10.3847/1538-4357/ac15f9},
  eid           = {120},
  eprint        = {2012.00169},
  primaryclass  = {astro-ph.HE},
  url           = {https://ui.adsabs.harvard.edu/abs/2021ApJ...920..120Y},
}

@Article{Porquet2005,
  author        = {Porquet, D. and Grosso, N. and B{\'e}langer, G. and Goldwurm, A. and Yusef-Zadeh, F. and Warwick, R.~S. and Predehl, P.},
  journal       = {\aap},
  title         = {Discovery of X-ray eclipses from the transient source CXOGC J174540.0-290031 with XMM-Newton},
  year          = {2005},
  month         = nov,
  number        = {2},
  pages         = {571-579},
  volume        = {443},
  archiveprefix = {arXiv},
  doi           = {10.1051/0004-6361:20053214},
  eprint        = {astro-ph/0507283},
  primaryclass  = {astro-ph},
  url           = {https://ui.adsabs.harvard.edu/abs/2005A&A...443..571P},
}

@Article{Zdziarski2026,
  author        = {Zdziarski, Andrzej A. and Chand, Swadesh and Dewangan, Gulab and Misra, Ranjeev and Szanecki, Micha{\l} and You, Bei and Parra, Maxime and Marcel, Gr{\'e}goire},
  journal       = {\apjl},
  title         = {The Strong Fe K Line and Spin of the Black Hole X-Ray Binary MAXI J1631─479},
  year          = {2026},
  month         = feb,
  number        = {2},
  pages         = {L37},
  volume        = {998},
  archiveprefix = {arXiv},
  doi           = {10.3847/2041-8213/ae3e8b},
  eid           = {L37},
  eprint        = {2511.03386},
  primaryclass  = {astro-ph.HE},
  url           = {https://ui.adsabs.harvard.edu/abs/2026ApJ...998L..37Z},
}

@Article{Shang2019,
  author        = {Shang, J.-R. and Debnath, D. and Chatterjee, D. and Jana, A. and Chakrabarti, S.~K. and Chang, H.-K. and Yap, Y.-X. and Chiu, C.-L.},
  journal       = {\apj},
  title         = {Evolution of X-Ray Properties of MAXI J1535-571: Analysis with the TCAF Solution},
  year          = {2019},
  month         = apr,
  number        = {1},
  pages         = {4},
  volume        = {875},
  archiveprefix = {arXiv},
  doi           = {10.3847/1538-4357/ab0c1e},
  eid           = {4},
  eprint        = {1806.07147},
  primaryclass  = {astro-ph.HE},
  url           = {https://ui.adsabs.harvard.edu/abs/2019ApJ...875....4S},
}

@Article{Lutovinov2003,
  author        = {Lutovinov, A.~A. and Revnivtsev, M.~G.},
  journal       = {Astronomy Letters},
  title         = {RXTE Observations of the Transient Sources IGR J17091-3624 and IGR J18539+0727},
  year          = {2003},
  month         = nov,
  pages         = {719-723},
  volume        = {29},
  archiveprefix = {arXiv},
  doi           = {10.1134/1.1624457},
  eprint        = {astro-ph/0306406},
  primaryclass  = {astro-ph},
  url           = {https://ui.adsabs.harvard.edu/abs/2003AstL...29..719L},
}

@Article{Ribo2017,
  author        = {Rib{\'o}, M. and Munar-Adrover, P. and Paredes, J.~M. and Marcote, B. and Iwasawa, K. and Mold{\'o}n, J. and Casares, J. and Migliari, S. and Paredes-Fortuny, X.},
  journal       = {\apjl},
  title         = {The First Simultaneous X-Ray/Radio Detection of the First Be/BH System MWC 656},
  year          = {2017},
  month         = feb,
  number        = {2},
  pages         = {L33},
  volume        = {835},
  archiveprefix = {arXiv},
  doi           = {10.3847/2041-8213/835/2/L33},
  eid           = {L33},
  eprint        = {1701.07265},
  primaryclass  = {astro-ph.HE},
  url           = {https://ui.adsabs.harvard.edu/abs/2017ApJ...835L..33R},
}

@Article{Kuulkers1999,
  author        = {Kuulkers, Erik and Fender, R.~P. and Spencer, Ralph E. and Davis, Richard J. and Morison, Ian},
  journal       = {\mnras},
  title         = {Multiple ejections during the 1975 outburst of A0620-00},
  year          = {1999},
  month         = jul,
  number        = {4},
  pages         = {919-925},
  volume        = {306},
  archiveprefix = {arXiv},
  doi           = {10.1046/j.1365-8711.1999.02578.x},
  eprint        = {astro-ph/9902360},
  primaryclass  = {astro-ph},
  url           = {https://ui.adsabs.harvard.edu/abs/1999MNRAS.306..919K},
}

@Article{Han1992,
  author   = {Han, Xiaohong and Hjellming, R.~M.},
  journal  = {\apj},
  title    = {Radio Observations of the 1989 Transient Event in V404 Cygni (= GS 2023+338)},
  year     = {1992},
  month    = nov,
  pages    = {304},
  volume   = {400},
  doi      = {10.1086/171996},
  url      = {https://ui.adsabs.harvard.edu/abs/1992ApJ...400..304H},
}

@Article{Hjellming1996,
  author   = {Hjellming, R.~M. and Rupen, M.~P. and Shrader, C.~R. and Campbell-Wilson, D. and Hunstead, R.~W. and McKay, D.~J.},
  journal  = {\apjl},
  title    = {Radio and X-Ray Flaring Events in X-Ray Nova Ophiuchi 1993},
  year     = {1996},
  month    = oct,
  pages    = {L105},
  volume   = {470},
  doi      = {10.1086/310312},
  url      = {https://ui.adsabs.harvard.edu/abs/1996ApJ...470L.105H},
}

@Article{Wu2002,
  author        = {Wu, K. and Soria, R. and Campbell-Wilson, D. and Hannikainen, D. and Harmon, B.~A. and Hunstead, R. and Johnston, H. and McCollough, M. and McIntyre, V.},
  journal       = {\apj},
  title         = {The 1998 Outburst of XTE J1550-564: A Model Based on Multiwavelength Observations},
  year          = {2002},
  month         = feb,
  number        = {2},
  pages         = {1161-1168},
  volume        = {565},
  archiveprefix = {arXiv},
  doi           = {10.1086/324328},
  eprint        = {astro-ph/0109222},
  primaryclass  = {astro-ph},
  url           = {https://ui.adsabs.harvard.edu/abs/2002ApJ...565.1161W},
}

@Article{Hjellming1975,
  author   = {Hjellming, R.~M. and Gibson, D.~M. and Owen, F.~N.},
  journal  = {\nat},
  title    = {Another major change in the radio source associated with Cyg X-1},
  year     = {1975},
  month    = jul,
  number   = {5513},
  pages    = {111-112},
  volume   = {256},
  doi      = {10.1038/256111a0},
  url      = {https://ui.adsabs.harvard.edu/abs/1975Natur.256..111H},
}

@Article{Motta2021,
  author        = {Motta, S.~E. and Kajava, J.~J.~E. and Giustini, M. and Williams, D.~R.~A. and Del Santo, M. and Fender, R. and Green, D.~A. and Heywood, I. and Rhodes, L. and Segreto, A. and Sivakoff, G. and Woudt, P.~A.},
  journal       = {\mnras},
  title         = {Observations of a radio-bright, X-ray obscured GRS 1915+105},
  year          = {2021},
  month         = may,
  number        = {1},
  pages         = {152-161},
  volume        = {503},
  archiveprefix = {arXiv},
  doi           = {10.1093/mnras/stab511},
  eprint        = {2101.01187},
  primaryclass  = {astro-ph.HE},
  url           = {https://ui.adsabs.harvard.edu/abs/2021MNRAS.503..152M},
}

@Article{Tremou2026,
  author        = {Tremou, E. and Corbel, S. and Fender, R. and Woudt, P. and Miller-Jones, J.~C.~A. and Heywood, I. and Carotenuto, F. and Motta, S. and Tzioumis, A. and Groot, P.~J. and Russell, D.~M. and Crook-Mansour, J. and Saikia, P. and Yu, W. and van den Eijnden, J. and van der Horst, A.~J. and Williams-Baldwin, D.~R.~A. and Zhang, X.},
  journal       = {\mnras},
  title         = {The odyssey of the black hole low-mass X-ray binary GX 339{\ensuremath{-}}4: five years of dense multiwavelength monitoring},
  year          = {2026},
  month         = mar,
  number        = {3},
  pages         = {stag139},
  volume        = {546},
  archiveprefix = {arXiv},
  doi           = {10.1093/mnras/stag139},
  eid           = {stag139},
  eprint        = {2601.13360},
  primaryclass  = {astro-ph.HE},
  url           = {https://ui.adsabs.harvard.edu/abs/2026MNRAS.546ag139T},
}

@Article{Paizis2011,
  author        = {Paizis, A. and Nowak, M.~A. and Wilms, J. and Chaty, S. and Corbel, S. and Rodriguez, J. and Del Santo, M. and Ubertini, P. and Chini, R.},
  journal       = {\apj},
  title         = {Unveiling the Nature of IGR J17177-3656 with X-Ray, Near-infrared, and Radio Observations},
  year          = {2011},
  month         = sep,
  number        = {2},
  pages         = {183},
  volume        = {738},
  archiveprefix = {arXiv},
  doi           = {10.1088/0004-637X/738/2/183},
  eid           = {183},
  eprint        = {1107.0201},
  primaryclass  = {astro-ph.HE},
  url           = {https://ui.adsabs.harvard.edu/abs/2011ApJ...738..183P},
}

@Article{Grollimund2025,
  author   = {Grollimund, Noa and Corbel, Stephane and Bahramian, Arash and Carotenuto, Francesco and Heywood, Ian and Cowie, Fraser and Hughes, Andrew and Fender, Rob and Motta, Sara},
  journal  = {The Astronomer's Telegram},
  title    = {Discovery of the radio counterpart of MAXI J1744-294 with MeerKAT},
  year     = {2025},
  month    = feb,
  pages    = {1},
  volume   = {17045},
  url      = {https://ui.adsabs.harvard.edu/abs/2025ATel17045....1G},
}

@Article{Marti2026,
  author        = {Mart{\'\i}, Josep and Luque-Escamilla, Pedro L.},
  journal       = {\mnras},
  title         = {Alignment of radio jets in the microquasar V4641 Sagittarii with its high-energy structures},
  year          = {2026},
  month         = jan,
  number        = {3},
  pages         = {staf2104},
  volume        = {545},
  archiveprefix = {arXiv},
  doi           = {10.1093/mnras/staf2104},
  eid           = {staf2104},
  eprint        = {2511.19695},
  primaryclass  = {astro-ph.HE},
  url           = {https://ui.adsabs.harvard.edu/abs/2026MNRAS.545f2104M},
}

@article{SafiHarb2022,
    title = {Hard x-ray emission from the eastern jet of {SS} 433 powering the {W50} “manatee” nebula: {Evidence} for particle reacceleration},
    volume = {935},
    url = {https://ui.adsabs.harvard.edu/abs/2022ApJ...935..163S},
    doi = {10.3847/1538-4357/ac7c05},
    number = {2},
    journal = {ApJ},
    author = {Safi-Harb, Samar and Mac Intyre, Brydyn and Zhang, Shuo and Pope, Isaac and Zhang, Shuhan and Saffold, Nathan and Mori, Kaya and Gotthelf, Eric V. and Aharonian, Felix and Band, Matthew and Braun, Chelsea and Fang, Ke and Hailey, Charles and Nynka, Melania and Rho, Chang D.},
    month = aug,
    year = {2022},
    pages = {163},
}

\begin{appendix} 
\nolinenumbers
\twocolumn
\section{Source sample}
\label{app:cat}
Table~\ref{tab:sources} presents the black-hole \xray binary constructed as described in Section~\ref{sec:the_cat}. The table only includes the most relevant quantities, and in particular, does not include columns taken directly from the WATCHDOG catalogue.

\subsection{Conflicting estimates}
As reported in the text, in the event of a conflict when merging the different catalogues, we generally favoured the most recently reported value. However, in most cases the discrepancies were small and only for a  small subset of sources was this step necessary. We detail those cases below. Due to the low number of objects with discrepancies and the fact that in almost all cases, the value we adopted was derived with additional observational constraints, we are confident that our choices did not impact the conclusions on the paper. 
 
\subsubsection{Distances}
As detailed in the main text, we favoured distances estimates collected by BlackCat and WATCHDOG rather than those in XRBCat since the latter relies almost exclusively on Gaia while the first two take into account source-specific studies. There are 4 systems for which the distance provided in BlackCAT and WATCHDOG disagree by more than 10\%:
\begin{itemize}
    \item XTE~J1859+226: WATCHDOG combines different distance estimates, whereas BlackCat uses the distance estimate derived in~\citet{CorralSantana2011}. Because the values from the inclination, masses and period all are taken from~\citet{CorralSantana2011} for both WATCHDOG and BlackCat, we adopted their distance estimate for consistency. This is the only source among the four for which a $>$100 TeV upper limit exists and thus the only source included in the results presented in Section~\ref{sec:the_corr}.
    \item XTE J1752-223: The distance for this source is highly uncertain, and while WATCHDOG relies on~\citet{Shaposhnikov2010}, which provided an estimate with relatively low uncertainty (3.5$\pm$0.4~kpc) while later work by~\citet{Ratti2012} indicate that the distance could be anywhere in the range from 3.5 to 8~kpc. We adopted the more recent (and uncertain) result.
    \item 1A 1524-62: WATCHDOG combines several distance literature estimates. We instead follow BlackCAT and use the updated value from~\citet{YanesRizo2024} which characterized the orbit and optical properties which high accuracy (which we also adopt).
    \item MAXI J1659-152: WATCHDOG combines several literature estimates. We instead follow BlackCAT and use the value from~\citep{Kuulkers2013} for consistency with the estimates of inclination and mass used.
\end{itemize}

\subsubsection{Orbital period}
The same is true about the orbital periods, with even smaller spreads. with the exception of three objects:
\begin{itemize}
    \item MAXI J1910.2-0546: BlackCAT reports a period estimate based on an early Atel~\citep{Casares2012}, which was later refined in a detailed study by~\citet{CorralSantana2025}.
    \item GRS 1716-249: WATCHDOG relies on the early work by ~\citet{Masetti1996}, which has now been superseded by an updated study with significantly more data~\citep{Casares2023}
    \item GRS 1758-258:  the value of 18 days reported in WATCHDOG was updated by~\citet{Marti2016}, who argue that this value corresponds to the precession period of the disk, with the actual orbital period being much shorter ($\sim$1 day).
\end{itemize}

\subsubsection{Black hole mass and inclination}
The black hole masses and inclinations reported in the different catalogues are all within 10-30\% (even before taking into account uncertainties).

\section{Fits to gamma-ray spectra}
\label{app:gamma}
\begin{figure*}[h]
	\centering
		\includegraphics[width=0.95\linewidth]{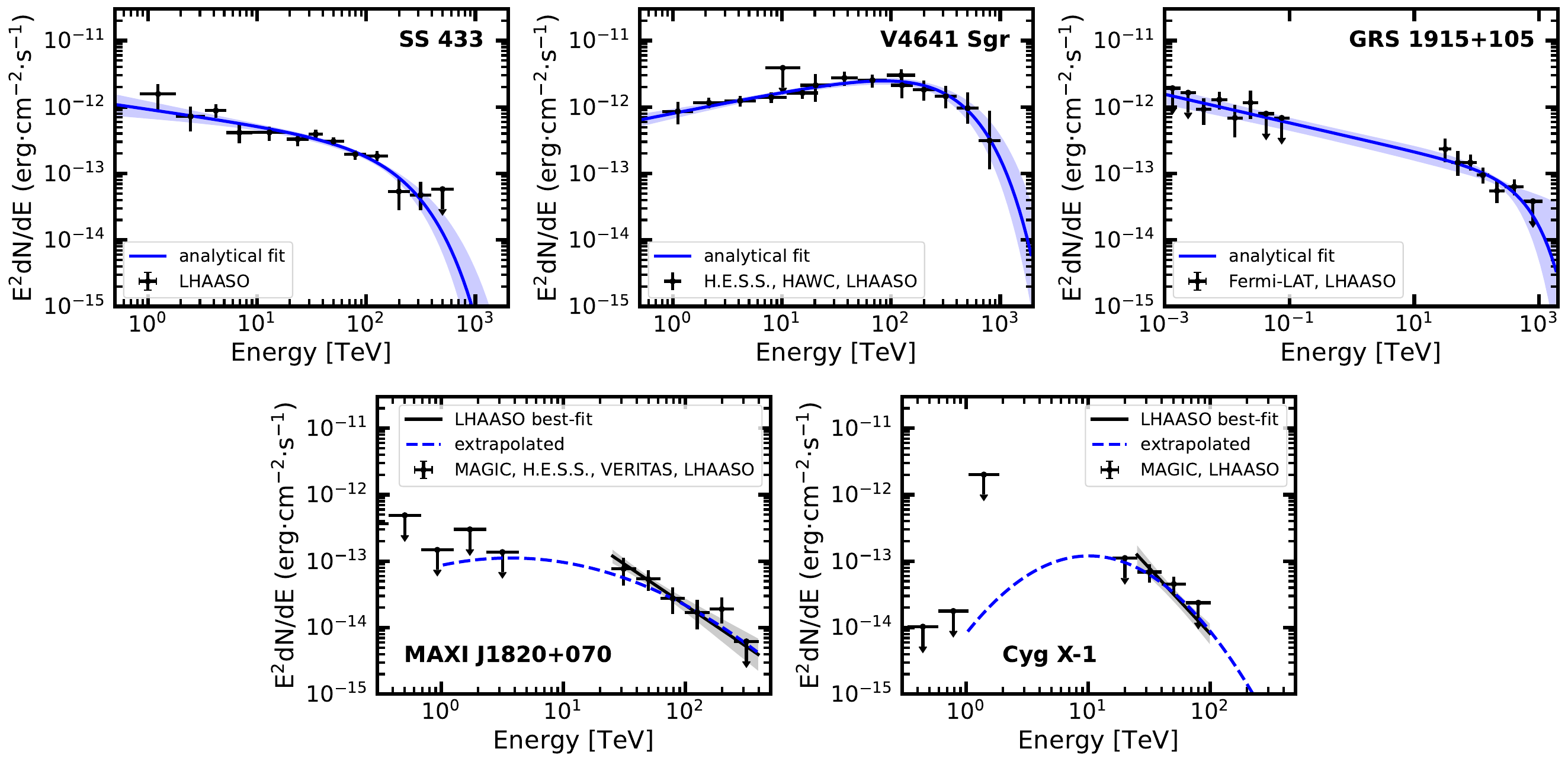} 
	\caption{\label{fig:spectra} Gamma-ray spectra of the sources with a persistent gamma-ray counterpart. Black points represent the measured spectra and blue lines the best-fit analytical model used to derive the quantities presented in Table~\ref{tab:detected}. For \maxi and \cyg, the best-fit model derived at energies $>25$~TeV by~\citet{LhaasoCollaboration2025} is shown in black, with a blue dashed line indicating an example extrapolation to lower energies, which is only shown for reference.}
\end{figure*}

As detailed in Section~\ref{sec:the_guys}, we have collected the available gamma-ray spectra for all the detected systems. To combine measurements from different instruments and determine the spectral shape across the gamma-ray range, we have fit simple analytical models to these spectra. The spectra and best-fit models are shown in Figure~\ref{fig:spectra}.

\section{Comparisons at lower energies}
\label{app:above1TeV}
We include below the same quantities shown in Figures~\ref{fig:arch} and~\ref{fig:power} but plotted against the gamma-ray luminosity above 1~TeV.

\begin{table*}
\caption{\label{tab:sources} List of black-hole (candidate) \xray binaries and their properties. }
\resizebox{\textwidth}{!}{\begin{tabular}{lcccccccccc}
 \hline \hline
 \noalign{\vskip 1mm} 
  Name  & l & b & D & $M_{\mathrm{BH}}$ & P$_{\mathrm{orb}}$ & i & L$_{\mathrm{5~GHz}}$ & L$_{>\mathrm{100~TeV}}$ & Jet P.A. &  Refs \\
      & (deg) & (deg) & (kpc) &(M$_{\odot}$) & (h) & (deg) & (erg s$^{-1}$) &  (erg s$^{-1}$)& (deg) & - \\
\noalign{\vskip 1mm} 
\hline
\noalign{\vskip 1mm} 
\noalign{\vskip 1mm} GRS 1716-249 & 0.142 & 6.991 & 6.9 $ \pm $ 1.1 & 6.4 $^{+3.2}_{-2.0}$ & 6.7 & 61.0 $^{+15.0}_{-15.0}$ & (4.41 $^{+1.52}_{-1.31}) \cdot 10^{30}$ &  - &  - & [1,2,3,4,5-8]\\ \noalign{\vskip 0.5mm} KS 1732-273 & 0.161 & 2.59 & 5.0 $ \pm $ 3.0 &  - &  - &  - &  - &  - &  - & [1,2,3,4]\\ \noalign{\vskip 0.5mm} GRS 1739-278 & 0.672 & 1.176 & 7.25 $ \pm $ 1.25 & 9.5 &  - & 70.0 $^{+5.0}_{-11.0}$ & (3.14 $^{+1.13}_{-1.0}) \cdot 10^{30}$ &  - &  - & [1,2,3,4, 9]\\ \noalign{\vskip 0.5mm} XTE J1748-288 & 0.676 & -0.222 & 5.0 $ \pm $ 3.0 &  - &  - &  - & (7.22 $^{+7.14}_{-4.68}) \cdot 10^{31}$ &  - &  - & [1,2,3,4,10-11]\\ \noalign{\vskip 0.5mm} IGR J17497-2821 & 0.953 & -0.453 & 5.0 $ \pm $ 3.0 &  - &  - &  - &  - &  - &  - & [1,2,3,4, 12]\\ \noalign{\vskip 0.5mm} MAXI J1803-298 & 1.147 & -3.728 & 5.0 $ \pm $ 3.0 & 6.0 $^{+0.4}_{-0.4}$ &  - & $>$70.0 & (3.08 $^{+3.04}_{-2.01}) \cdot 10^{30}$ &  - & 313.9 & [2,13-15]\\ \noalign{\vskip 0.5mm} Swift J174510.8-262411 & 2.111 & 1.403 & 5.0 $ \pm $ 3.0 &  - & $<$21.0 &  - & (2.49 $^{+2.39}_{-1.62}) \cdot 10^{30}$ &  - &  - & [1,2,3,4]\\ \noalign{\vskip 0.5mm} Swift J1753.7-2544 & 3.648 & 0.103 & 5.0 $ \pm $ 3.0 &  - &  - &  - &  - &  - &  - & [2]\\ \noalign{\vskip 0.5mm} CXOU J174805.0-244643 & 3.84 & 1.686 & 5.5 $ \pm $ 0.9 &  - &  - &  - &  - &  - &  - & [3,4]\\ \noalign{\vskip 0.5mm} GRS 1758-258 & 4.508 & -1.361 & 5.0 $ \pm $ 3.0 & 10.0 & 24.0 & 67.0 $^{+8.0}_{-13.0}$ & (7.48 $^{+7.38}_{-4.87}) \cdot 10^{28}$ &  - & 13.0 & [1,3,4,16-19]\\ \noalign{\vskip 0.5mm} MAXI J1727-203 & 5.179 & 7.849 & 5.0 $ \pm $ 3.0 & $>$11.5 &  - &  - &  - &  - &  - & [3,4]\\ \noalign{\vskip 0.5mm} MAXI J1659-152 & 5.515 & 16.526 & 8.6 $ \pm $ 3.7 & 5.7 & 2.414 $ \pm $ 0.005 & 72.5 $^{+7.5}_{-7.5}$ & (2.18 $^{+2.24}_{-1.48}) \cdot 10^{30}$ &  - &  - & [1,2,3,4,20-21]\\ \noalign{\vskip 0.5mm} XTE J1752-223 & 6.423 & 2.114 & 6.0 $ \pm $ 2.0 & 9.6 $^{+0.9}_{-0.9}$ &  - & 35.0 $^{+4.0}_{-4.0}$ & (4.31 $^{+3.44}_{-2.34}) \cdot 10^{30}$ &  - & -51.0 & [1,2,3,4,22-26]\\ \noalign{\vskip 0.5mm} V4641 Sgr & 6.774 & -4.789 & 6.2 $ \pm $ 0.7 & 6.4 $^{+0.6}_{-0.6}$ & 67.6 & 72.3 $^{+4.1}_{-4.1}$ & (9.66 $^{+2.33}_{-2.12}) \cdot 10^{31}$ & (1.56 $^{+0.48}_{-0.4}) \cdot 10^{34}$ & 162.0 & [1,2,3,4, 27]\\ \noalign{\vskip 0.5mm} XTE J1818-245 & 7.443 & -4.192 & 3.55 $ \pm $ 0.75 &  - &  - &  - & (2.06 $^{+0.94}_{-0.78}) \cdot 10^{30}$ &  - &  - & [1,2,3,4, 28]\\ \noalign{\vskip 0.5mm} MAXI J1828-249 & 8.115 & -6.545 & 5.0 $ \pm $ 3.0 & 4.0 &  - &  - & (2.06 $^{+2.03}_{-1.35}) \cdot 10^{29}$ &  - &  - & [2,3,4, 29]\\ \noalign{\vskip 0.5mm} Swift J1727.8-1613 & 8.642 & 10.255 & 5.5 $ \pm $ 1.4 & $>$3.12 & 10.8 & $<$74.0 & (1.73 $^{+0.99}_{-0.76}) \cdot 10^{31}$ & $< 1.89 \cdot 10^{32}$ & -0.6 & [2,30-32]\\ \noalign{\vskip 0.5mm} MAXI J1810-222 & 8.77 & -1.981 & 5.0 $ \pm $ 3.0 &  - &  - &  - & (8.53 $^{+8.24}_{-5.45}) \cdot 10^{28}$ &  - &  - & [2,3,4,33-34]\\ \noalign{\vskip 0.5mm} XTE J1812-182 & 12.358 & 0.034 & 5.0 $ \pm $ 3.0 &  - &  - &  - &  - &  - &  - & [1]\\ \noalign{\vskip 0.5mm} MAXI J1836-194 & 13.945 & -5.355 & 7.0 $ \pm $ 3.0 & 7.0 & $<$4.9 & 9.5 $^{+5.5}_{-5.5}$ & (8.38 $^{+8.69}_{-5.51}) \cdot 10^{30}$ &  - &  - & [1,2,3,4,35-36]\\ \noalign{\vskip 0.5mm} IGR J18175-1530 & 15.312 & 0.247 & 5.0 $ \pm $ 3.0 &  - &  - &  - &  - &  - &  - & [1,3,4]\\ \noalign{\vskip 0.5mm} MAXI J1813-095 & 20.112 & 3.948 & 5.0 $ \pm $ 3.0 & 7.4 $^{+1.5}_{-1.5}$ &  - & 36.5 $^{+8.5}_{-8.5}$ & (1.03 $^{+1.03}_{-0.67}) \cdot 10^{29}$ &  - &  - & [2,37-39]\\ \noalign{\vskip 0.5mm} Swift J1842.5-1124 & 21.727 & -3.179 & 5.0 $ \pm $ 3.0 &  - &  - &  - & (3.42 $^{+3.39}_{-2.23}) \cdot 10^{28}$ &  - &  - & [1,2,3,4, 40]\\ \noalign{\vskip 0.5mm} Swift J1753.5-0127 & 24.898 & 12.186 & 3.9 $ \pm $ 0.7 & 8.8 $^{+1.3}_{-1.3}$ & 3.2 $ \pm $ 0.02 & 79.0 $^{+5.0}_{-5.0}$ & (2.33 $^{+0.92}_{-0.75}) \cdot 10^{28}$ &  - &  - & [1,2,3,4]\\ \noalign{\vskip 0.5mm} MAXI J1910.2-0546 & 29.902 & -6.844 & 5.0 $ \pm $ 3.0 & 9.0 $^{+1.5}_{-1.5}$ & $<$4.5 & 15.5 $^{+2.5}_{-2.5}$ & (3.74 $^{+3.76}_{-2.45}) \cdot 10^{29}$ &  - &  - & [1,2,41-43]\\ \noalign{\vskip 0.5mm} EXO 1846-031 & 29.958 & -0.919 & 5.0 $ \pm $ 3.0 & 12.0 &  - & 62.0 $^{+9.0}_{-10.0}$ & (1.03 $^{+1.03}_{-0.66}) \cdot 10^{30}$ &  - & 99.0 & [1,2,3,4, 44]\\ \noalign{\vskip 0.5mm} MAXI J1848-015 & 31.308 & -0.091 & 3.4 $ \pm $ 1.0 &  - &  - & 76.0 $^{+2.0}_{-2.0}$ & (3.46 $^{+2.35}_{-1.72}) \cdot 10^{27}$ &  - & 144.9 & [3,4, 45]\\ \noalign{\vskip 0.5mm} XTE J1901+014 & 35.381 & -1.622 & 5.0 $ \pm $ 3.0 &  - &  - &  - &  - &  - &  - & [1]\\ \noalign{\vskip 0.5mm} MAXI J1820+070 & 35.854 & 10.159 & 2.96 $ \pm $ 0.33 & 8.48 $^{+0.72}_{-0.79}$ & 16.45 & 63.0 $^{+3.0}_{-3.0}$ & (6.19 $^{+1.41}_{-1.29}) \cdot 10^{29}$ & (1.72 $^{+0.93}_{-0.63}) \cdot 10^{31}$ & 25.1 & [2,3,4,46-50]\\ \noalign{\vskip 0.5mm} XTE J1856+053 & 38.258 & 1.249 & 5.0 $ \pm $ 3.0 &  - &  - &  - &  - &  - &  - & [1]\\ \noalign{\vskip 0.5mm} SS 433 & 39.694 & -2.245 & 5.5 $ \pm $ 0.2 & 15.0 $^{+2.0}_{-2.0}$ & 314.4 & 78.8 $^{+0.0}_{-0.1}$ & (2.71 $^{+0.21}_{-0.19}) \cdot 10^{32}$ & (4.81 $^{+1.02}_{-0.76}) \cdot 10^{32}$ & 80.0 & [1,3,4, 51]\\ \noalign{\vskip 0.5mm} IGR J18539+0727 & 39.847 & 2.846 & 5.0 $ \pm $ 3.0 &  - &  - &  - & (7.85 $^{+7.71}_{-5.04}) \cdot 10^{29}$ &  - &  - & [1,3,4,52-53]\\ \noalign{\vskip 0.5mm} XTE J1908+094 & 43.263 & 0.434 & 6.5 $ \pm $ 3.5 &  - &  - & $>$79.0 & (3.26 $^{+4.53}_{-2.54}) \cdot 10^{29}$ &  - & -11.0 & [1,2,3,4,54-55]\\ \noalign{\vskip 0.5mm} GRS 1915+105 & 45.366 & -0.219 & 9.4 $ \pm $ 0.8 & 12.4 $^{+2.0}_{-1.8}$ & 739.0 & 70.0 $^{+2.0}_{-2.0}$ & (7.19 $^{+1.29}_{-1.16}) \cdot 10^{32}$ & (1.56 $^{+0.92}_{-0.37}) \cdot 10^{33}$ & 149.7 & [1,2,3,4,56-64]\\ \noalign{\vskip 0.5mm} 4U 1957+115 & 51.308 & -9.33 & 7.8 $ \pm $ 3.0 & 3.0 & 9.33 & $>$50.0 &  - &  - &  - & [1,3,4]\\ \noalign{\vskip 0.5mm} XTE J1859+226 & 54.046 & 8.608 & 12.5 $ \pm $ 1.5 & 10.83 $^{+4.67}_{-4.67}$ & 6.6 $ \pm $ 0.05 & 60.0 $^{+3.0}_{-3.0}$ & (1.03 $^{+0.26}_{-0.23}) \cdot 10^{32}$ & $< 8.11 \cdot 10^{32}$ &  - & [1,2,3,4,65-67]\\ \noalign{\vskip 0.5mm} GS 2000+25 & 63.366 & -2.999 & 2.7 $ \pm $ 0.7 & 8.37 $^{+1.3}_{-1.3}$ & 8.3 & 60.0 $^{+5.0}_{-5.0}$ & (2.14 $^{+1.44}_{-1.14}) \cdot 10^{29}$ & $< 1.51 \cdot 10^{31}$ &  - & [1,2,3,4, 68]\\ \noalign{\vskip 0.5mm} Cyg X-1 & 71.335 & 3.067 & 1.86 $ \pm $ 0.12 & 14.81 $^{+0.98}_{-0.98}$ & 134.4 & 27.06 $^{+0.76}_{-0.76}$ & (9.31 $^{+2.39}_{-2.35}) \cdot 10^{29}$ & (1.56 $^{+1.03}_{-0.62}) \cdot 10^{30}$ &  - & [1,3,4, 69]\\ \noalign{\vskip 0.5mm} V404 Cyg & 73.119 & -2.091 & 2.39 $ \pm $ 0.14 & 7.15 $^{+0.35}_{-0.35}$ & 155.3 & 80.1 $^{+5.1}_{-5.1}$ & (5.6 $^{+0.72}_{-0.69}) \cdot 10^{31}$ & $< 5.75 \cdot 10^{30}$ &  - & [1,2,3,4,70-72]\\ \noalign{\vskip 0.5mm} XTE J2012+381 & 75.388 & 2.247 & 5.0 $ \pm $ 3.0 &  - &  - & 68.0 $^{+6.0}_{-11.0}$ & (4.49 $^{+4.39}_{-2.92}) \cdot 10^{29}$ &  - &  - & [1,2,3,4]\\ \noalign{\vskip 0.5mm} Cyg X-3 & 79.845 & 0.7 & 8.3 $ \pm $ 1.6 & 2.4 $^{+2.1}_{-1.1}$ & 4.8 & 43.0 $^{+11.0}_{-9.0}$ & (8.24 $^{+3.41}_{-2.83}) \cdot 10^{34}$ & $< 1.94 \cdot 10^{34}$ &  - & [1,3,4]\\ \noalign{\vskip 0.5mm} MWC 656 & 100.175 & -12.398 & 2.6 $ \pm $ 0.6 & 5.35 $^{+1.55}_{-1.55}$ & 1448.9 & $>$66.0 & (5.74 $^{+3.21}_{-2.58}) \cdot 10^{26}$ &  - &  - & [1,3,4,73-74]\\ \noalign{\vskip 0.5mm} Cl Cam & 149.177 & 4.133 & 5.0 $ \pm $ 3.0 &  - & 465.84 &  - & (1.5 $^{+1.45}_{-0.98}) \cdot 10^{32}$ &  - &  - & [1,75-76]\\ \noalign{\vskip 0.5mm} AT 2019wey & 151.161 & 5.3 & 5.0 $ \pm $ 3.0 &  - & $<$8.2 & $<$30.0 & (1.97 $^{+1.9}_{-1.28}) \cdot 10^{29}$ &  - & 122.0 & [2,3,4,77-78]\\ \noalign{\vskip 0.5mm} XTE J1118+480 & 157.661 & 62.32 & 1.72 $ \pm $ 0.1 & 7.3 $^{+0.73}_{-0.73}$ & 4.1 & 75.0 $^{+7.0}_{-7.0}$ & (9.13 $^{+1.17}_{-1.12}) \cdot 10^{28}$ & $< 3.75 \cdot 10^{30}$ &  - & [1,2,3,4,79-80]\\ \noalign{\vskip 0.5mm} GRO J0422+32 & 165.881 & -11.913 & 2.49 $ \pm $ 0.3 & 3.69 $^{+0.41}_{-0.41}$ & 5.1 & 63.7 $^{+5.2}_{-5.2}$ & (3.71 $^{+0.93}_{-0.85}) \cdot 10^{28}$ & $< 1.17 \cdot 10^{31}$ &  - & [1,2,3,4,81-82]\\ \noalign{\vskip 0.5mm} 1A 0620-00 & 209.338 & -6.223 & 1.06 $ \pm $ 0.1 & 5.86 $^{+0.25}_{-0.25}$ & 7.8 & 51.0 $^{+0.9}_{-0.9}$ & (2.02 $^{+0.67}_{-0.65}) \cdot 10^{30}$ & $< 1.11 \cdot 10^{30}$ &  - & [1,2,3,4,83-86]\\ \noalign{\vskip 0.5mm} MAXI J0637-430 & 251.518 & -20.671 & 8.0 $ \pm $ 1.0 & 5.0 & 2.2 &  - & (2.53 $^{+0.7}_{-0.66}) \cdot 10^{28}$ &  - &  - & [2,3,4,87-90]\\ \noalign{\vskip 0.5mm} LMC X-3 & 273.576 & -32.082 & 48.1 $ \pm $ 2.2 & 6.95 $^{+0.33}_{-0.33}$ & 40.9 & 69.65 $^{+0.56}_{-0.56}$ &  - &  - &  - & [1, 91]\\ \noalign{\vskip 0.5mm} GRS 1009-45 & 275.879 & 9.346 & 3.8 $ \pm $ 0.3 & 5.95 $^{+0.89}_{-0.89}$ & 6.8 & 62.0 $^{+5.1}_{-5.1}$ &  - &  - &  - & [1,2,3,4]\\ \noalign{\vskip 0.5mm} LMC X-1 & 280.203 & -31.516 & 50.0 $ \pm $ 2.3 & 10.91 $^{+1.41}_{-1.41}$ & 93.8 & 36.38 $^{+1.92}_{-1.92}$ &  - &  - &  - & [1]\\ \noalign{\vskip 0.5mm} IGR J11321-5311 & 291.087 & 7.854 & 5.0 $ \pm $ 3.0 &  - &  - &  - &  - &  - &  - & [1,3,4]\\ \noalign{\vskip 0.5mm} GRS 1124-68 & 295.3 & -7.073 & 4.95 $ \pm $ 0.65 & 11.0 $^{+2.1}_{-1.4}$ & 10.4 & 42.2 $^{+2.1}_{-2.7}$ & (2.51 $^{+0.68}_{-0.61}) \cdot 10^{31}$ &  - &  - & [1,2,3,4,92-94]\\ \noalign{\vskip 0.5mm} MAXI J1305-704 & 304.238 & -7.619 & 7.5 $ \pm $ 1.5 & 8.9 & 9.74 $ \pm $ 0.1 & 67.5 $^{+7.5}_{-7.5}$ & (1.55 $^{+0.69}_{-0.55}) \cdot 10^{29}$ &  - &  - & [1,2,3,4, 95]\\ \noalign{\vskip 0.5mm} 47 Tuc-X9 & 305.897 & -44.888 & 4.53 $ \pm $ 0.01 &  - & 0.45 &  - & (5.16 $^{+0.49}_{-0.47}) \cdot 10^{27}$ &  - &  - & [3,4]\\ \noalign{\vskip 0.5mm} MAXI J1348-630 & 309.264 & -1.103 & 2.2 $ \pm $ 0.6 & 8.7 &  - & $<$46.0 & (1.41 $^{+0.86}_{-0.66}) \cdot 10^{31}$ &  - & 33.2 & [2,3,4,96-97]\\ \noalign{\vskip 0.5mm} BW Cir & 309.977 & -2.78 & 43.0 $ \pm $ 18.0 & 7.47 & 61.1 & 54.0 $^{+26.8}_{-26.8}$ & (3.1 $^{+3.11}_{-2.02}) \cdot 10^{31}$ &  - &  - & [1,2,3,4,98-100]\\ \noalign{\vskip 0.5mm} 1A 1524-62 & 320.319 & -4.427 & 8.0 $ \pm $ 0.9 &  - & 6.2 $ \pm $ 0.2 &  - &  - &  - &  - & [1,2,3,4, 101]\\ \noalign{\vskip 0.5mm} Swift J1539.2-6227 & 321.019 & -5.643 & 5.0 $ \pm $ 3.0 &  - &  - &  - &  - &  - &  - & [1,2,3,4]\\ \noalign{\vskip 0.5mm} Swift J151857.0-572147 & 321.816 & -0.003 & 5.8 $ \pm $ 2.5 & 9.5 $^{+1.0}_{-1.0}$ &  - & 40.5 $^{+5.5}_{-5.5}$ & (3.42 $^{+3.58}_{-2.27}) \cdot 10^{32}$ &  - & 135.0 & [2,102-104]\\ \noalign{\vskip 0.5mm} MAXI J1535-571 & 323.724 & -1.129 & 4.1 $ \pm $ 0.2 & 8.9 $^{+1.0}_{-1.0}$ &  - & $<$45.0 & (1.81 $^{+0.18}_{-0.17}) \cdot 10^{31}$ &  - & 124.7 & [2,3,4,105-106]\\ \noalign{\vskip 0.5mm} MAXI J1543-564 & 325.085 & -1.121 & 10.0 $ \pm $ 3.0 &  - &  - &  - & (2.27 $^{+1.36}_{-0.5}) \cdot 10^{29}$ &  - &  - & [1,2,3,4, 107]\\ \noalign{\vskip 0.5mm} XTE J1550-564 & 325.882 & -1.827 & 4.5 $ \pm $ 0.5 & 10.39 $^{+2.26}_{-2.26}$ & 37.0 & 67.4 $^{+9.7}_{-9.7}$ & (4.54 $^{+1.06}_{-0.94}) \cdot 10^{31}$ &  - & 6.5 & [1,2,3,4,108-114]\\ \noalign{\vskip 0.5mm} Swift J1357.2-0933 & 328.702 & 50.004 & 3.9 $ \pm $ 2.4 &  - & 2.8 $ \pm $ 0.3 & $>$80.0 &  - &  - &  - & [1,2,3,4,115-118]\\ \noalign{\vskip 0.5mm} 4U 1543-475 & 330.918 & 5.426 & 7.5 $ \pm $ 0.5 & 9.4 $^{+2.0}_{-2.0}$ & 26.8 & 30.0 $^{+6.0}_{-6.0}$ & (1.01 $^{+0.13}_{-0.13}) \cdot 10^{31}$ &  - &  - & [1,2,3,4, 119]\\ \noalign{\vskip 0.5mm} XTE J1637-498 & 335.426 & -1.783 & 5.0 $ \pm $ 3.0 &  - &  - &  - &  - &  - &  - & [1,120-121]\\ \noalign{\vskip 0.5mm} MAXI J1631-479 & 336.287 & 0.312 & 5.0 $ \pm $ 3.0 & 9.2 $^{+1.3}_{-1.0}$ &  - & 22.0 $^{+10.0}_{-12.0}$ & (5.59 $^{+5.5}_{-3.62}) \cdot 10^{30}$ &  - &  - & [2,3,4,122-125]\\ \noalign{\vskip 0.5mm} XTE J1650-500 & 336.718 & -3.427 & 2.6 $ \pm $ 0.7 & 4.72 $^{+2.16}_{-2.16}$ & 7.7 $ \pm $ 0.02 & 75.2 $^{+5.9}_{-5.9}$ & (2.02 $^{+1.19}_{-0.92}) \cdot 10^{29}$ &  - &  - & [1,2,3,4, 126]\\ \noalign{\vskip 0.5mm} 4U 1630-472 & 336.911 & 0.25 & 8.1 $ \pm $ 3.5 &  - &  - & 46.4 $^{+0.2}_{-3.0}$ & (2.02 $^{+1.32}_{-0.99}) \cdot 10^{30}$ &  - &  - & [1,2,3,4,127-128]\\ \noalign{\vskip 0.5mm} GX 339-4 & 338.939 & -4.326 & 10.0 $ \pm $ 2.0 & $>$3.0 & 42.1 & 50.0 $^{+10.0}_{-10.0}$ & (5.31 $^{+2.35}_{-1.88}) \cdot 10^{31}$ &  - & -62.0 & [1,2,3,4,129-134] \\ 
\noalign{\vskip 1mm} 
\hline
\end{tabular}}

\end{table*}

\setcounter{table}{0}

\begin{table*}
\caption{List of black-hole (candidate) \xray binaries and their properties. }
\resizebox{\textwidth}{!}{\begin{tabular}{lcccccccccc}
 \hline \hline
 \noalign{\vskip 1mm} 
  Name  & l & b & D & $M_{\mathrm{BH}}$ & P$_{\mathrm{orb}}$ & i & L$_{\mathrm{5~GHz}}$ & L$_{>\mathrm{100~TeV}}$ & Jet P.A. &  Refs \\
      & (deg) & (deg) & (kpc) &(M$_{\odot}$) & (h) & (deg) & (erg s$^{-1}$) &  (erg s$^{-1}$)& (deg) & - \\
\noalign{\vskip 1mm} 
\hline
\noalign{\vskip 1mm} 
\noalign{\vskip 0.5mm} XTE J1652-453 & 340.53 & -0.787 & 5.0 $ \pm $ 3.0 &  - &  - &  - & (1.42 $^{+1.41}_{-0.93}) \cdot 10^{29}$ &  - &  - & [1,2,3,4, 135]\\ \noalign{\vskip 0.5mm} XTE J1727-476 & 342.203 & -6.923 & 5.0 $ \pm $ 3.0 &  - &  - &  - &  - &  - &  - & [1,2,3,4]\\ \noalign{\vskip 0.5mm} Swift J1658.2-4242 & 343.251 & 0.053 & 10.0 $ \pm $ 1.0 & 14.0 & 4.0 $ \pm $ 0.4 & 64.0 $^{+2.0}_{-3.0}$ & (1.41 $^{+0.31}_{-0.28}) \cdot 10^{30}$ &  - &  - & [2,3,4,136-138]\\ \noalign{\vskip 0.5mm} GRO J1655-40 & 344.982 & 2.456 & 3.2 $ \pm $ 0.5 & 5.3 $^{+0.5}_{-0.5}$ & 62.9 & 70.2 $^{+1.9}_{-1.9}$ & (3.37 $^{+1.1}_{-0.96}) \cdot 10^{32}$ &  - & 47.0 & [1,2,3,4,139-144]\\ \noalign{\vskip 0.5mm} Swift J1713.4-4219 & 345.242 & -1.96 & 5.0 $ \pm $ 3.0 &  - &  - &  - &  - &  - &  - & [1, 145]\\ \noalign{\vskip 0.5mm} SAX J1711.6-3808 & 348.441 & 0.798 & 5.0 $ \pm $ 3.0 &  - &  - &  - &  - &  - &  - & [1,2,3,4]\\ \noalign{\vskip 0.5mm} IGR J17091-3624 & 349.525 & 2.213 & 14.0 $ \pm $ 3.0 & 8.7 &  - & 35.0 $^{+5.0}_{-5.0}$ & (1.79 $^{+0.57}_{-0.49}) \cdot 10^{30}$ &  - &  - & [1,2,3,4, 146]\\ \noalign{\vskip 0.5mm} IGR J17098-3628 & 349.554 & 2.074 & 10.5 $ \pm $ 3.0 &  - &  - &  - & (2.24 $^{+1.53}_{-1.15}) \cdot 10^{29}$ &  - &  - & [1,2,3,4, 147]\\ \noalign{\vskip 0.5mm} IGR J17177-3656 & 350.107 & 0.509 & 23.0 $ \pm $ 3.0 &  - &  - & 67.5 $^{+7.5}_{-7.5}$ & (7.6 $^{+2.82}_{-2.6}) \cdot 10^{29}$ &  - &  - & [3,4,148-149]\\ \noalign{\vskip 0.5mm} XMM SL1J171900.4-35321 & 351.399 & 1.098 & 5.0 $ \pm $ 3.0 &  - &  - &  - &  - &  - &  - & [1,3,4]\\ \noalign{\vskip 0.5mm} Swift J1728.9-3613 & 351.955 & -0.966 & 8.4 $ \pm $ 0.8 & 4.6 &  - &  - & (4.6 $^{+0.99}_{-0.87}) \cdot 10^{30}$ &  - &  - & [2,3,4,150-151]\\ \noalign{\vskip 0.5mm} XTE J1720-318 & 354.624 & 3.101 & 6.5 $ \pm $ 3.5 &  - &  - &  - & (1.24 $^{+1.64}_{-0.97}) \cdot 10^{30}$ &  - &  - & [1,2,3,4,152-153]\\ \noalign{\vskip 0.5mm} Swift J1729.5-3223 & 355.249 & 1.069 & 5.0 $ \pm $ 3.0 &  - &  - &  - & (2.69 $^{+2.63}_{-1.76}) \cdot 10^{28}$ &  - &  - & [3,4, 154]\\ \noalign{\vskip 0.5mm} GRS 1730-312 & 356.725 & 0.944 & 5.0 $ \pm $ 3.0 &  - &  - &  - &  - &  - &  - & [1,2,3,4]\\ \noalign{\vskip 0.5mm} XTE J1719-291 & 356.745 & 4.754 & 5.0 $ \pm $ 3.0 &  - &  - &  - &  - &  - &  - & [1, 155]\\ \noalign{\vskip 0.5mm} SLX 1746-331 & 356.807 & -2.973 & 5.0 $ \pm $ 3.0 & $>$3.28 &  - & 55.0 $^{+5.0}_{-5.0}$ &  - &  - &  - & [1,2,3,4, 156]\\ \noalign{\vskip 0.5mm} 4U 1755-33 & 357.215 & -4.872 & 6.5 $ \pm $ 2.5 &  - & 4.4 &  - &  - &  - & 137.1 & [1,2,3,4,157-158]\\ \noalign{\vskip 0.5mm} H 1743-322 & 357.255 & -1.833 & 10.4 $ \pm $ 2.9 & 15.0 &  - & 75.0 $^{+3.0}_{-3.0}$ & (1.57 $^{+0.97}_{-0.75}) \cdot 10^{30}$ &  - & 90.0 & [1,2,3,4]\\ \noalign{\vskip 0.5mm} GRS 1737-31 & 357.588 & -0.099 & 5.0 $ \pm $ 3.0 &  - &  - &  - &  - &  - &  - & [1,2,3,4]\\ \noalign{\vskip 0.5mm} IGR J17285-2922 & 357.651 & 2.898 & 5.0 $ \pm $ 3.0 &  - & 9.3 & $>$18.0 & (1.76 $^{+1.76}_{-1.14}) \cdot 10^{31}$ &  - &  - & [1,3,4]\\ \noalign{\vskip 0.5mm} XTE J1755-324 & 358.039 & -3.631 & 5.0 $ \pm $ 3.0 &  - &  - &  - &  - &  - &  - & [1,2]\\ \noalign{\vskip 0.5mm} KS 1739-304 & 358.327 & -0.292 & 5.0 $ \pm $ 3.0 &  - &  - &  - &  - &  - &  - & [3,4]\\ \noalign{\vskip 0.5mm} H 1705-250 & 358.587 & 9.057 & 8.6 $ \pm $ 2.1 & 8.92 $^{+3.56}_{-3.56}$ & 12.5 $ \pm $ 0.03 & 64.0 $^{+16.0}_{-16.0}$ &  - &  - &  - & [1,2,3,4]\\ \noalign{\vskip 0.5mm} IGR J17451-3022 & 358.712 & -0.658 & 5.0 $ \pm $ 3.0 &  - & 6.284 & 73.5 $^{+2.5}_{-2.5}$ &  - &  - &  - & [2,159-160]\\ \noalign{\vskip 0.5mm} 1E 1740.7-2942 & 359.116 & -0.106 & 5.0 $ \pm $ 3.0 &  - & 305.52 & 65.0 $^{+0.0}_{-0.0}$ & (7.03 $^{+6.67}_{-4.6}) \cdot 10^{28}$ &  - & 339.0 & [1,3,4,161-164]\\ \noalign{\vskip 0.5mm} MAXI J1744-294 & 359.443 & -0.041 & 8.0 $ \pm $ 0.3 &  - &  - & 28.0 $^{+4.0}_{-3.0}$ & (4.21 $^{+1.93}_{-1.92}) \cdot 10^{31}$ &  - &  - & [2,165-166]\\ \noalign{\vskip 0.5mm} IGR J17454-2919 & 359.644 & -0.177 & 5.0 $ \pm $ 3.0 &  - &  - &  - &  - &  - &  - & [1,2,3,4, 167]\\ \noalign{\vskip 0.5mm} XTE J1817-330 & 359.817 & -7.996 & 5.5 $ \pm $ 4.5 &  - &  - &  - & (1.5 $^{+3.52}_{-1.32}) \cdot 10^{29}$ &  - &  - & [1,2,3,4,168-169]\\ \noalign{\vskip 0.5mm} 1A 1742-289 & 359.929 & -0.042 & 5.0 $ \pm $ 3.0 &  - &  - &  - & (3.22 $^{+3.14}_{-2.09}) \cdot 10^{31}$ &  - &  - & [1, 170]\\ \noalign{\vskip 0.5mm} CXOGC J174540.0-290031 & 359.944 & -0.047 & 5.0 $ \pm $ 3.0 &  - & 7.9 & 82.5 $^{+0.0}_{-7.5}$ & (4.19 $^{+4.07}_{-2.73}) \cdot 10^{30}$ &  - & 190.0 & [1,2,3,4,171-172]\\ \noalign{\vskip 0.5mm}
\noalign{\vskip 1mm} 
\hline
\end{tabular}}
\tablefoot{All uncertanties are 1$\sigma$. All values of L$_{>\mathrm{100~TeV}}$ are derived from~\citet{LhaasoCollaboration2025}. The uncertainties on P$_{\mathrm{orb}}$ are only indicated if they are greater than $10^{-3}$P$_{\mathrm{orb}}$. The full machine-readable version of the table is available in \url{https://github.com/LauraOlivera/x-ray-binary-catalog}.\\
\textbf{References:} {\tiny [1]~\citet{Tetarenko2016}; [2]~\citet{CorralSantana2016}; [3-4]~\citet{Neumann2023,Avakyan2023}; [5]~\citet{Casares2023}; [6]~\citet{Saikia2022}; [7]~\citet{Bassi2019}; [8]~\citet{Hjellming1996}; [9]~\citet{Xie2020}; [10]~\citet{Hjellming1998}; [11]~\citet{2007MNRAS.378.1111B}; [12]~\citet{Paizis2007}; [13]~\citet{MataSanchez2022}; [14]~\citet{Wood2023}; [15]~\citet{Shidatsu2022}; [16]~\citet{Marti2017}; [17]~\citet{Mariani2025}; [18]~\citet{2016A&A...596A..46M}; [19]~\citet{Soria2011}; [20]~\citet{Torres2021}; [21]~\citet{Jonker2012}; [22]~\citet{Shaposhnikov2010}; [23]~\citet{Ratti2012}; [24]~\citet{Brocksopp2013}; [25]~\citet{Garcia2018}; [26]~\citet{Yang2011}; [27]~\citet{Marti2026}; [28]~\citet{CadolleBel2009}; [29]~\citet{Corbel2014}; [30]~\citet{MataSanchez2025}; [31]~\citet{Burridge2025}; [32]~\citet{Wood2024}; [33]~\citet{DelSanto2023}; [34]~\citet{2022MNRAS.513.6196R}; [35]~\citet{Russell2013}; [36]~\citet{Russell2015}; [37]~\citet{ArmasPadilla2019}; [38]~\citet{Russell2018}; [39]~\citet{Jana2021a}; [40]~\citet{Zhang2022}; [41]~\citet{CorralSantana2025}; [42]~\citet{King2012}; [43]~\citet{Saikia2023}; [44]~\citet{Williams2022}; [45]~\citet{Bahramian2023}; [46]~\citet{Atri2020}; [47]~\citet{Bright2025}; [48]~\citet{Poutanen2022}; [49]~\citet{Torres2020}; [50]~\citet{Mikolajewska2022}; [51]~\citet{Blundell2011}; [52]~\citet{Pandey2006}; [53]~\citet{Lutovinov2003}; [54]~\citet{Rushton2017}; [55]~\citet{Chaty2006}; [56]~\citet{Reid2023}; [57]~\citet{Fender1999}; [58]~\citet{Chaty2001}; [59]~\citet{Mirabel1994}; [60]~\citet{Tetarenko2018}; [61]~\citet{Zdziarski2014}; [62]~\citet{Fender2000}; [63]~\citet{2025A&A...696A.222M}; [64]~\citet{Motta2021}; [65]~\citet{CorralSantana2011}; [66]~\citet{Yamaoka2025}; [67]~\citet{Brocksopp2002}; [68]~\citet{Hjellming1988}; [69]~\citet{Hjellming1975}; [70]~\citet{Fender2023}; [71]~\citet{MillerJones2019}; [72]~\citet{Han1992}; [73]~\citet{Dzib2015}; [74]~\citet{Ribo2017}; [75]~\citet{Bartlett2013}; [76]~\citet{Clark2000}; [77]~\citet{Yadlapalli2021}; [78]~\citet{Yao2021}; [79]~\citet{Chatterjee2019}; [80]~\citet{Fender2001}; [81]~\citet{Shrader1994}; [82]~\citet{Fender2001a}; [83]~\citet{Harrison2007}; [84]~\citet{Grunsven2017}; [85]~\citet{Gallo2006}; [86]~\citet{Kuulkers1999}; [87]~\citet{Jana2021}; [88]~\citet{2022MNRAS.515.3105S}; [89]~\citet{2019ATel13275....1R}; [90]~\citet{Tetarenko2021}; [91]~\citet{Orosz2014}; [92]~\citet{Plotkin2021}; [93]~\citet{Ball1995}; [94]~\citet{Wu2016}; [95]~\citet{Coriat2012}; [96]~\citet{Chauhan2021}; [97]~\citet{Carotenuto2021}; [98]~\citet{Plotkin2021}; [99]~\citet{Brocksopp2001}; [100]~\citet{Koljonen2016}; [101]~\citet{YanesRizo2024}; [102]~\citet{Peng2024}; [103]~\citet{Carotenuto2024a}; [104]~\citet{Mondal2024}; [105]~\citet{2019ApJ...883..198R}; [106]~\citet{Shang2019}; [107]~\citet{MillerJones2011}; [108]~\citet{Rodriguez2003}; [109]~\citet{Hannikainen2001}; [110]~\citet{Kaaret2003}; [111]~\citet{2011ApJ...730...75O}; [112]~\citet{Hannikainen2009}; [113]~\citet{2017MNRAS.472..141M}; [114]~\citet{Wu2002}; [115]~\citet{Charles2019}; [116]~\citet{Russell2018a}; [117]~\citet{2015MNRAS.454.2199M}; [118]~\citet{Plotkin2016}; [119]~\citet{Zhang2025}; [120]~\citet{Curran2011}; [121]~\citet{Markwardt2008}; [122]~\citet{Rout2023}; [123]~\citet{Russell2019}; [124]~\citet{Monageng2021}; [125]~\citet{Zdziarski2026}; [126]~\citet{Corbel2004}; [127]~\citet{Zhang2026}; [128]~\citet{Kourmpetis2026}; [129]~\citet{Gallo2004}; [130]~\citet{Corbel2000}; [131]~\citet{Corbel2002}; [132]~\citet{Casella2010}; [133]~\citet{Zdziarski2019}; [134]~\citet{Tremou2026}; [135]~\citet{Calvelo2009}; [136]~\citet{Russell2018b}; [137]~\citet{Mondal2023}; [138]~\citet{Xu2018}; [139]~\citet{Hjellming1995}; [140]~\citet{Tingay1995}; [141]~\citet{Migliari2007}; [142]~\citet{Greene2001}; [143]~\citet{Motta2014}; [144]~\citet{Brocksopp2005}; [145]~\citet{Onori2021}; [146]~\citet{Rodriguez2011}; [147]~\citet{Rupen2005}; [148]~\citet{Ma2012}; [149]~\citet{Paizis2011}; [150]~\citet{Balakrishnan2023}; [151]~\citet{Heinke2023}; [152]~\citet{Brocksopp2005a}; [153]~\citet{Chaty2006a}; [154]~\citet{Russell2021}; [155]~\citet{ArmasPadilla2011}; [156]~\citet{Peng2023}; [157]~\citet{Kaaret2006}; [158]~\citet{Angelini2003}; [159]~\citet{Bozzo2016}; [160]~\citet{Zdziarski2016}; [161]~\citet{LuqueEscamilla2015}; [162]~\citet{Mirabel1992}; [163]~\citet{DelSanto2005}; [164]~\citet{Gallo2002}; [165]~\citet{2025arXiv250914465M}; [166]~\citet{Grollimund2025}; [167]~\citet{Paizis2015}; [168]~\citet{Sala2007}; [169]~\citet{Rupen2006}; [170]~\citet{Fender2001b}; [171]~\citet{Bower2005}; [172]~\citet{Porquet2005};}}
\end{table*}

\begin{figure*}[t]
    \centering
    \begin{subfigure}{0.33\textwidth}
        \centering
        \includegraphics[width=0.9\textwidth]{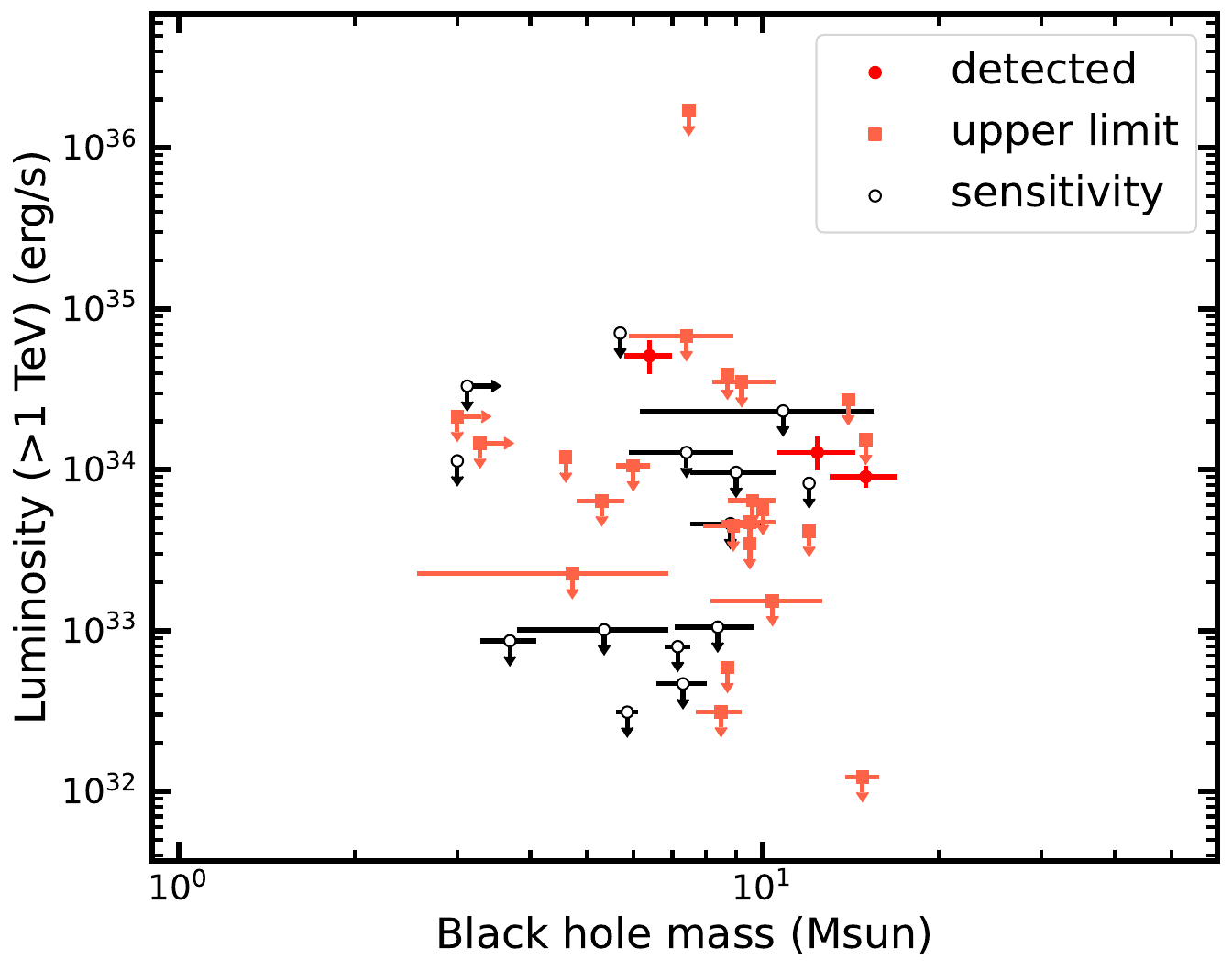} 
    \end{subfigure}
    \begin{subfigure}{0.33\textwidth}
        \centering
		\includegraphics[width=0.9\textwidth]{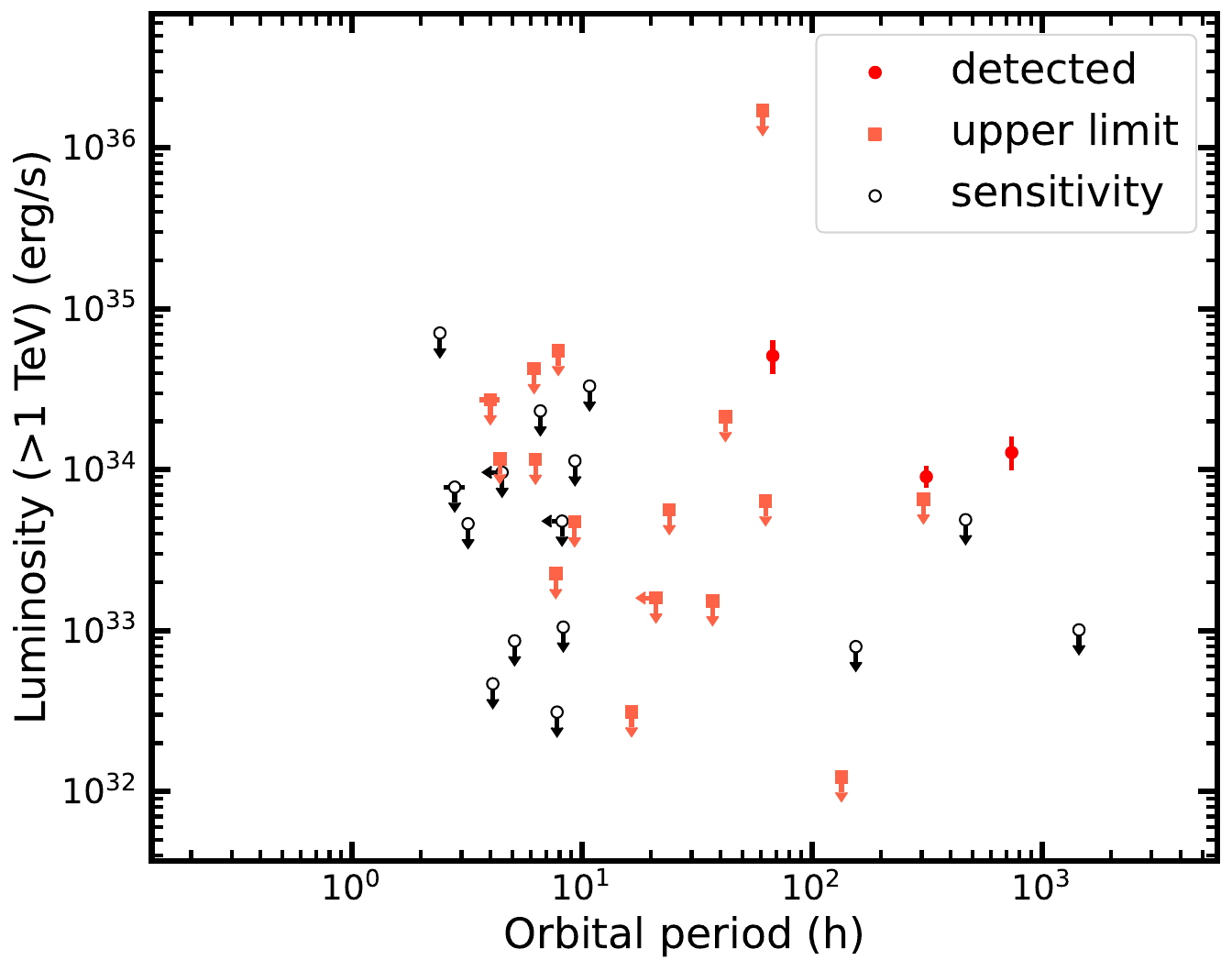} 
    \end{subfigure}
    \begin{subfigure}{0.33\textwidth}
        \centering
        \includegraphics[width=0.9\textwidth]{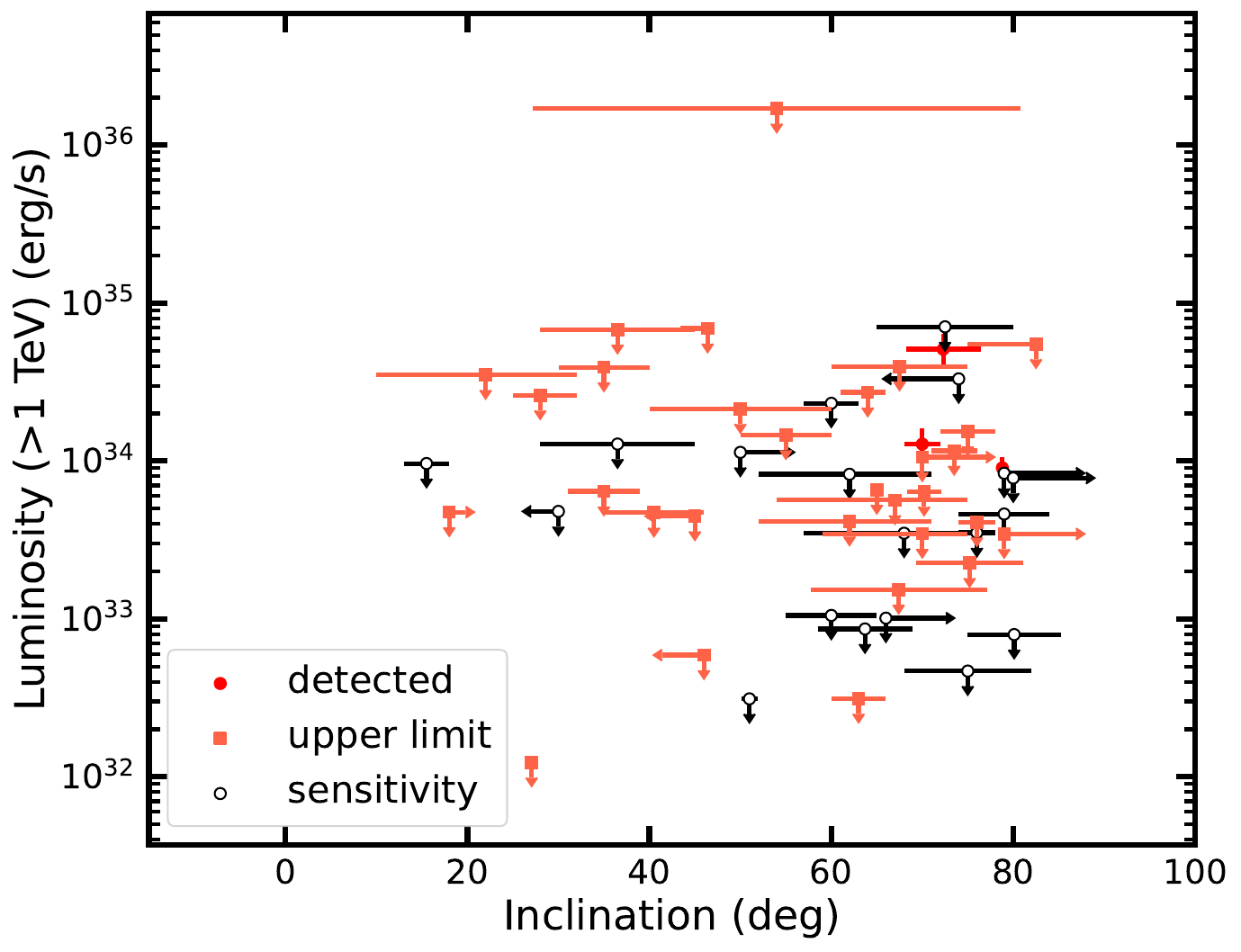} 
    \end{subfigure}
    \centering
    \begin{subfigure}{0.33\textwidth}
        \centering
        \includegraphics[width=0.9\textwidth]{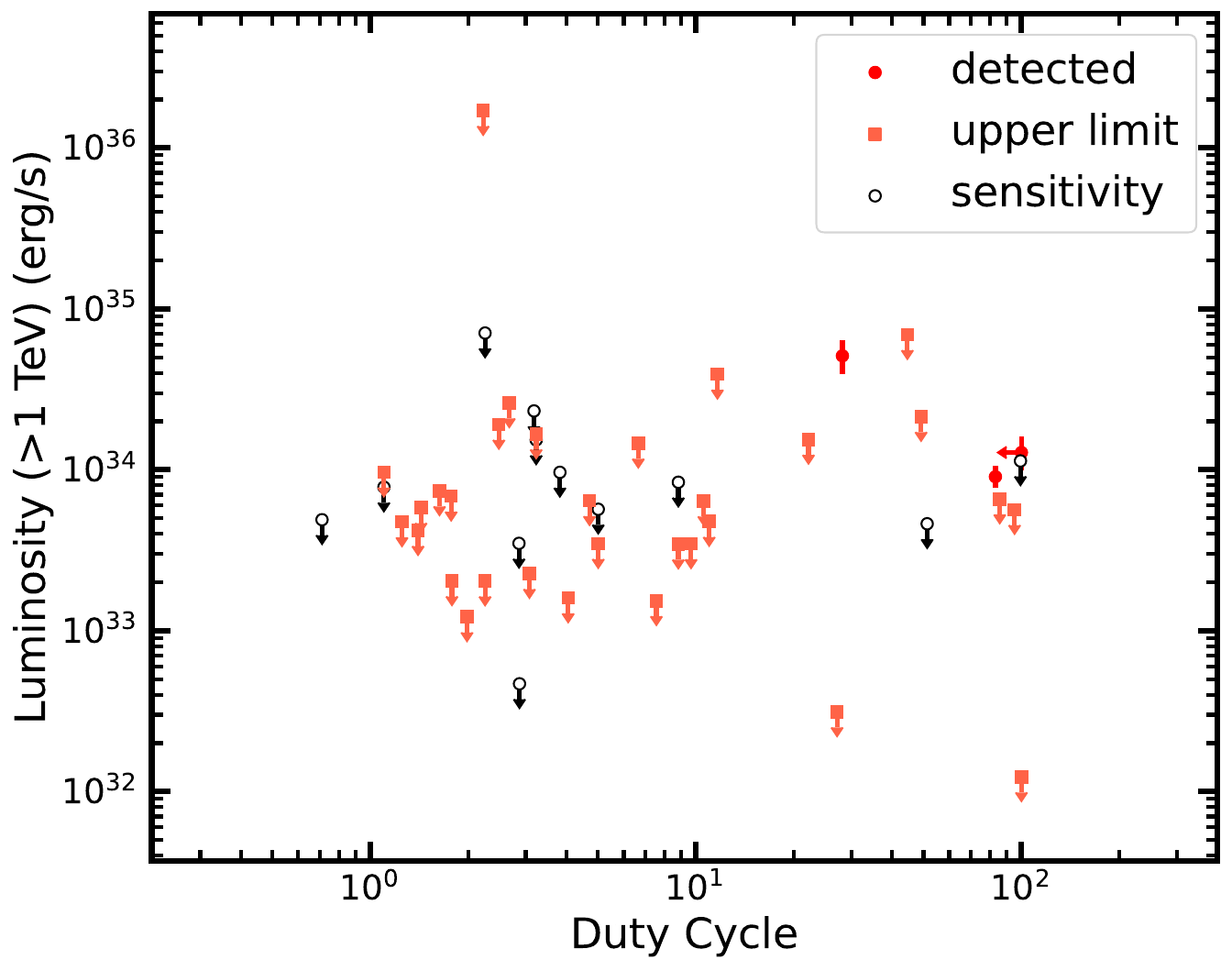} 
    \end{subfigure}
    \begin{subfigure}{0.33\textwidth}
        \centering
		\includegraphics[width=0.9\textwidth]{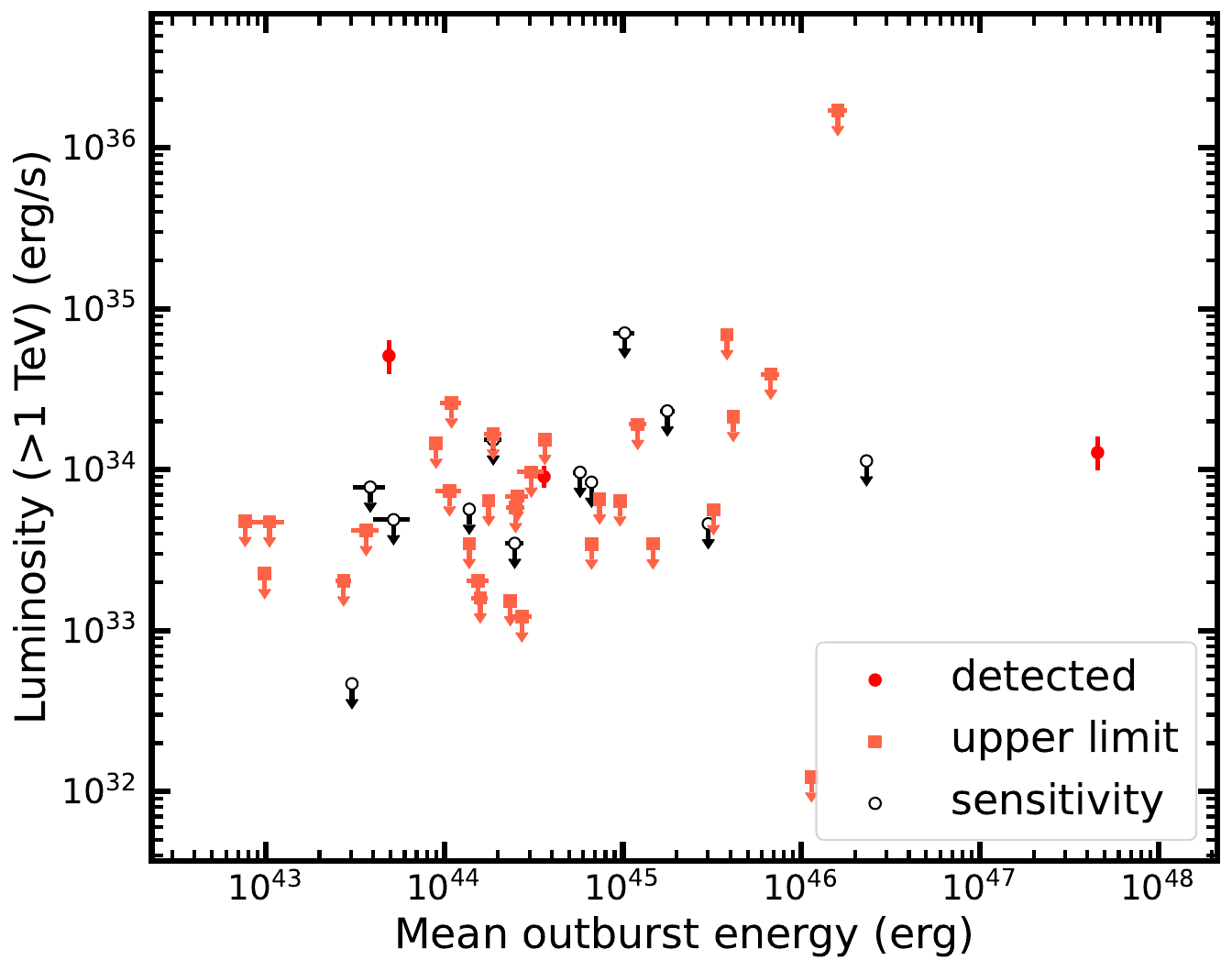} 
    \end{subfigure}
    \begin{subfigure}{0.33\textwidth}
        \centering
		\includegraphics[width=0.9\textwidth]{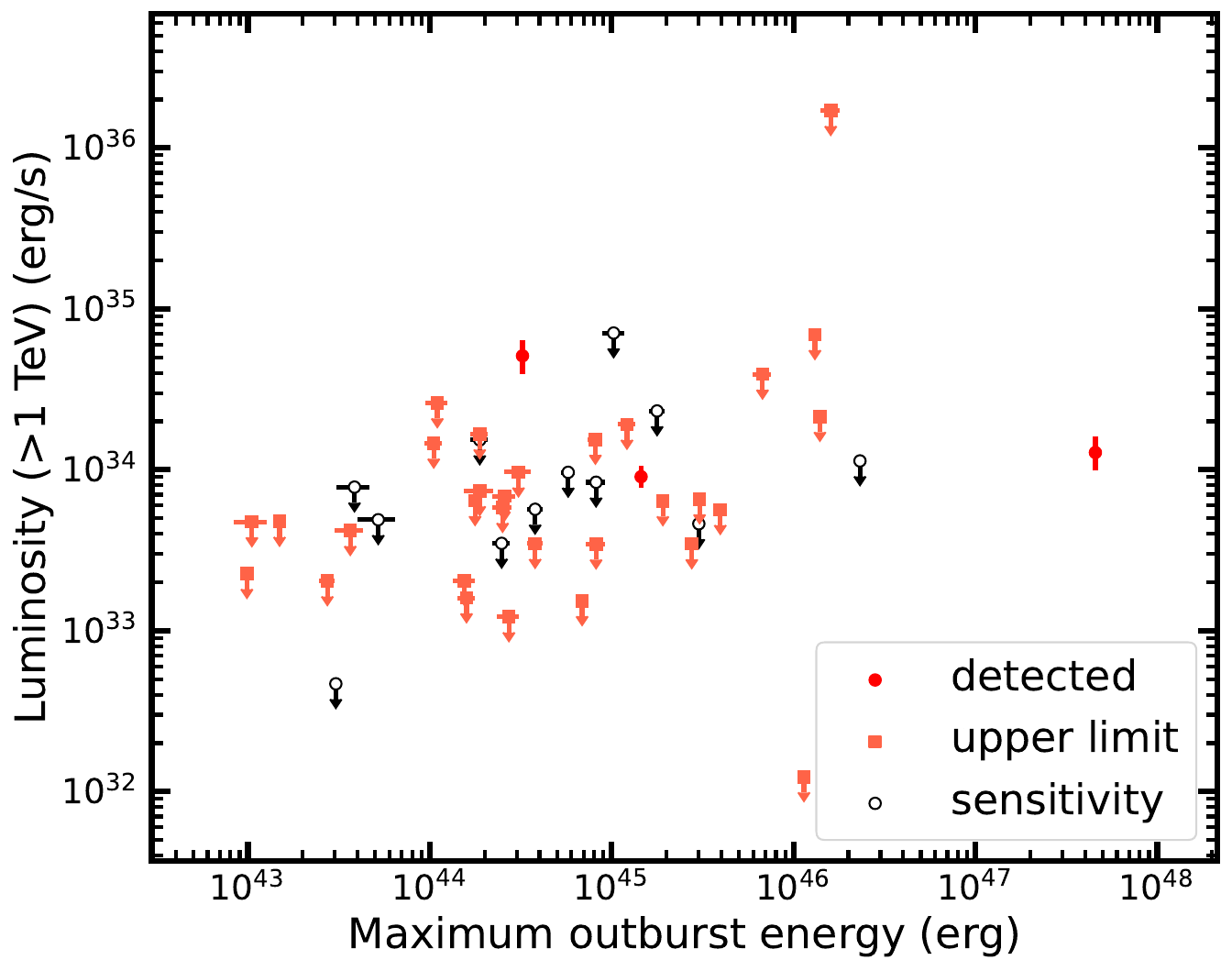} 
    \end{subfigure}
    \begin{subfigure}{0.33\textwidth}
        \centering
        \includegraphics[width=0.9\textwidth]{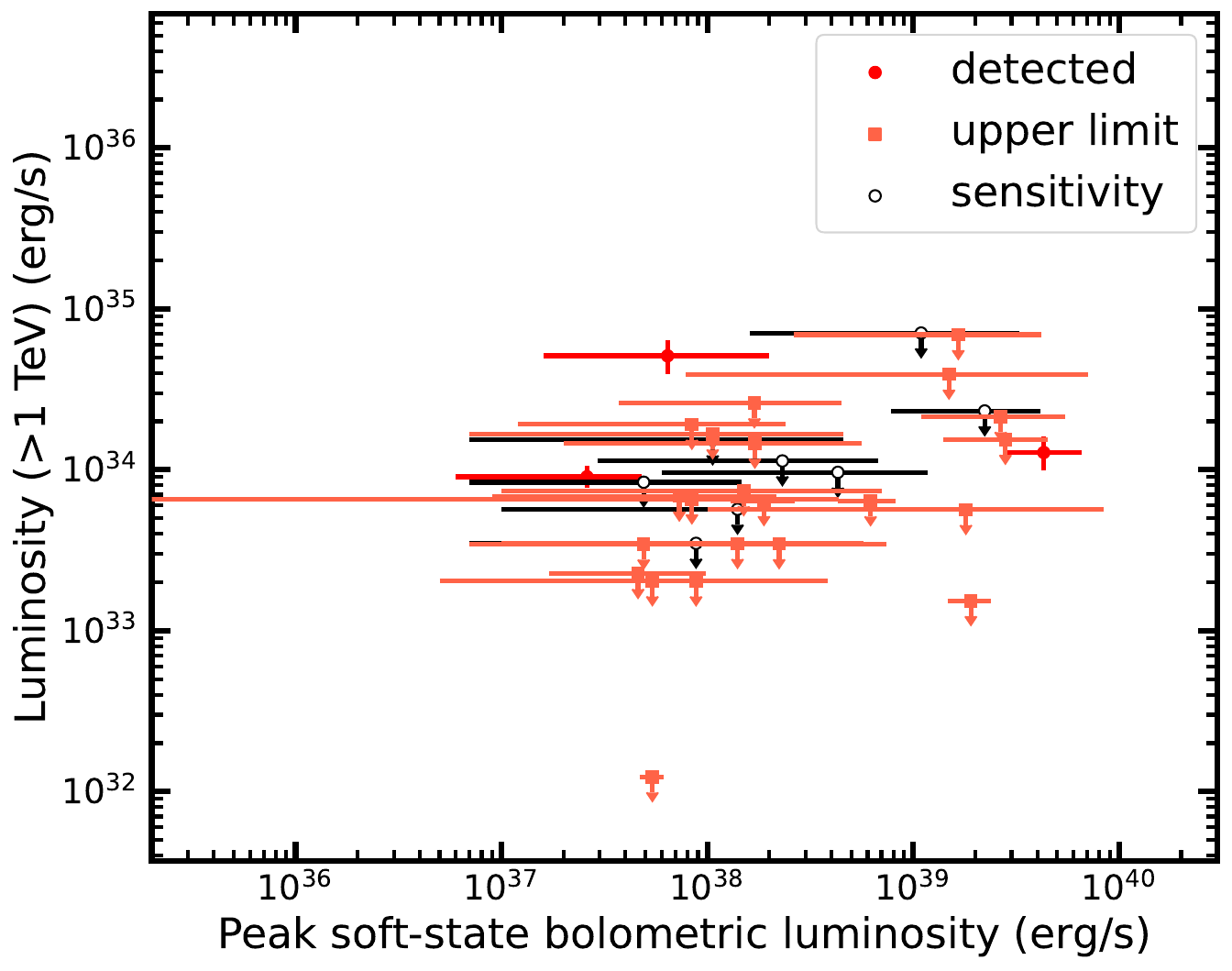} 
    \end{subfigure}
    \begin{subfigure}{0.33\textwidth}
        \centering
		\includegraphics[width=0.9\textwidth]{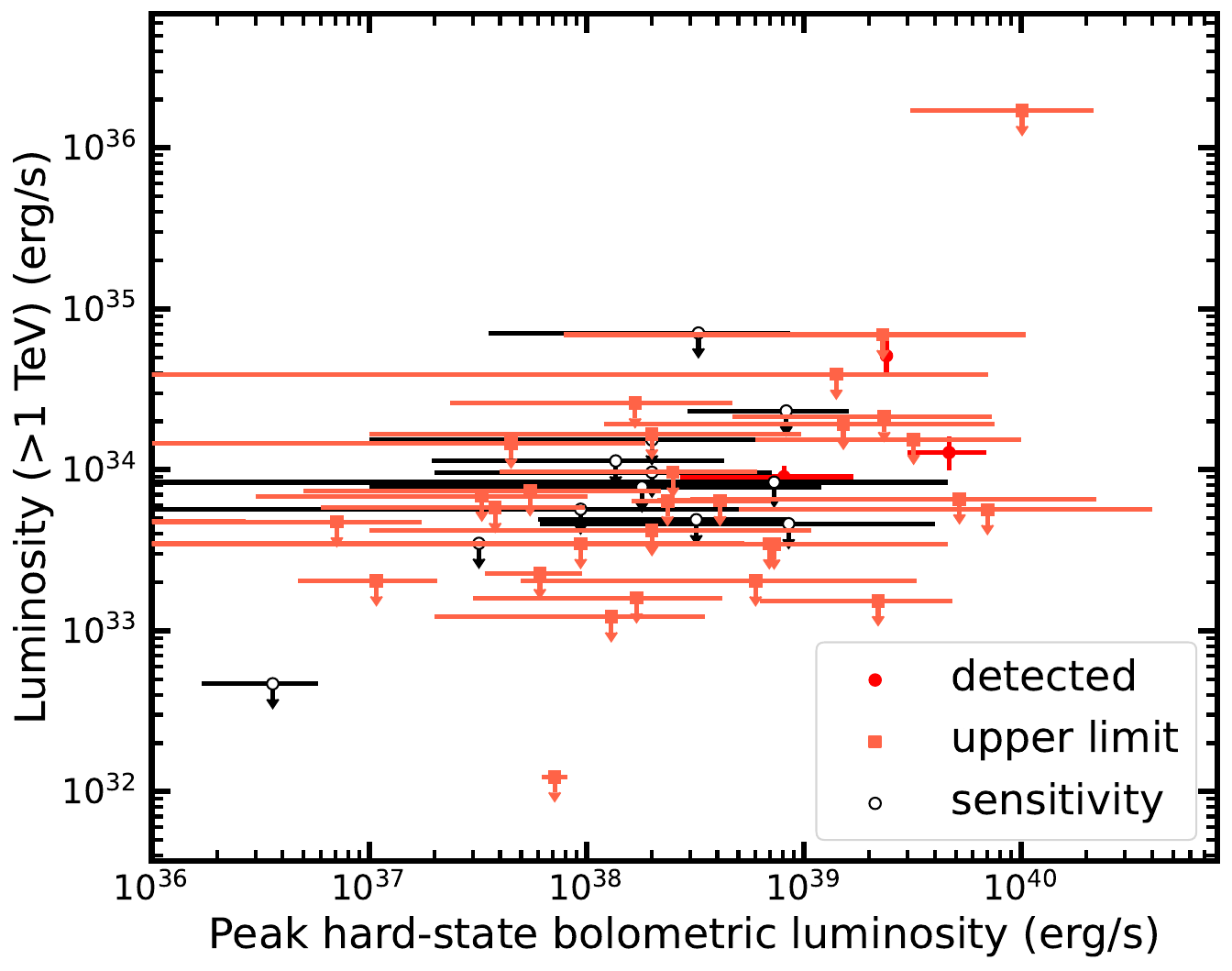} 
    \end{subfigure}
    \begin{subfigure}{0.33\textwidth}
        \centering
		\includegraphics[width=0.9\textwidth]{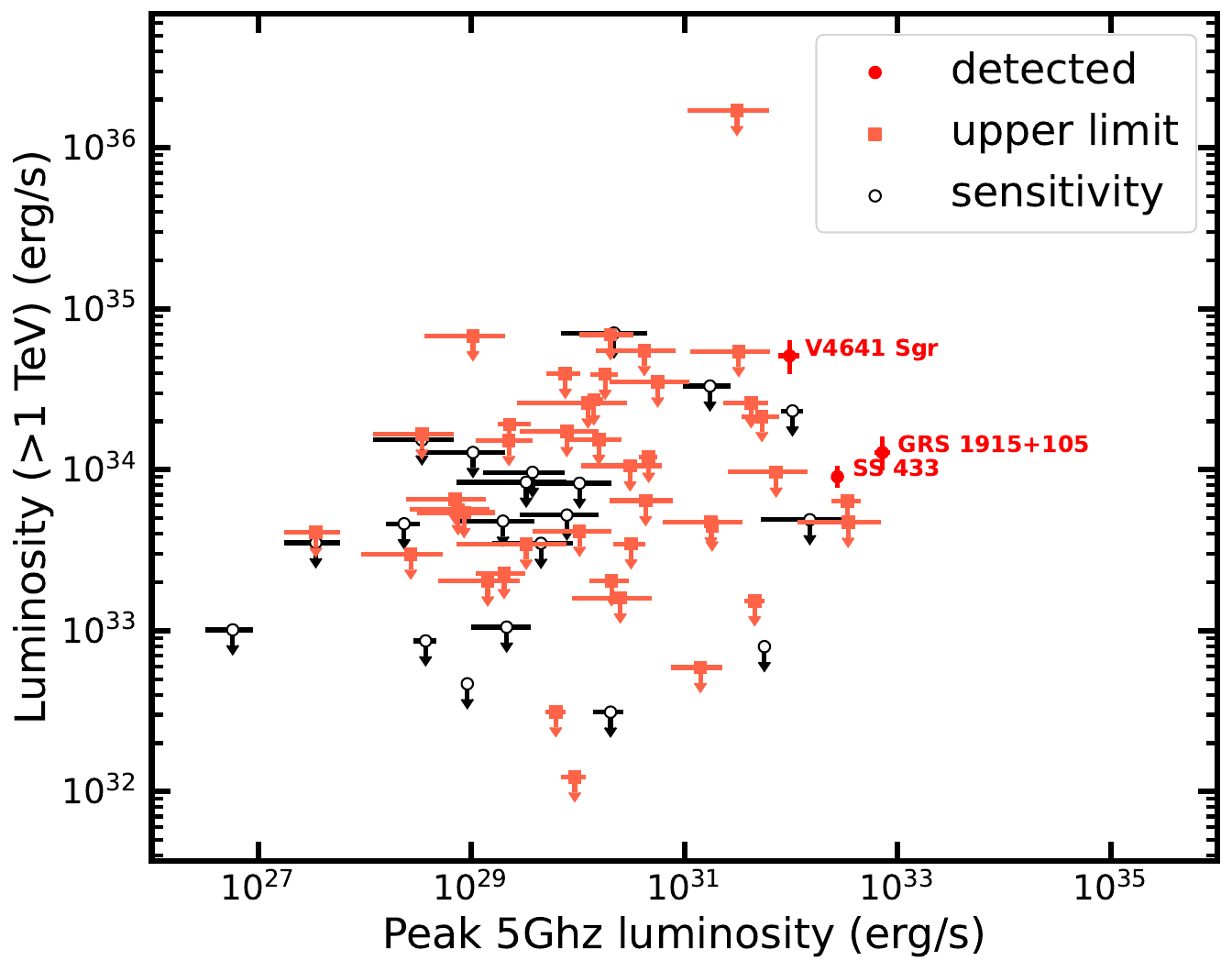} 
    \end{subfigure}
    \caption{\label{fig:1TeV} Same as Figures~\ref{fig:arch} and~\ref{fig:power} using the gamma-ray luminosity above 1~TeV.}
\end{figure*}

\begin{figure*}[h]
	\centering
		\includegraphics[width=0.95\linewidth]{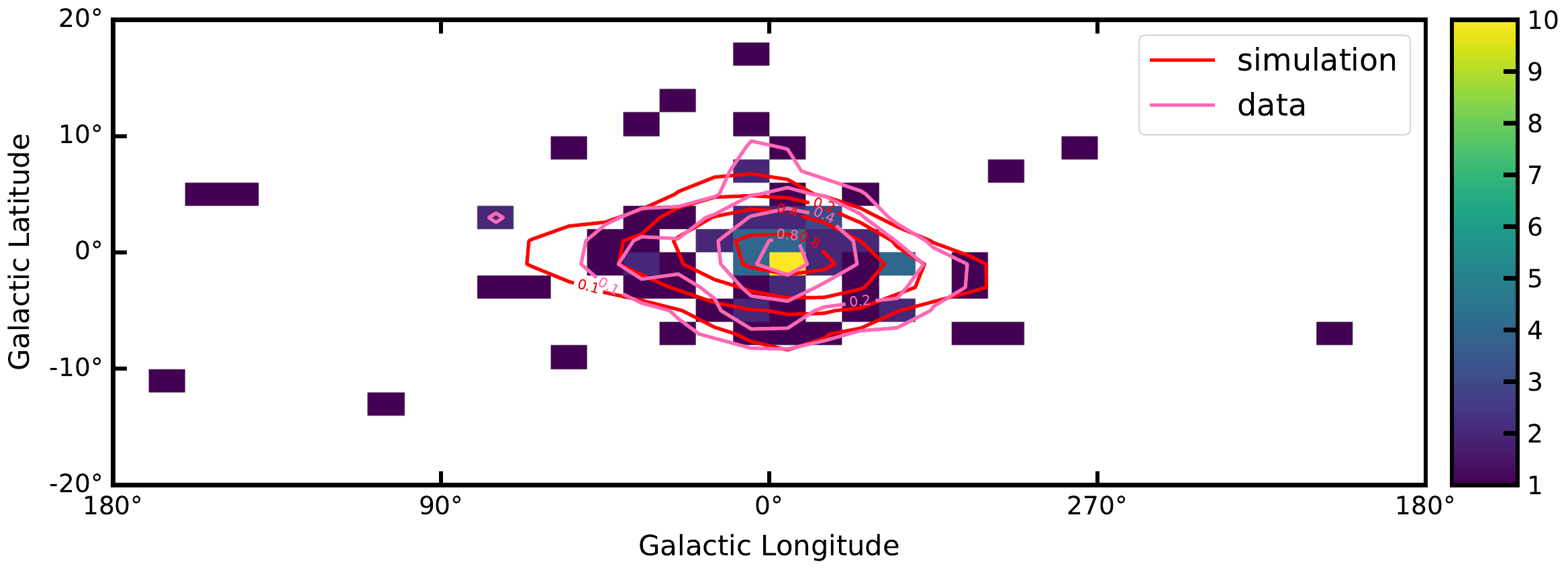} 
	\caption{\label{fig:sim2d} Real and synthetic distribution of \xray binaries as used in Section~\ref{sec:the_cat}. The 2D histogram shows the Galactic distribution of \xray binaries in the sample. The red lines show the contours for the synthetic distribution. The pink lines are the contours for the real distribution, smoothed with a 5$\degree$ Gaussian kernel.}
\end{figure*}
\end{appendix}
\end{document}